\documentclass[11pt]{article}
\usepackage{bm}
\usepackage[font=small]{caption}
\RequirePackage{amsthm,amsmath,amsfonts,amssymb}
\usepackage{omega}
\usepackage[a4paper,left=2.5cm,right=2.5cm,top=3cm,bottom=3cm]{geometry}

\usepackage{bm}
\usepackage{natbib}
\usepackage[ruled]{algorithm2e}
\usepackage[dvipsnames]{xcolor}
\usepackage[colorlinks = true,
            linkcolor = NavyBlue,
            urlcolor  = NavyBlue,
            citecolor = NavyBlue,
            anchorcolor = NavyBlue]{hyperref}

\usepackage{url}
\usepackage[graphicx]{realboxes}
\usepackage{enumerate}
\usepackage{comment}
\usepackage{subfigure}
\usepackage[title]{appendix}
\usepackage{chngcntr}
\RequirePackage{subfigure}
\RequirePackage{algorithmic}

\author{}

\numberwithin{equation}{section}

\theoremstyle{plain}
\newtheorem{remark1}{Remark}
\newtheorem{theorem1}{Theorem}
\newtheorem{lemma1}{Lemma}

\usepackage{setspace}

\newcommand{\ee}{\end{aligned} \end{equation}}
\newcommand{\eq}{\end{quote}}

\newcommand{\ep}{\end{parts}}

\newcommand{\bqp}{\begin{quote}\begin{parts}}

\newcommand{\epq}{\end{parts}\end{quote}}

\DeclareMathOperator*{\argmin}{argmin}
\newcommand{\Rom}[1]{\text{\uppercase\expandafter{\romannumeral #1\relax}}}
\newcommand{\bee}{\begin{equation}\begin{aligned}}

\newcommand{\emm}{\end{bmatrix}}

\newcommand{\argmax}{\operatornamewithlimits{argmax}}
\numberwithin{equation}{section}
\newcommand{\vertiii}[1]{{\vert\kern-0.25ex\vert\kern-0.25ex\vert #1 
    \vert\kern-0.25ex\vert\kern-0.25ex\vert}}
    
\newcommand\ma{\mathbf{A}}
\newcommand\mb{\mathbf{B}}

\newcommand\me{\mathbf{E}}
\newcommand\mf{\mathbf{F}}
\newcommand\mh{\mathbf{H}}
\newcommand\mi{\mathbf{I}}

\newcommand\mr{\mathbf{R}}

\newcommand\mx{\mathbf{X}}

\newcommand\mz{\mathbf{Z}}
\newcommand\mw{\mathbf{W}}

\newcommand\mm{\mathbf{M}}

\newcommand\mt{\mathbf{T}}

\newcommand\mpp{\mathbf{P}}
\newcommand\mo{\mathbf{O}}
\newcommand\my{\mathbf{Y}}
\newcommand\muu{\mathbf{U}}

\newcommand\mLambda{\bm{\Lambda}}

\newcommand{\bea}{\begin{eqnarray}}
	\newcommand{\eea}{\end{eqnarray}}
\newcommand{\beas}{\begin{eqnarray*}}
	\newcommand{\eeas}{\end{eqnarray*}}

\title{Detection and estimation of vertex-wise latent position shifts across networks}
\begin{document}

\author{Runbing Zheng
\thanks{Department of Applied Mathematics and Statistics, Johns Hopkins University}
}

\date{}

\maketitle

\begin{abstract}
Pairwise network comparison is essential for various applications, including neuroscience, disease research, and dynamic network analysis. While existing literature primarily focuses on comparing entire network structures, we address a vertex-wise comparison problem where two random networks share the same set of vertices but allow for structural variations in some vertices, enabling a more detailed and flexible analysis of network differences. In our framework, some vertices retain their latent positions between networks, while others undergo shifts. 
To identify the shifted and unshifted vertices and estimate their latent position shifts, we propose a method that first derives vertex embeddings in a low-rank Euclidean space for each network, then aligns these estimated vertex latent positions into a common space to resolve potential non-identifiability, and finally tests whether each vertex is shifted or not and estimates the vertex shifts.
Our theoretical results establish the test statistic for the algorithms, guide parameter selection, and provide performance guarantees. 
Simulation studies and real data applications, including a case-control study in disease research and dynamic network analysis, demonstrate that the proposed algorithms are both computationally efficient and effective in extracting meaningful insights from network comparisons.

\end{abstract}

\noindent%
\vfill

\vfill

\begin{sloppypar}

\counterwithout{figure}{section}
\counterwithout{theorem}{section}
\counterwithout{definition}{section}

\theoremstyle{definition}
\newtheorem{rema}{Remark}

\section{Introduction}

Pairwise network comparison is of both theoretical and practical importance, with applications across various fields including neuroscience \citep{zalesky2010network,van2010comparing,tang2017semiparametric}, 
case-control studies in disease research \citep{segerstolpe2016single}, and dynamic network analysis \citep{liu2018global}.
For instance, comparing the brain networks of individuals in case and control groups for a particular disease can help identify neurological characteristics associated with the disease and provide insights into the underlying mechanisms.
For dynamic network analysis and its related challenges, such as understanding or visualizing its overall evolution patterns and detecting change points, we frequently need to study the similarity between two networks constructed using disjoint time intervals. 
For example, when analyzing the trading network of a specific product among different countries over several years, comparing the trading networks for each pair of years enables the study of the product's trading network evolution over time.

Existing literature on pairwise network comparison primarily focuses on testing whether the \textit{entire} underlying network structures of two independent networks on the same set of vertices are equivalent.
\cite{ghoshdastidar2017two} considers two networks generated from two distributions, and provide a framework for formulating the two-sample hypothesis testing problem to determine whether the generating distributions of two networks are sufficiently similar with minimal structural assumptions. 
\cite{ghoshdastidar2020two} studies two inhomogeneous random networks with population Bernoulli probability matrices $\mpp^{(1)}$ and $\mpp^{(2)}$, where multiple independent and identically distributed networks are available for each $\mpp^{(i)}$, and analyze the minimax separation rate $\rho^\star$ for the threshold $\rho$ in testing $\mpp^{(1)} = \mpp^{(2)}$ against $d(\mpp^{(1)},\mpp^{(2)}) > \rho$ for some common choices of the distance function $d(\cdot, \cdot)$.
\cite{tang2017nonparametric,tang2017semiparametric} consider the problem of comparing two random dot product graphs (RDPGs) \citep{athreya2018statistical,young2007random} with probability matrices $\mpp^{(1)} = \mx^{(1)}\mx^{(1)\top}$ and $\mpp^{(2)} = \mx^{(2)}\mx^{(2)\top}$, where the $k$th row of $\mx^{(i)}$, $\mathbf{x}^{(i)}_k$, represents the latent position of vertex $k$ in network $i$. 
\cite{tang2017nonparametric} supposes that the latent positions of all vertices in each network are i.i.d. following distributions $F^{(1)}$ and $F^{(2)}$, respectively, i.e., $\{\mathbf{x}^{(1)}_k\}_k \overset{\text{i.i.d.}}{\sim} F^{(1)}$ and $\{\mathbf{x}^{(2)}_k\}_k \overset{\text{i.i.d.}}{\sim} F^{(2)}$, and propose a hypothesis test to determine whether $F^{(1)} = F^{(2)} \circ U$ for some unitary operator $U$, which is equivalent to testing whether the two RDPGs are identically distributed.
\cite{tang2017semiparametric} studies a more general problem by relaxing the i.i.d. assumption for vertices and testing whether the latent position matrices $\mx^{(1)}$ and $\mx^{(2)}$ are the same up to an orthogonal transformation, i.e., $\mx^{(1)} = \mx^{(2)}\mw^{(1,2)}$ for some orthogonal $\mw^{(1,2)}$, which is equivalent to testing whether the probability matrices $\mpp^{(1)}$ and $\mpp^{(2)}$ are identical. 
\cite{li2018two} considers two networks generated from weighted stochastic block models and propose a hypothesis test to assess whether the community assignments of all vertices in the two networks are the same.
\cite{jin2024optimal} studies networks generated from the degree-corrected mixed-membership model allowing degree heterogeneity and mixed memberships, and propose the interlacing balance measure to test whether the underlying Bernoulli probability matrices $\mpp^{(1)}$ and $\mpp^{(2)}$ of the two networks are the same.

There are also existing studies on the comparison of multiple networks.
Existing multiple network models typically assume that networks share common structures, but these common structures are usually \textit{not} vertex-wise.
Consider $m$ networks with observed adjacency matrices $\{\ma^{(i)}\}_{i=1}^m$ and underlying probability matrices $\{\mpp^{(i)}\}_{i=1}^m$.
For networks with $\mpp^{(i)}\equiv \mx\mx^{\top}$, \cite{levin2017central} proposes the omnibus embedding estimation, in which multiple graphs are jointly embedded by $\tilde{\mathbf{A}}$ with $\tilde{\mathbf{A}}_{i,j}=(\ma^{(i)}+\ma^{(j)})/2$,  and  apply it to testing whether $\{\mx^{(i)}\}_{i=1}^m$ are all equivalent for networks with $\mpp^{(i)}=\mx^{(i)}\mx^{(i)\top}$.
\cite{jones2020multilayer,gallagher2021spectral} assume the networks share the left subspace with $\mpp^{(i)}=\mx\mr^{(i)}\my^{(i)}$, where $\mr^{(i)}$ are low-dimensional matrices, and propose parameter estimation using spectral embedding of the row concatenation $[\ma^{(1)}|\dots|\ma^{(m)}]$.
The multilayer stochastic block model \citep{holland1983stochastic, han2015consistent, paul2020spectral, lei2023bias, lei2024computational} assumes that vertices share common community assignments across different layers while allowing for layer-specific block probabilities.
Multilayer eigenscaling models \citep{nielsen2018multiple, wang2019joint, draves2020bias, weylandt2022multivariate} assume that the networks share common subspaces and constrain the low-dimensional heterogeneous structure to be diagonal, i.e., $\mpp^{(i)}=\muu\mr^{(i)}\muu^\top$, where $\muu$ represents the common subspace across networks and the heterogeneous matrices $\mr^{(i)}$ are diagonal, while the common subspace independent edge model \citep{arroyo2021inference} also assumes that networks share common subspaces but imposes no specific constraint on $\mr^{(i)}$.
The MultiNeSS model \citep{macdonald2022latent} assumes $\mpp^{(i)}=\mx^{(i)}\mi^\star\mx^{(i)\top}$, where $\mi^\star$ is a diagonal matrix with $+1$s and $-1$s on the diagonal, and assumes the latent position matrices can be written as $\mx^{(i)}=[\mx_c|\mx_s^{(i)}]$, where $\mx_c$ gives a matrix of common latent position coordinates and $\mx_s^{(i)}$ are individual latent position coordinates.

Other problems related to pairwise network comparison include change point detection in dynamic networks and the graph matching problem.
Dynamic network change point detection typically assumes a segmented time structure, where network evolution is stable within each segment but undergoes abrupt changes at segment boundaries, and such changes typically affect the \textit{entire} probability matrix.
\cite{wang2021optimal} assumes that the probability matrix remains unchanged within each segment and undergoes changes only at change points.
\cite{bhattacharjee2020change} studies dynamic stochastic block models and assumes that community assignments and block probabilities change only at change points.
\cite{marenco2021online,marenco2022online,larroca2021change,padilla2022change} study dynamic RDPGs with different modeling assumptions. 
\cite{marenco2021online,marenco2022online} assume vertex latent positions change only at change points.
\cite{larroca2021change} assumes that latent positions are regenerated from new distributions at change points, while maintaining the same underlying distribution within each segment (without necessarily being re-drawn).
\cite{padilla2022change} also assumes that different segments have different latent position distributions, and specifically models temporal dependence within segments: vertex latent positions either remain unchanged from the previous time step with probability $\rho$ or are independently resampled from the segment-specific distribution with probability $1-\rho$. Notably, while this model assumes vertex-wise changes in latent positions within segments, the updated latent positions are constrained to follow the same segment-specific distribution, and the change point detection framework primarily focuses on identifying the locations of change points, without providing estimates for which vertices within segments have changed and how they have changed.
In the aforementioned problems, networks are defined on the same vertex set with known vertex correspondences across networks, while the graph matching problem assumes unknown vertex correspondences and seeks to recover the correct alignment under which the \textit{entire} network structures are identical or highly similar.
For instance, the graph isomorphism \cite{fortin1996graph} assumes that two networks have identical adjacency matrices under the correct vertex permutation.
\cite{qi2025asymptotically} assumes identical probability matrices under proper alignment, which in the RDPG framework implies that corresponding vertices share the same latent positions across networks.
\cite{sussman2019matched} considers networks of different sizes, where the smaller network corresponds to a subgraph of the larger one under correct vertex alignment, with their probability matrices exhibiting this inclusion relationship.

In this paper, instead of comparing the \textit{entire} underlying structures of  networks, we conduct a \textit{vertex-wise} comparison between two random networks on the same set of vertices while allowing for structural variations in some vertices.
Specifically we consider two independent RDPGs on the same set of vertices, and suppose that some vertices remain latent positions while others experience latent position shifts.
The subset of unshifted vertices, denoted as $\mathcal{U}$, retains the same latent positions between the two networks (up to an orthogonal transformation considering the orthogonal nonidentifiability in RDPGs), i.e., $\mx^{(1)}_{\mathcal{U}} = \mx^{(2)}_{\mathcal{U}}\mw^{(1,2)}$ for some orthogonal $\mw^{(1,2)}$.
The remaining vertices in $\mathcal{U}^\mathcal{C}$ undergo shifts in their latent positions, given by $\mx^{(1)}_{\mathcal{U}^\mathcal{C}} = \mx^{(2)}_{\mathcal{U}^\mathcal{C}}\mw^{(1,2)} + \my_{\mathcal{U}^\mathcal{C}}$ where the shift matrix $\my_{\mathcal{U}^\mathcal{C}}$ has no zero rows. Notice that the two networks preserve the pairwise connection probabilities within $\mathcal{U}$, i.e., $\mpp^{(1)}_{\mathcal{U},\mathcal{U}} = \mpp^{(2)}_{\mathcal{U},\mathcal{U}}$, and the probabilities outside this submatrix typically change.
Given the observed adjacency matrices of the pair of networks, we aim to (1) identify the unshifted and shifted vertices, i.e., determine $\mathcal{U}$, and (2) estimate the latent position shifts in $\my_{\mathcal{U}^\mathcal{C}}$. 
To the best of our knowledge, this model allowing for partial vertex-wise latent position shifts is novel.

\begin{figure}[htbp!]
\centering
\subfigure
{\includegraphics[height=10cm]{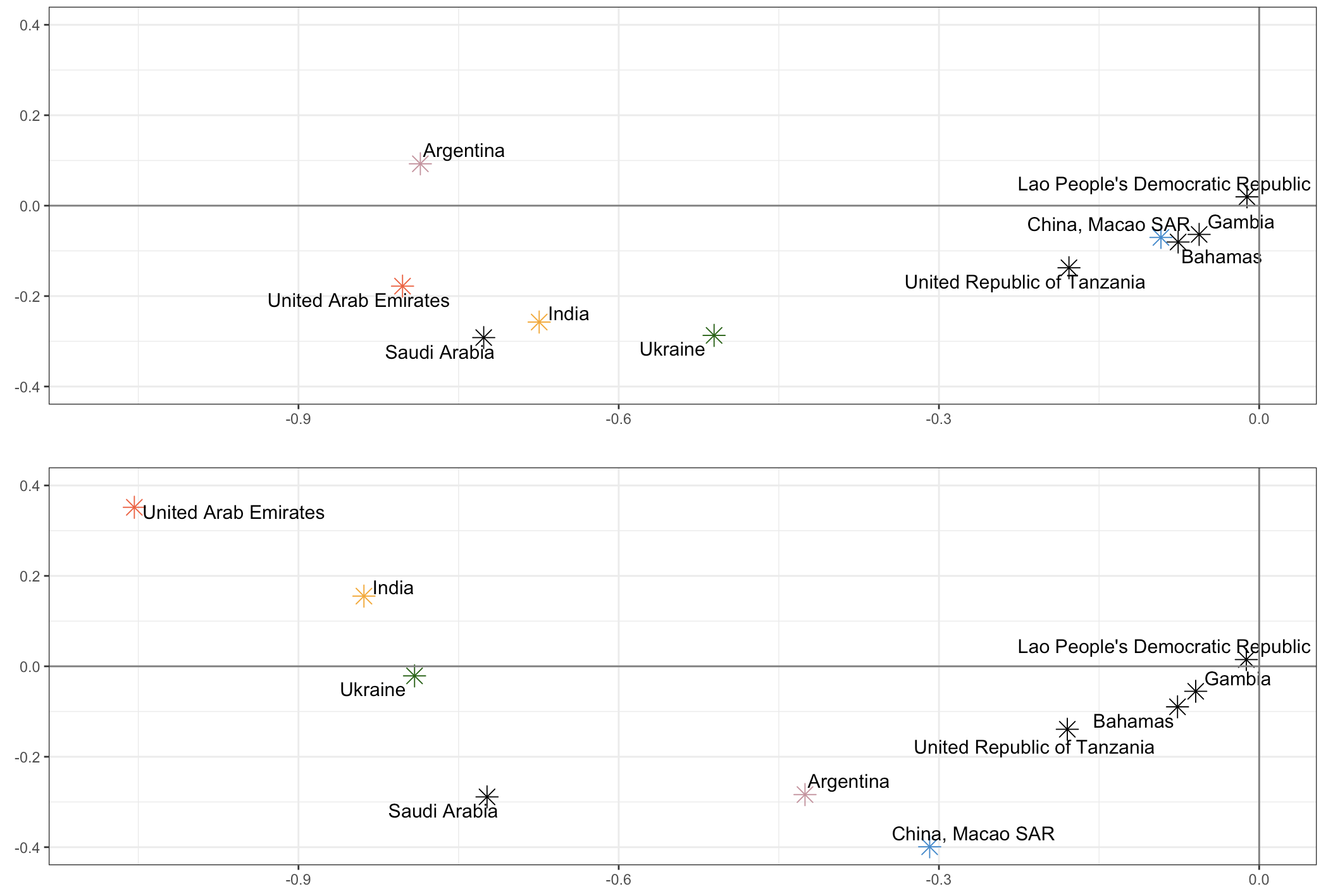}}
\caption{\footnotesize Estimated latent positions of $10$ countries in the chocolate-trading networks in year $2010$ (upper panel) and year $2022$ (lower panel), with 5 countries (black) having non-significant latent position shifts and 5 countries (colored) having significant latent position shifts.
}
\label{fig:food_country 15}
\end{figure}

A motivating example provides real-data evidence for the presence of both shifted and unshifted vertices in network comparisons. We compare the estimated vertex latent positions of two chocolate trading networks from the years $2010$ and $2022$ (with further details provided in Section~\ref{sec:chocolatenetwork}). Figure~\ref{fig:food_country 15} presents the $2$-dimensional latent position estimates of $10$ countries, where black stars represent the $5$ countries with non-significant shifts, and colored stars represent the $5$ countries exhibiting significant shifts. This real-world example naturally motivates our problem formulation.

Another motivating example for our model arises from stochastic block models (SBMs) \citep{holland1983stochastic} with block assignment switching. More specifically, we considers SBMs that share the same block probability matrix but allow some vertices to switch their block assignments between the two networks. In this case identifying the vertices with shifted latent positions is equivalent to detecting those that have changed their block assignments (see Section~\ref{sec:simu_noseeds} for details on the SBM setting and related experiments).

The vertex-wise comparison between networks can have significant practical implications in various real-world problems. For instance, in disease research if we only conduct an entire-network-level comparison between the brain networks of individuals in the case and control groups, we can merely determine whether the disease is generally associated with the overall brain network structure. But by performing a vertex-wise comparison, we can identify specific brain regions associated with the disease, which is crucial for more detailed analysis and understanding of its mechanisms.
Another example is in trading networks. As shown in Figure~\ref{fig:food_country 15}, a vertex-wise comparison allows us to identify specific countries whose trading patterns have changed between the two years. Consequently, when considering a dynamic trading network over a sequence of years, we can analyze how a specific country's trading pattern evolves over time. The vertex-wise comparison model thus provides a new perspective for studying the evolution of dynamic networks by tracking structural changes at the level of individual vertex shifts.
 

Based on existing entire-network comparison methods \citep{ghoshdastidar2017two,ghoshdastidar2020two,tang2017nonparametric,tang2017semiparametric,li2018two,jin2024optimal}, a natural idea to identifying the unshifted vertices would be to examine all possible subsets of vertices and apply an existing algorithm to test whether the corresponding subnetworks share the same underlying structure between the two networks. However, this approach incurs a computational cost of approximately  $2^n$ times the complexity of the original algorithm, rendering it infeasible for large networks.
Thus, there is a need to develop specialized and efficient algorithms tailored to the vertex-wise network comparison problem.
We first consider a simple scenario where a (possibly very limited) subset of unshifted vertices, termed the seed set, is known, and propose a seed-based algorithm that aligns the estimated latent positions between the two networks according to the seed set to identify the entire set of unshifted vertices $\mathcal{U}$ and estimate the latent position shifts in $\my_{\mathcal{U}^\mathcal{C}}$. Building on the seed-based algorithm, we further develop a seed-free algorithm that does not rely on any prior knowledge of unshifted vertices, making it more practical for real-world applications. The key idea is to randomly generate some small subsets of vertices as candidate seed sets, apply the seed-based algorithm to each candidate, and ultimately select the candidate seed set that achieves the best alignment, i.e., yielding the largest number of unshifted vertices, as the most likely correct seed set.

Our theoretical results derive the test statistic used in the algorithms, guide the choice of parameters, and provide theoretical guarantees for the performance of the algorithms.
In particular, we derive an error bound in the two-to-infinity norm for the estimated shift matrix, and using this error bound, we show that for each vertex, the estimated shift vector is approximately normally distributed around the true shift vector. 
Leveraging the normal approximation result, we construct a test statistic for each vertex, and show that the limiting distribution of this test statistic converges to a central $\chi^2$ distribution under the null hypothesis of no shift and to a non-central $\chi^2$ distribution under a local alternative hypothesis where a shift exists. The test statistic is used in our proposed algorithms to identify unshifted and shifted vertices. 
The error bound result also shows that the algorithms achieve consistent estimation even with only a minimal number of seed vertices, providing insights into parameter choice for the algorithms to achieve strong performance while maintaining computational efficiency.

The structure of our paper is as follows. In Section~\ref{sec:model} we introduce the model for a pair of independent RDPGs with partially shifted vertices. 
In Section~\ref{sec:alg}, we first propose the seed-based algorithm and then build upon it to develop the seed-free algorithm. We also analyze the computational complexity of the algorithms.
Theoretical results for our algorithms are presented in Section~\ref{sec:thm}. 
We have numerical simulations in Section~\ref{sec:sim}, and have real data experiments, including a brain network comparison study and an analysis of the dynamic chocolate trading network, in Section~\ref{sec:real}.
In the supplementary material, we integrate vertex-wise comparison with the mirror idea from \cite{athreya2024discovering} to analyze dynamic network evolution patterns, and extend vertex-wise comparison from RDPGs to generalized RDPGs (GRDPGs) \citep{rubin2017statistical,young2007random} where the probability matrices take the more flexible form $\mpp^{(i)} = \mx^{(i)}\mathbf{I}_{d_+,d_-}\mx^{(i)\top}$ with diagonal matrix $\mathbf{I}_{d_+,d_-}$ containing $d_+$ entries of $+1$ and $d_-$ entries of $-1$, and to networks of different ranks. We also provide additional experimental results and detailed proofs of all stated results in the supplementary material.

\subsection{Notations}
\label{sec:notations}

We summarize some notations used in this paper. 
For any positive integer $n$, we denote by $[n]$ the set $\{1,2,\dots, n\}$. 
For two non-negative sequences
$\{a_n\}_{n \geq 1}$ and $\{b_n\}_{n \geq 1}$, we write $a_n \lesssim
b_n$ (resp. $a_n \gtrsim b_n$) if there exists some constant $C>0$
such that $a_n \leq C b_n$ (resp. $a_n \geq C b_n$) for all  $n \geq 1$, and we write $a_n \asymp b_n$ if $a_n\lesssim b_n$ and $a_n\gtrsim b_n$.
The notation $a_n \ll b_n$ (resp. $a_n \gg b_n$) means that there exists some sufficiently small (resp. large) constant $C>0$ such that $a_n \leq Cb_n$ (resp. $a_n \geq Cb_n$).
If $a_n/b_n$ stays bounded away from $+\infty$, we write $a_n=O(b_n)$ and $b_n=\Omega(a_n)$, and we use the notation $a_n=\Theta(b_n)$ to indicate that $a_n=O(b_n)$ and $a_n=\Omega(b_n)$.
If $a_n/b_n\to 0$, we write $a_n=o(b_n)$ and $b_n=\omega(a_n)$.
We say a sequence of events $\mathcal{A}_n$ holds with high probability if for any $c > 0$ there exists a finite constant $n_0$ depending only on $c$ such that $\mathbb{P}(\mathcal{A}_n)\geq 1-n^{-c}$ for all $n \geq n_0$.
We write $a_n = O_p(b_n)$ (resp. $a_n = o_p(b_n)$) to denote that $a_n = O(b_n)$ (resp. $a_n = o(b_n)$) holds with high probability.
We denote by $\mathcal{O}_d$ the set of $d \times d$ orthogonal
matrices.
For any matrix $\mm\in \mathbb{R}^{A\times B}$ and index sets $\mathcal{A}\subseteq [A]$, $\mathcal{B}\subseteq [B]$, we denote by $\mm_{\mathcal{A},\mathcal{B}}\in \mathbb{R}^{|\mathcal{A}|\times |\mathcal{B}|}$ the submatrix of $\mm$ formed from rows $\mathcal{A}$ and columns $\mathcal{B}$, and we denote by $\mm_{\mathcal{A}}\in\mathbb{R}^{|\mathcal{A}|\times B}$ the submatrix of $\mm$ consisting of the rows indexed by $\mathcal{A}$. 
Given a matrix $\mathbf{M}$, we denote
its spectral, Frobenius, and infinity norms by $\|\mathbf{M}\|$, 
$\|\mathbf{M}\|_{F}$, and $\|\mathbf{M}\|_{\infty}$, respectively. 
We also denote the maximum entry (in modulus) of $\mathbf{M}$ by $\|\mathbf{M}\|_{\max}$ and the two-to-infinity norm of
$\mathbf{M}$ by
$\|\mathbf{M}\|_{2 \to \infty} = \max_{\|\bm{x}\| = 1} \|\mathbf{M}
\bm{x}\|_{\infty} = \max_{i} \|\mathbf{m}_i\|,$
where $\mathbf{m}_i$ denotes the $i$th row of $\mathbf{M}$, i.e., $\|\mathbf{M}\|_{2 \to \infty}$ is the maximum of the $\ell_2$
norms of the rows of $\mathbf{M}$. 
Perturbation bounds using the two-to-infinity norm for the eigenvectors and/or
singular vectors of a noisily observed matrix had recently
attracted interests from the statistics community, see e.g.,
\cite{chen2021spectral,cape2019two,lei2019unified,damle,fan2018eigenvector,abbe2020entrywise}
and the references therein.

\section{Model}\label{sec:model}

Our model for pairs of random networks is based on the concept of the random dot product graph (RDPG) \citep{athreya2018statistical,young2007random}.

\begin{definition}[Random dot product graph (RDPG)]
Let $d\geq 1$ be given and let $\mathcal{X}$ be a subset of $\mathbb{R}^d$ such that $x^\top y\in[0,1]$ for any $x,y\in\mathcal{X}$.
For a given $n\geq 1$, let $\mx=[\mathbf{x}_1,\mathbf{x}_2,\dots,\mathbf{x}_n]^\top$ be a $n\times d$ matrix with $\mathbf{x}_k\in\mathcal{X}$ for all $k\in[n]$. A random network $G$ is said to be a random dot product graph with latent positions of the vertices in $\mx$, where each row $\mathbf{x}_k\in\mathbb{R}^d$ denotes the latent position for the $k$th vertex, if the adjacency matrix $\ma$ of $G$ is a symmetric matrix whose upper triangular entries $\{\ma_{s,t}\}_{s<t}$ are independent Bernoulli random variables with
$$
\ma_{s,t}\sim \operatorname{Bernoulli}(\mathbf{x}_s^\top \mathbf{x}_t).
$$
We define $\mpp:=\mx\mx^\top$ as the connection probability matrix of $G$ and denote such a graph by RDPG$(\mx\mx^\top)$.
In this case, the success probabilities of $\{\ma_{s,t}\}_{s<t}$ are given by  $\mpp_{s,t}=\mathbf{x}_s^\top \mathbf{x}_t$.
\end{definition} 

\begin{remark}[Orthogonal nonidentifiability in RDPGs]\label{rm:noni RDPG}
Note that if $G \sim \operatorname{RDPG}(\mx\mx^\top)$ with the latent position matrix $\mx \in \mathbb{R}^{n \times d}$, then for any orthogonal matrix $\mw \in \mathcal{O}_d$, $\mx\mw$ also gives rise to an RDPG with the same probability distribution, as the connection probability matrix $\mpp = \mx\mx^\top = (\mx\mw)(\mx\mw)^\top$ remains unchanged. Thus, the RDPG model is non-identifiable up to orthogonal transformations.
\end{remark}

The RDPG is a special case of latent
position graphs or graphons
\citep{Hoff2002,diaconis08:_graph_limit_exchan_random_graph,lovasz12:_large}.
In the general latent position graph model, each vertex $s$ is associated with a latent or unobserved vector $\mathbf{x}_s\in\mathcal{X}$ where $\mathcal{X}$ is some latent space such as $\mathbb{R}^d$, and given the collection of latent vectors $\{\mathbf{x}_s\}$, the edges are conditionally independent Bernoulli random variables with success probabilities $\mpp_{s,t} = \kappa(\mathbf{x}_s, \mathbf{x}_t)$ for some kernel function $\kappa:\mathcal{X}\times \mathcal{X}\to[0,1]$. For RDPGs, $\kappa$ is the inner product.

We consider a pair of networks $G^{(1)}$ and $G^{(2)}$ and assume they are independent RDPGs with latent position matrices $\mx^{(1)}$ and $\mx^{(2)}$, respectively, i.e., $G^{(i)} \sim \operatorname{RDPG}(\mx^{(i)}\mx^{(i)\top})$ for $i \in \{1, 2\}$. Furthermore, we assume that a subset of vertices shares latent positions between the two networks, up to some orthonormal transformation. We denote the set of unshifted vertices as $\mathcal{U} \subseteq [n]$ and the set of shifted vertices as its complement, $\mathcal{U}^\mathcal{C}$.
That is, there exists $\mw^{(1,2)} \in \mathcal{O}_d$ such that
$$
\mx^{(2)}_{\mathcal{U}} = \mx_{\mathcal{U}}^{(1)}\mw^{(1,2)}.
$$
Notice that the probability submatrices associated with $\mathcal{U}$ are also identical across the two networks, i.e., $\mpp^{(1)}_{\mathcal{U},\mathcal{U}}=\mpp^{(2)}_{\mathcal{U},\mathcal{U}}$.
We denote the vertex-wise changes in latent positions by $\my \in \mathbb{R}^{n \times d}$, satisfying
\begin{equation}\label{eq:X2=X1W+Y}
\mx^{(2)} = \mx^{(1)}\mw^{(1,2)} + \my,
\end{equation}
where $\my_{\mathcal{U}} = \mathbf{0}$ for the unshifted vertices, and $\my_{\mathcal{U}^\mathcal{C}}$ captures the shifts for the shifted vertices.

Given the observed adjacency matrices $\ma^{(1)}$ and $\ma^{(2)}$, our goal is to estimate the set of unshifted vertices $\mathcal{U}$ (with the set of shifted vertices $\mathcal{U}^{\mathcal{C}}$) and the shifts in $\my$.
The corresponding algorithms are provided in Section~\ref{sec:alg}.

\section{Algorithm}\label{sec:alg}

Given each observed adjacency matrix $\ma^{(i)}$ for $i \in \{1,2\}$, we first estimate the underlying latent position matrix $\mx^{(i)}$ using the scaled leading eigenvectors $\hat\mx^{(i)} = \hat\muu^{(i)}(\hat\mLambda^{(i)})^{1/2}$, where $\hat\mLambda^{(i)}\in \mathbb{R}^{d\times d}$ is a diagonal matrix whose diagonal entries are the leading $d$ eigenvalues of $\ma^{(i)}$ in descending order, and the orthonormal columns of $\hat\muu^{(i)}\in\mathbb{R}^{n\times d}$ constitute the corresponding eigenvectors.
We now propose algorithms to align $\hat\mx^{(1)}$ and $\hat\mx^{(2)}$ in order to estimate the set of unshifted vertices $\mathcal{U}$ and the shifts in $\my$.

\subsection{Seed-based algorithm}

We begin with the case where a seed set $\mathcal{S} \subset \mathcal{U}$ of unshifted vertices is available, with $|\mathcal{S}| \geq d$. Using this seed set, the orthogonal matrix $\hat{\mathbf{W}}^{(1,2)}$ aligning the estimated latent position matrices $\hat{\mathbf{X}}^{(1)}$ and $\hat{\mathbf{X}}^{(2)}$ can be determined by
$$\hat{\mathbf{W}}^{(1,2)}= \argmin\limits_{\mathbf{O} \in \mathcal{O}_d} \|\hat{\mathbf{X}}^{(1)}_{\mathcal{S}}\mathbf{O}-\hat{\mathbf{X}}^{(2)}_{\mathcal{S}}\|_F.$$
The shifts in $\mathbf{Y}$ are then estimated as
$$
\hat{\mathbf{Y}} = \hat{\mathbf{X}}^{(2)} - \hat{\mathbf{X}}^{(1)}\hat{\mathbf{W}}^{(1,2)}.
$$
In Section~\ref{sec:thm}, Theorem~\ref{thm:CLT} shows that the estimated shifts in $\hat{\mathbf{Y}}$ are asymptotically normally distributed around the corresponding true shifts in $\mathbf{Y}$.
Using the test statistic $T_k$ proposed in Eq.~\eqref{eq:T_k}, we can test whether each vertex $k$ is unshifted or shifted, i.e., whether $\mathbf{y}_k = \mathbf{0}$ or not, where $\mathbf{y}_k$ is the $k$th row of $\mathbf{Y}$, representing the shift for vertex $k$. Based on the asymptotic distribution result for the test statistic $T_k$ given in Theorem~\ref{thm:HT}, we compute the corresponding p-values $\{p_k\}_{k\in[n]}$ and, to account for multiple testing across all $n$ vertices, apply the Benjamini-Hochberg procedure to control the false discovery rate (FDR), obtaining the estimated set of unshifted vertices $\hat{\mathcal{U}}$. 
See Algorithm~\ref{alg:with seeds} for the detailed procedure.


\begin{algorithm}[htbp!]
\small
\begingroup
\setlength{\baselineskip}{1\baselineskip}
\caption{Vertex-wise shift learning across networks with seeds and FDR control.}\label{alg:with seeds}
\begin{algorithmic}[1]
\REQUIRE{Adjacency matrices $\mathbf{A}^{(1)},\mathbf{A}^{(2)}\in\mathbb{R}^{n\times n}$, embedding dimension $d$, seed set $\mathcal{S}\subset [n]$ with $|\mathcal{S}|\geq d$, FDR level $\alpha$.}
\begin{enumerate}
	\item For each network $i \in \{1, 2\}$, compute the estimated latent position matrix
$
\hat{\mathbf{X}}^{(i)} = \hat{\mathbf{U}}^{(i)} (\hat{\mathbf{\Lambda}}^{(i)})^{1/2} \in \mathbb{R}^{n \times d},
$
where $\hat{\mathbf{\Lambda}}^{(i)} \in \mathbb{R}^{d \times d}$ is a diagonal matrix containing the leading $d$ eigenvalues of $\mathbf{A}^{(i)}$ in descending order, and $\hat{\mathbf{U}}^{(i)} \in \mathbb{R}^{n \times d}$ is the matrix whose columns are the corresponding orthonormal eigenvectors.
		
	\item Obtain the orthogonal transformation $\hat{\mathbf{W}}^{(1,2)}\in\mathcal{O}_d$ to align the estimated latent position matrices by solving the orthogonal Procrustes problem on the seed set $\mathcal{S}$:
$$
\hat{\mathbf{W}}^{(1,2)}=\argmin_{\mathbf{O}\in\mathcal{O}_d}\|\hat{\mathbf{X}}^{(2)}_{\mathcal{S}}-\hat{\mathbf{X}}^{(1)}_{\mathcal{S}}\mathbf{O}\|_\mathrm{F}.
$$
    
    \item Obtain the estimated shift matrix as $\hat{\mathbf{Y}} = \hat{\mathbf{X}}^{(2)} - \hat{\mathbf{X}}^{(1)}\hat{\mathbf{W}}^{(1,2)}\in\mathbb{R}^{n\times d}$.

    \item For each vertex $k \in [n]$, compute the test statistic $T_k$ as 
$$
T_k = \hat{\mathbf{y}}_k^\top (\hat{\mathbf{\Gamma}}^{(k)})^{-1} \hat{\mathbf{y}}_k,
$$
where $\hat{\mathbf{\Gamma}}^{(k)}$ is a $d \times d$ matrix defined by
$$
\begin{aligned}
	\hat{\mathbf{\Gamma}}^{(k)}& =
(\hat{\mathbf{X}}^{(2)\top}\hat{\mathbf{X}}^{(2)})^{-1}
\hat{\mathbf{X}}^{(2)\top}\hat{\mathbf{\Xi}}^{(k,2)}\hat{\mathbf{X}}^{(2)}
(\hat{\mathbf{X}}^{(2)\top}\hat{\mathbf{X}}^{(2)})^{-1}\\
&+		\hat{\mathbf{W}}^{(1,2)\top}
(\hat{\mathbf{X}}^{(1)\top}\hat{\mathbf{X}}^{(1)})^{-1}
\hat{\mathbf{X}}^{(1)\top}\hat{\mathbf{\Xi}}^{(k,1)}\hat{\mathbf{X}}^{(1)}
(\hat{\mathbf{X}}^{(1)\top}\hat{\mathbf{X}}^{(1)})^{-1}\hat{\mathbf{W}}^{(1,2)},
\end{aligned}
$$
and for $i = 1, 2$, $\hat{\mathbf{\Xi}}^{(k,i)}$ is an $n \times n$ diagonal matrix with diagonal entries
$$
\hat{\mathbf{\Xi}}^{(k,i)}_{\ell,\ell}
=\hat{\mathbf{P}}^{(i)}_{k\ell}(1-\hat{\mathbf{P}}^{(i)}_{k\ell}),
$$
where $\hat{\mathbf{P}}^{(i)}$ is the estimated connection probability matrix for network $i$, computed as $\hat{\mathbf{P}}^{(i)} = \hat{\mathbf{X}}^{(i)}\hat{\mathbf{X}}^{(i)\top}$, with entries adjusted to lie in $[0,1]$, i.e., $\hat{\mathbf{P}}^{(i)}_{k,\ell} = \min(\max(\hat{\mathbf{P}}^{(i)}_{k,\ell}, 0), 1)$.

Then compute the p-value $p_k$ as
$$
p_k = \mathbb{P}(\chi^2_d > T_k) = 1 - F_{\chi^2_d}(T_k),
$$
where $F_{\chi^2_d}(\cdot)$ is the cumulative distribution function of the $\chi^2_d$ distribution.

    \item Apply the Benjamini-Hochberg procedure (see Algorithm~\ref{alg:BH} for details) with p-values $\{p_k\}_{k=1}^n$ and the specified FDR level $\alpha$ to obtain the estimated set of invariant vertices $\hat{\mathcal{U}}$.
\end{enumerate} 

\ENSURE{The estimated set of invariant vertices $\hat{\mathcal{U}}$ and the estimated shift matrix $\hat{\mathbf{Y}}$.}

\end{algorithmic}

\endgroup
\end{algorithm}

\subsection{Seed-free algorithm}\label{sec:alg_no seed}

In many practical scenarios, a seed set is not available. Building on Algorithm~\ref{alg:with seeds}, we now propose an algorithm that eliminates the need for a known seed set.

When no seed set is available, our strategy is to guess a correct seed set. The key idea is that if the guessed seed set $\mathcal{S}$ is correct, i.e., $\mathcal{S} \subset \mathcal{U}$, then using Algorithm~\ref{alg:with seeds} with this correct $\mathcal{S}$ allows us to determine the accurate orthogonal transformation $\hat\mw^{(1,2)}$ between $\hat\mx^{(1)}$ and $\hat\mx^{(2)}$, and consequently, the estimated set of unshifted vertices $\hat{\mathcal{U}}$, consisting of the aligned vertices via $\hat\mw^{(1,2)}$, closely approximates the true $\mathcal{U}$.
Conversely, if the guessed seed set $\mathcal{S}$ is incorrect, i.e., $\mathcal{S} \not\subset \mathcal{U}$, Algorithm~\ref{alg:with seeds} with this incorrect $\mathcal{S}$ fails to identify the true $\mathcal{U}$, and in such cases, the number of vertices that can be aligned using an incorrect seed set is typically small,  resulting in a small size for the corresponding estimated unshifted set $|\hat{\mathcal{U}}|$.

\begin{figure}[htbp!]
\centering
\subfigure[\footnotesize $\{T_k\}$ obtained using a correct seed set]
{\includegraphics[height=5.5cm]{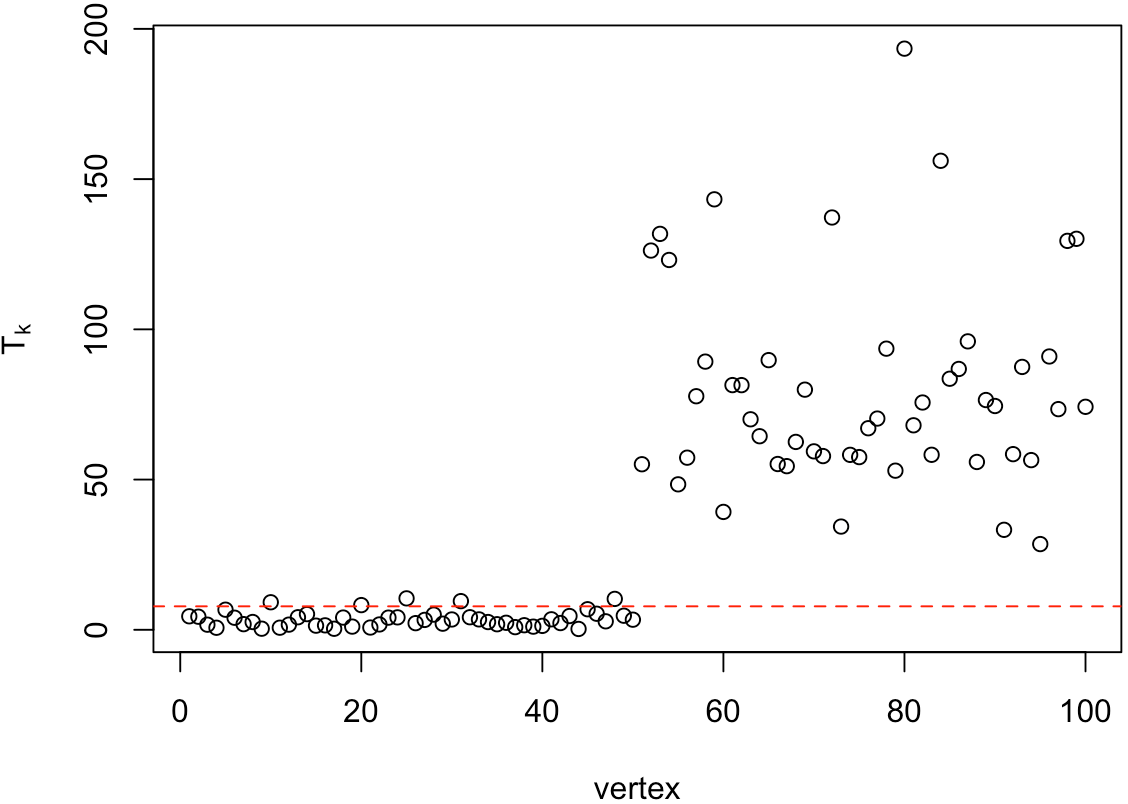}}
\subfigure[\footnotesize $\{T_k\}$ obtained using a incorrect seed set]
{\includegraphics[height=5.5cm]{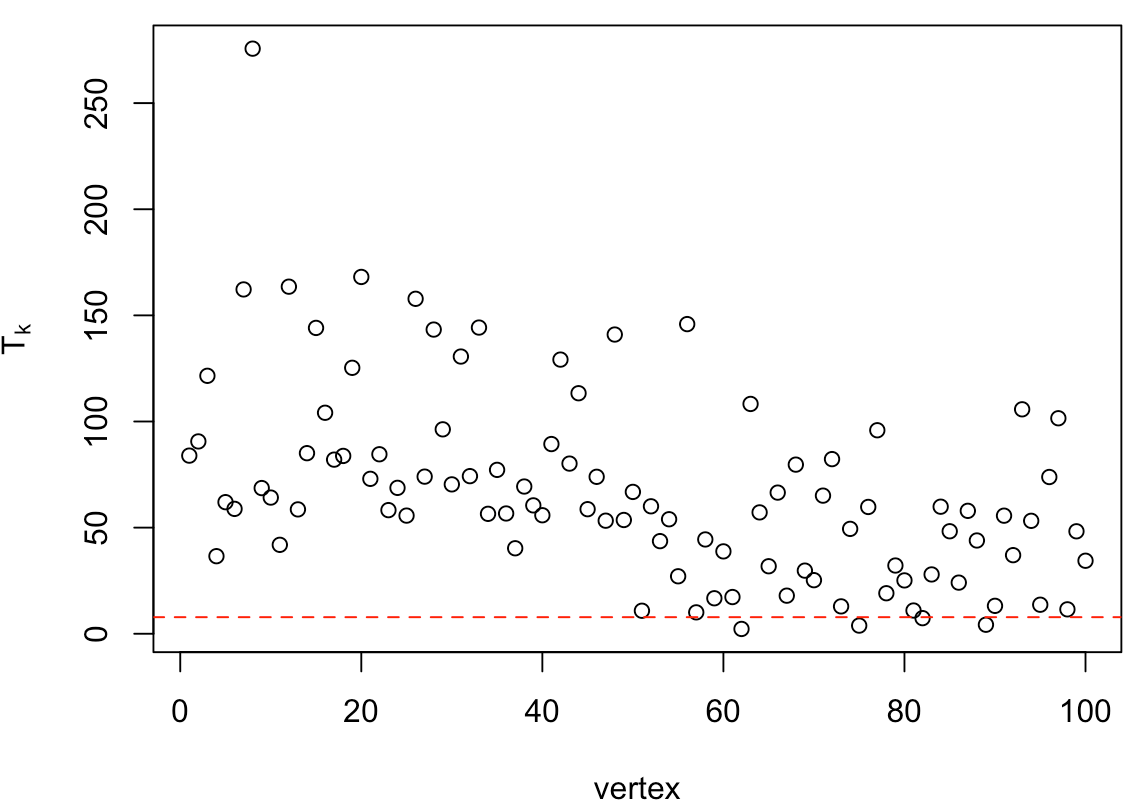}}
\caption{\footnotesize The test statistics $\{T_k\}_{k \in [n]}$ are obtained using either a correct seed set $\mathcal{S} = \{1, 2, 3\}$ or an incorrect seed set $\mathcal{S} = \{50, 51, 52\}$. 
Among the total of $n=100$ vertices, the first $50$ exhibit no shift while the last $50$ experience shifts (i.e., $\mathcal{U}=\{1,2,\dots,50\}$). 
The red horizontal dashed line indicates the critical value $c_{\alpha=0.05}$, used to reject the null hypothesis that there is no shift for each vertex based on individual testing.
}
\label{fig:correctS_wrongS}
\end{figure}

We illustrate the above idea with a simulation example. Consider two networks on 100 vertices, where the first 50 vertices exhibit no shift while the last 50 vertices experience shifts (i.e., $\mathcal{U} = \{1, 2, \dots, 50\}$); further details about this experiment are provided in Section~\ref{sec:simu_noseeds}.
For the case where a correct seed set, such as $\mathcal{S} = \{1, 2, 3\} \subset \mathcal{U}$, is used, the test statistics $\{T_k\}$ for all $n=100$ vertices, obtained by Algorithm~\ref{alg:with seeds}, are shown in the left panel of Figure~\ref{fig:correctS_wrongS}. The figure indicates that the test statistics $\{T_k\}$ for the first 50 vertices are mostly below the red dashed line, which represents the corresponding critical value $c_\alpha$. This demonstrates that most unshifted vertices can be successfully identified, and the estimated set of unshifted vertices $\hat{\mathcal{U}}$ closely approximates the true $\mathcal{U}$. (Note that in our actual algorithm, we apply multiple testing correction; here we show the results of testing each vertex individually for visualization purposes in this motivating example.)
Conversely, when an incorrect seed set, such as $\mathcal{S} = \{51, 52, 53\} \not\subset \mathcal{U}$, is used, the test statistics $\{T_k\}$ are shown in the right panel of Figure~\ref{fig:correctS_wrongS}. In this case, the true set $\mathcal{U}$ cannot be identified, and the number of aligned vertices, i.e., those with $T_k$ below the critical value $c_\alpha$, is very small-only 4 vertices in this experiment.

Following this idea, we randomly generate $M$ candidate seed sets, each of size $L \geq d$, and apply Algorithm~\ref{alg:with seeds} with each candidate seed set $m \in [M]$ to obtain the corresponding set of aligned vertices $\mathcal{U}^{[m]}$. The candidate seed set that yields the largest $\mathcal{U}^{[m]}$ is selected as the most likely correct seed set, denoted by $\mathcal{S}^\star$. Finally, based on $\mathcal{S}^\star$, we can obtain the estimated set of invariant vertices $\hat{\mathcal{U}}$ and the estimated shift matrix $\hat{\my}$. See Algorithm~\ref{alg:without seeds} for the detailed procedure.

\begin{algorithm}[htbp!]
\small
\begingroup
\setlength{\baselineskip}{1\baselineskip}
\caption{Vertex-wise shift learning across networks without seeds.}\label{alg:without seeds}
\begin{algorithmic}[1]
\REQUIRE{Adjacency matrices $\mathbf{A}^{(1)},\mathbf{A}^{(2)}\in\mathbb{R}^{n\times n}$, embedding dimension $d$, seed set size $L\geq d$, number of candidate seed sets $M$, FDR level $\alpha$, filtering significance level $\tilde{\alpha}$.}
\begin{enumerate}
    \item Apply Step~1 of Algorithm~\ref{alg:with seeds} to obtain the estimated latent position matrices
$
\hat{\mathbf{X}}^{(1)}, \hat{\mathbf{X}}^{(2)} \in \mathbb{R}^{n \times d}
$
from $\mathbf{A}^{(1)}$ and $\mathbf{A}^{(2)}$, respectively.

\item Preparation for candidate seed set filtering. Compute the estimated connection probability matrices $\hat{\mathbf{P}}^{(i)}$ and the diagonal matrices $\hat{\mathbf{\Xi}}^{(k,i)}$ for $i\in\{1,2\},k\in[n]$ using the formulas in Step~4 of Algorithm~\ref{alg:with seeds}. Compute $\hat{\Delta}=\hat{\mathbf{P}}^{(1)}-\hat{\mathbf{P}}^{(2)}$ and the $n\times n$ matrix $\hat{\mathbf{\Upsilon}}$ given by
$$
	 \hat{\mathbf{\Upsilon}}_{k,\ell} =
\begin{cases} 
    \hat{\mathbf{\Psi}}^{(k)}_{\ell,\ell} + \hat{\mathbf{\Psi}}^{(\ell)}_{k,k}
    +2\hat{\bm\Pi}^{(1)}_{k,k}\hat{\bm\Pi}^{(1)}_{\ell,\ell}\hat{\mathbf{\Xi}}^{(k,1)}_{\ell,\ell}
    +2\hat{\bm\Pi}^{(2)}_{k,k}\hat{\bm\Pi}^{(2)}_{\ell,\ell}\hat{\mathbf{\Xi}}^{(k,2)}_{\ell,\ell} , & \text{if } k \neq \ell, \\
    4 \hat{\mathbf{\Psi}}^{(k)}_{k,k}, & \text{if } k = \ell,
\end{cases}
$$
where $\hat{\bm\Pi}^{(i)}=\hat{\mathbf{X}}^{(i)}(\hat{\mathbf{X}}^{(i)\top}\hat{\mathbf{X}}^{(i)})^{-1}\hat{\mathbf{X}}^{(i)\top}$ and
$\hat{\mathbf{\Psi}}^{(k)}=\hat{\bm\Pi}^{(1)}\hat{\mathbf{\Xi}}^{(k,1)}\hat{\bm\Pi}^{(1)}+\hat{\bm\Pi}^{(2)}\hat{\mathbf{\Xi}}^{(k,2)}\hat{\bm\Pi}^{(2)}$. 
Then compute the $n\times n$ test statistic matrix $\tilde{\mathbf{T}}$ with entries $\tilde{\mathbf{T}}_{k,\ell}= \hat{\mathbf{\Upsilon}}_{k,\ell}^{-1/2}\hat{\Delta}_{k,\ell}$, and set the threshold $\tilde{c}_{\tilde{\alpha}_B/2}$ as the upper $\tilde{\alpha}_B/2$-quantile of the standard normal distribution, where $\tilde{\alpha}_B=\frac{\tilde{\alpha}}{L(L+1)/2}$ applies the Bonferroni correction.

    \item Randomly generate $M$ candidate seed sets, each $\mathcal{S}^{[m]}\subset [n]$ with size $|\mathcal{S}^{[m]}| = L$ for all $m \in [M]$.
    
    \item For each candidate seed set $m \in [M]$, if $\|\tilde{\mathbf{T}}_{\mathcal{S}^{[m]},\mathcal{S}^{[m]}}\|_{\max} \leq \tilde{c}_{\tilde{\alpha}_B/2}$, apply Steps~2-5 of Algorithm~\ref{alg:with seeds} with $\mathcal{S}^{[m]}$ to obtain $h^{[m]} = |\hat{\mathcal{U}}^{[m]}|$; otherwise, set $h^{[m]} = 0$.
    
    \item Select the most likely correct seed set $\mathcal{S}^\star = \mathcal{S}^{[m^\star]}$, where
    $$
    m^\star = \argmax_{m \in [M]} h^{[m]}.
    $$
    
    \item Apply Steps~2-5 of Algorithm~\ref{alg:with seeds} with seed set $\mathcal{S}^\star$ to obtain the estimated set of unshifted vertices $\hat{\mathcal{U}}^\star$. Update $\mathcal{S}^\star = \hat{\mathcal{U}}^\star$ and rerun Steps~2-5 of Algorithm~\ref{alg:with seeds} with the updated seed set to produce the final output.
	
\end{enumerate}

\ENSURE{The estimated set of invariant vertices $\hat{\mathcal{U}}$ and the estimated shift matrix $\hat{\mathbf{Y}}$.}

\end{algorithmic}

\endgroup
\end{algorithm}

To determine the seed set size $L$ and the number of candidate seed sets $M$ for Algorithm~\ref{alg:without seeds}, the theoretical result in Theorem~\ref{thm:X1W-X2} indicates that our algorithms remain effective even when the seed set size $L = |\mathcal{S}|$ is as small as the minimal dimension $d$; see Remark~\ref{rem:small S} for further discussion. Therefore, to reduce computational cost, we can simply set $L = d$.
Next, we consider the appropriate size for $M$. Suppose the proportion of unshifted vertices is $p$, i.e., $p=|\mathcal{U}| / n$. Then the probability that a randomly chosen seed vertex belongs to $\mathcal{U}$ is $p$, and the probability that a randomly generated seed set $\mathcal{S}$ of size $L=d$ is fully contained within $\mathcal{U}$ is approximately $p^d$.
To ensure that at least one candidate seed set is correct with a probability of at least $q$, it suffices to choose $M \geq \frac{\ln(1 - q)}{\ln(1 - p^d)}$.
For instance, when the latent position dimension $d = 3$ and the proportion of unshifted vertices $p = 1/4$, it is sufficient to set $M \geq 293$ to ensure that at least one candidate seed set is correct with a probability of at least $q = 0.99$.
Therefore, in many practical scenarios, choosing $M$ on the order of a few thousand is typically sufficient to guarantee robust performance.

To further reduce computational complexity, we incorporate a filtering step in Step~4 of Algorithm~\ref{alg:without seeds} for the randomly generated seed sets (with the preparation in Step~2). We observe that the underlying probability submatrices associated with the unshifted vertex set $\mathcal{U}$ are identical across the two networks, i.e., $\mpp^{(1)}_{\mathcal{U},\mathcal{U}} = \mpp^{(2)}_{\mathcal{U},\mathcal{U}}$, which implies $\Delta_{\mathcal{U},\mathcal{U}}=\mathbf{0}$ where $\Delta:=\mpp^{(1)}-\mpp^{(2)}$. Consequently, for any correct seed set $\mathcal{S} \subset \mathcal{U}$, we also have $\Delta_{\mathcal{S},\mathcal{S}} = \mathbf{0}$, and thus $\Delta_{k,\ell}=0$ for all vertex pairs $(k,\ell)\in\mathcal{S}\times\mathcal{S}$.
We estimate $\Delta$ with $\hat{\Delta} = \hat{\mpp}^{(1)} - \hat{\mpp}^{(2)}$. Using the test statistic $\tilde{T}_{k,\ell}$ proposed in Eq.~\eqref{eq:tildeT_kl} based on $\hat{\Delta}_{k,\ell}$, we can test the null hypothesis $\Delta_{k,\ell}=0$ for any vertex pair $(k,\ell)\in[n]\times [n]$, and Theorem~\ref{thm:max_HT} provides the asymptotic distribution result for $\tilde{T}_{k,\ell}$. For a given seed set candidate $\mathcal{S}$, we apply Bonferroni correction across all distinct pairs in $\mathcal{S}\times\mathcal{S}$ (note that $\hat{\Delta}_{\mathcal{S},\mathcal{S}}$ is symmetric) with the critical value $\tilde{c}_{\tilde{\alpha}_B/2}$. We retain only those seed sets $\mathcal{S}$ for which we fail to reject $\Delta_{k,\ell}=0$ for all vertex pairs $(k,\ell) \in \mathcal{S}\times\mathcal{S}$, i.e., $\|\tilde{\mt}_{\mathcal{S},\mathcal{S}}\|_{\max}\leq \tilde{c}_{\tilde{\alpha}_B/2}$. Seed sets satisfying this criterion are retained for subsequent computation of $h^{[m]}$. This filtering step substantially reduces the number of candidate seed sets requiring full evaluation, thereby enhancing computational efficiency while preserving accuracy. See Section~\ref{sec:filtering} for the corresponding theoretical results and further details.

Now we provide a more detailed discussion of the required number of seed set candidates $M$, which involves considering the cases where $p$ is too small or $d$ is too large.
For $p$, note that our setting always assumes a substantial number of unshifted vertices, meaning that $p$ is not too small. As demonstrated in the two real data examples in Section~\ref{sec:real}, the proportion $p$ of detected unshifted vertices even exceeds $0.5$ in both examples. If $p$ is too small---for example, if among $1000$ total vertices only $1\%$ are unshifted, yielding merely $10$ unshifted vertices---then these $10$ unshifted vertices do not constitute a meaningful pattern in the data but rather appear as a coincidence, and thus our model becomes unsuitable in such cases. Therefore, within the applicable scope of our model, $p$ is not expected to be very small.
For $d$, we first note that we assume a low-rank model, so $d$ is bounded and will not be too large. When $d$ is small, $M$ remains manageable, and after generating candidates, we perform the filtering step and only candidates that pass the filtering undergo subsequent computations. The primary concern now arises when $d$ is bounded but still relatively large. For such cases, rather than randomly generating a large number of candidates and then filtering out most of them using the filtering step, we provide an alternative algorithm that directly generates random seed set candidates within the feasible region such that $|\tilde{T}_{k,\ell}| \leq \tilde{c}_{\tilde{\alpha}_B/2}$ (see Algorithm~\ref{alg:feasible_sampling} in Section~\ref{sec:filtering} for details). This direct sampling approach can significantly reduce the required $M$, and makes it essentially not dependent on the relatively large $d$. Further detailed discussion of this direct generation algorithm and this scenario is provided in Section~\ref{sec:filtering}.

Note that in Step~5 of Algorithm~\ref{alg:without seeds}, we expand the selected seed set of size $L$ to a larger seed set before producing the final output. This step is motivated by the theoretical result in Theorem~\ref{thm:X1W-X2}, which indicates that while the minimal size $d$ is sufficient to achieve the good estimate error rate, an increased seed set size can further improve the estimates; see Remark~\ref{rem:small S} for more details.

Finally, we discuss how to choose the embedding dimension $d$ in practice.
We can examine the eigenvalues of the adjacency matrices, and an ubiquitous and principled method is to examine the so-called scree plot and look for ``elbow" and ``knees" defining the cut-off between the top (signal) $d$ dimension and the noise dimensions. 
\cite{zhu2006automatic} provide an automatic dimensionality selection procedure to look for the ``elbow'' by maximizing a profile likelihood function.
 \cite{han2023universal} suggest another universal approach to rank inference via residual subsampling for estimating the rank.
 Other related methods include the eigenvalue ratio test \citep{ahn2013eigenvalue} and techniques based on the empirical distribution of eigenvalues \citep{onatski2010determining}. 
 See Section~\ref{sec:different rank} for more discussion on embedding dimension selection.
 
\subsection{Computational complexity}

We now describe the computational complexity of our proposed algorithms.
For Algorithm~\ref{alg:with seeds}, the embedding step in Step~1 has a complexity of roughly  $O(n^2d)$  \citep{van1990partial};
the orthogonal Procrustes problem on the seed set of size  $|\mathcal{S}|$ in Step~2 has a complexity of approximately  $O(|\mathcal{S}| d^2)$;
Steps~3-5 require some matrix operations and its complexity is roughly  $O(n^2d)$.
Thus the overall computational complexity of Algorithm~\ref{alg:with seeds} is $O(n^2d)$. Note that the embedding dimension $d$ is typically bounded while the number of vertices $n$ determines the primary computational cost.

For Algorithm~\ref{alg:without seeds}, its Steps~1 and 2 involves Steps~1 and 4 of Algorithm~\ref{alg:with seeds}, with a complexity of is $O(n^2d)$; 
the random generation procedure in its Step~3 has a complexity of $O(LM)$;
Step~4 involves applying Algorithm~\ref{alg:with seeds} and has a complexity of $O(Mn^2d)$;
the selection of the optimal seed set in Step~5 has a complexity of $O(M)$;
Step~6 again applies Algorithm~\ref{alg:with seeds} and has a complexity of $O(n^2d)$.
Hence the overall computational complexity of Algorithm~\ref{alg:without seeds} is $O(Mn^2d)$. And according to the discussion in Section~\ref{sec:alg_no seed}, the number of candidate seed sets $M$ is typically sufficient in the range of several hundred to a few thousand.

Empirically, we record the running time of the simulation experiments described in Section~\ref{sec:simu_noseeds} for $n$ ranging from $50$ to $800$ vertices, as shown in the right panel of Figure~\ref{fig:simulation_noseeds}. From Figure~\ref{fig:simulation_noseeds}, the running time of Algorithm~\ref{alg:without seeds} scales approximately quadratically with $n$, consistent with the theoretical analysis above. Furthermore, even for $n=800$, the running time on a standard computer is only about $4.5$ seconds per Monte Carlo replicate.

\section{Theoretical Results}\label{sec:thm}

For each rank-$d$ probability matrix $\mpp^{(i)} = \mx^{(i)}\mx^{(i)\top}$ for $i = 1, 2$, we denote the eigen-decomposition of $\mpp^{(i)}$ by $\mpp^{(i)} =\muu^{(i)}\mLambda^{(i)}\muu^{(i)\top}$, where $\mLambda^{(i)} \in \mathbb{R}^{d \times d}$ is a diagonal matrix containing the $d$ eigenvalues of $\mpp^{(i)}$ in descending order, and $\muu^{(i)} \in \mathbb{R}^{n \times d}$ is a matrix whose orthonormal columns are the corresponding eigenvectors. 
For ease of exposition, in the theoretical analysis we assume $\mx^{(i)} = \muu^{(i)}(\mLambda^{(i)})^{1/2}$. Generally $\mx^{(i)}$ may differ from $\muu^{(i)}(\mLambda^{(i)})^{1/2}$ by an orthogonal matrix, which only involves some book-keeping but does not affect the essence of any theoretical results.

We shall make the following assumptions on the probability matrices $\mpp^{(1)},\mpp^{(2)}\in\mathbb{R}^{n\times n}$.
We emphasize that, because our theoretical results address either large-sample approximations or limiting distributions, these assumptions should be interpreted in the regime where $n$ is arbitrarily large and/or $n \rightarrow \infty$.

\begin{assumption}\label{assum:main}
The following conditions hold for sufficiently large $n$.
	\begin{itemize}
	    \item $\{\muu^{(i)}\}$ are $n\times d$ matrices with bounded coherence, i.e., $$\|\muu^{(i)}\|_{2\to\infty}\lesssim d^{1/2}n^{-1/2}.$$
	    \item There exists a factor $\rho_n\in[0,1]$ depending on $n$ such that for each $i=1,2$, $\|\mLambda^{(i)}\|=\Theta(n\rho_n)$ where $n\rho_n=\Omega(\log n)$. We interpret $n\rho_n$ as the growth rate for the average degree of the networks $\{\ma^{(i)}\}$ generated from $\{\mpp^{(i)}\}$.
	    \item $\{\mpp^{(i)}\}$ have bounded condition numbers, i.e., there exists a finite constant $M>0$ such that $$\max_{i=1,2}\frac{\lambda_{1}(\mLambda^{(i)})}{\lambda_{d}(\mLambda^{(i)})}\leq M,$$ where $\lambda_{1}(\mLambda^{(i)})$ and $\lambda_{d}(\mLambda^{(i)})$ denote the largest and smallest non-zero eigenvalues of $\mpp^{(i)}$, respectively.
		\item For the seed set $\mathcal{S}$, the latent position matrix $\mx^{(2)}_{\mathcal{S}}=\mx^{(1)}_{\mathcal{S}}\mw^{(1,2)}$ achieves rank $d$, and we suppose 
		$$(\sigma_d(\mx^{(1)}_{\mathcal{S}}))^2=\lambda_{d}(\mpp^{(1)}_{\mathcal{S},\mathcal{S}})
		\gtrsim \frac{|\mathcal{S}|}{n}\lambda_{d}(\mpp^{(1)});$$
		see \ref{rm:lambda} for more discussion for this condition.
	\end{itemize}
\end{assumption}

\begin{remark1}\label{rm:lambda}
The assumption on $\lambda_{d}(\mpp_{\mathcal{S},\mathcal{S}})$ holds with high probability when $\mathcal{S}$ is drawn uniformly at random from $\mathcal{U}$; see Lemma~D.11 in \cite{zheng2024chain}, consistent with the seed set generation method in Algorithm~\ref{alg:without seeds}.
Theoretical results under weaker assumptions on it or even without such assumptions can also be derived following our analysis; see Section~\ref{sec:weaker} for details.
\end{remark1}

\begin{remark1}
  \label{rem:assumptions}
  The remaining conditions in Assumption~\ref{assum:main} are quite mild and typically seen in the literature.
  The first condition on bounded coherence and the third condition on bounded condition number are prevalent and typically mild in random networks and many other high-dimensional statistics inference problems including matrix completion, covariance estimation, and subspace estimation, see e.g., \cite{candes_recht,fan2018eigenvector,lei2019unified,abbe2020entrywise,cape2019two,cai2021subspace,chen2021spectral}.
  The second condition $\|\mpp^{(i)}\| \asymp n \rho_n = \Omega(\log n)$ implies that the average degree of each network $\ma^{(i)}$ grows poly-logarithmically in $n$, and this semi-sparse regime $n \rho_n = \Omega(\log n)$ is generally necessary for spectral methods to work \citep{xie2024entrywise}. 
  \end{remark1}
  
We now establish uniform error bounds and provide normal approximations for the row-wise fluctuations of the estimated shift matrix $\hat\my$ obtained based on a given seed set $\mathcal{S}$ around the true shift matrix $\my$.

\begin{theorem1}\label{thm:X1W-X2}
Let the adjacency matrices $\ma^{(1)} \sim \operatorname{RDPG}(\mx^{(1)}\mx^{(1)\top})$ and $\ma^{(2)} \sim \operatorname{RDPG}(\mx^{(2)}\mx^{(2)\top})$, where $\mx^{(2)} = \mx^{(1)}\mw^{(1,2)} + \my$ with some $\mw^{(1,2)} \in \mathcal{O}_d$ and the shift matrix $\my$ satisfying $\my_{\mathcal{U}} = \mathbf{0}$ for the set of unshifted vertices $\mathcal{U}$.
Let the probability matrices $\mpp^{(i)}=\mx^{(i)}\mx^{(i)\top}$ and the seed set $\mathcal{S} \subset \mathcal{U}$ satisfy the conditions in Assumption~\ref{assum:main}.
We obtain the estimated orthogonal transformation as
$$
\hat\mw^{(1,2)}= \argmin\limits_{\mo \in \mathcal{O}_d} \|\hat\mx^{(1)}_{\mathcal{S}}\mo-\hat\mx^{(2)}_{\mathcal{S}}\|_F,
$$
and then compute the estimated shift matrix as
$$
\hat\my=\hat\mx^{(2)}-\hat\mx^{(1)}\hat\mw^{(1,2)}.
$$
	Then there exists $\mw\in\mathcal{O}_d$ such that
	$$
	\hat\my\mw-\my=\me^{(2)}\mx^{(2)}(\mx^{(2)\top}\mx^{(2)})^{-1}-\me^{(1)}\mx^{(1)}(\mx^{(1)\top}\mx^{(1)})^{-1}\mw^{(1,2)}+\mr,
	$$
	where $\me^{(i)}=\ma^{(i)}-\mpp^{(i)}$ and $\mr$ is a $n\times d$ matrix satisfying
	$$
	\|\mr\|_{2\to\infty}\lesssim |\mathcal{S}|^{-1/2}n^{-1/2}\log^{1/2} n
    + n^{-1/2}(n\rho_n)^{-1/2}\log n
    $$
	with high probability,
	and we further have
	$$
	\|\hat\my\mw-\my\|_{2\to\infty}\lesssim  n^{-1/2}\log^{1/2} n
	$$
	with high probability.
\end{theorem1}

\begin{remark1}\label{rem:small S}
We first note that Theorem~\ref{thm:X1W-X2} imposes no restrictions on the size of the seed set $|\mathcal{S}|$. Then regardless of the value of $|\mathcal{S}|$, we have 
$
\|\hat{\my} \mw - \my\|_{2\to\infty} \leq C n^{-1/2} \log^{1/2} n
$
with high probability, where $C$ is a constant not depending on $|\mathcal{S}|$.
More specifically, the estimation error $\hat{\my} \mw - \my$ consists of two terms:
$
\big[\me^{(2)}\mx^{(2)}(\mx^{(2)\top} \mx^{(2)})^{-1} - \me^{(1)}\mx^{(1)}(\mx^{(1)\top} \mx^{(1)})^{-1} \mw^{(1,2)}\big]$ and $\mr.
$
The first term does not dependent on the seed set $\mathcal{S}$ and has a row-wise upper bound of $n^{-1/2} \log^{1/2} n$, while the second term $\mathbf{R}$ has a row-wise upper bound of 
$
|\mathcal{S}|^{-1/2} n^{-1/2} \log^{1/2} n + n^{-1/2} (n \rho_n)^{-1/2} \log n.
$
Therefore, under the mild conditions $n \rho_n = \Omega(\log n)$ and $|\mathcal{S}| \geq d$, the total estimation error for $\hat{\my} \mw - \my$ is $n^{-1/2} \log^{1/2} n$.
This result implies that when $|\mathcal{S}|$ is very small, such as the minimal value $d$ (the smallest seed set size required to align the $d$-dimensional latent positions), the estimation error remains of the same order as when $|\mathcal{S}|$ is very large; see Section~\ref{sec:simu_small seeds} for the corresponding simulation.
Therefore, in Algorithm~\ref{alg:without seeds}, when proposing seed set candidates, we recommend selecting a small seed set size $|\mathcal{S}|$, as the loss of estimation accuracy is limited and the computational cost can be significantly reduced.

On the other hand, an increase in $|\mathcal{S}|$ can indeed reduce part of the estimation error due to $\|\mathbf{R}\|_{2\to\infty}$ (although this improvement is limited). This motivates an enhancement to the final step of Algorithm~\ref{alg:without seeds}: using all detected unshifted vertices as a new, larger seed set to replace the original one, and then rerunning Algorithm~\ref{alg:with seeds} to obtain the final estimate.

In summary, Theorem~\ref{thm:X1W-X2} demonstrates that the estimation error obtained for any $|\mathcal{S}|$ is of the same order. Therefore, considering computational costs, we recommend using a small $|\mathcal{S}|$ in the seed set candidate selection step of Algorithm~\ref{alg:without seeds}, as this still provides an accurate estimate. At the same time, the estimate can be further improved with a larger seed set, which motivates an enhancement to the final step of Algorithm~\ref{alg:without seeds}.

\end{remark1}

\begin{theorem1}\label{thm:CLT}
Consider the setting in Theorem~\ref{thm:X1W-X2}.
Further assume that $\ma^{(1)}$ and $\ma^{(2)}$ are independent, $n\rho_n=\omega(\log^2 n)$, and $|\mathcal{S}|=\omega(\log n)$.
Then for any $k\in[n]$, let $\mathbf{\Xi}^{(k,i)}$ for $i=1,2$ be $n\times n$ diagonal matrix whose diagonal entries are of the form
\begin{equation}\label{eq:Xi}
	\mathbf{\Xi}^{(k,i)}_{\ell,\ell}=\mpp^{(i)}_{k,\ell}(1-\mpp^{(i)}_{k,\ell}).
\end{equation}
Define $\mathbf{\Gamma}^{(k)}$ as the $d\times d$ symmetric matrix
$$
\begin{aligned}
	\mathbf{\Gamma}^{(k)}
:&=(\mx^{(2)\top}\mx^{(2)})^{-1}\mx^{(2)\top}\mathbf{\Xi}^{(k,2)}\mx^{(2)}(\mx^{(2)\top}\mx^{(2)})^{-1}\\
&+\mw^{(1,2)\top}(\mx^{(1)\top}\mx^{(1)})^{-1}\mx^{(1)\top}\mathbf{\Xi}^{(k,1)}\mx^{(1)}(\mx^{(1)\top}\mx^{(1)})^{-1}\mw^{(1,2)}.
\end{aligned}
$$
Note that $\|\mathbf{\Gamma}^{(k)}\|\lesssim n^{-1}$. Further suppose $\lambda_{\min}(\mathbf{\Gamma}^{(k)})\gtrsim n^{-1}$.
Then for the $k$th rows $\hat{\mathbf{y}}_k$ of $\hat\my$ and ${\mathbf{y}}_k$ of $\my$ we have
$$
(\mathbf{\Gamma}^{(k)})^{-1/2}(\mw^\top \hat{\mathbf{y}}_k-{\mathbf{y}}_k)\rightsquigarrow \mathcal{N}(\mathbf{0}, \mathbf{I})
$$
as $n\to\infty$.
Furthermore, for any pair of indices $k\neq k'$, the vectors $(\mw^\top \hat{\mathbf{y}}_k-{\mathbf{y}}_k)$ and $(\mw^\top \hat{\mathbf{y}}_{k'}-{\mathbf{y}}_{k'})$ are asymptotically independent.
\end{theorem1}

\begin{remark1}
The theoretical results and corresponding algorithms can be extended to the setting where $\ma^{(1)}$ and $\ma^{(2)}$ are correlated. In this case, the expansion in Theorem~\ref{thm:X1W-X2} remains valid, and some normal approximation result like Theorem~\ref{thm:CLT} can still hold but the covariance matrix $\mathbf{\Gamma}^{(k)}$ should be adjusted to account for the correlation between $\me^{(1)}$ and $\me^{(2)}$.
\end{remark1}

\begin{remark1}\label{rm:cov}
The row-wise normal approximations in
Theorem~\ref{thm:CLT}
assumes
that the minimum eigenvalue of $\mathbf{\Gamma}^{(k)}$ grows at rate $n^{-1}$,
and this condition holds whenever the entries of
$\mpp^{(i)}$ are homogeneous, e.g., 
suppose $\min_{k\ell} \mpp^{(i)}_{k \ell} \asymp
\max_{k\ell} \mpp^{(i)}_{k \ell} \asymp \rho_n$, then 
we have $\min_{k,i,\ell} \mathbf{\Xi}^{(k,i)}_{\ell,\ell} \gtrsim \rho_n$ and hence 
$$
\begin{aligned}
	&\lambda_{\min}\big((\mx^{(2)\top}\mx^{(2)})^{-1}\mx^{(2)\top}\mathbf{\Xi}^{(k,2)}\mx^{(2)}(\mx^{(2)\top}\mx^{(2)})^{-1}\big)\\
&\geq \min_{\ell\in[n]}(\mathbf{\Xi}^{(k,i)}_{\ell,\ell})\cdot  \lambda_{\min}\big((\mx^{(2)\top}\mx^{(2)})^{-1}\mx^{(2)\top}
\mx^{(2)}(\mx^{(2)\top}\mx^{(2)})^{-1}
\big)\\
&\geq \min_{\ell\in[n]}(\mathbf{\Xi}^{(k,i)}_{\ell,\ell})
 \cdot \lambda_{\min}\big((\mx^{(2)\top}\mx^{(2)})^{-1}
\big)\\
&\geq \min_{\ell\in[n]}(\mathbf{\Xi}^{(k,i)}_{\ell,\ell})
 \cdot \lambda_{\max}^{-1}\big(\mpp^{(2)}
\big)
 \gtrsim \rho_n\cdot (n\rho_n)^{-1} \gtrsim n^{-1}.
\end{aligned}
$$
Similarly we also have 
$$\lambda_{\min}\big(\mw^{(1,2)\top}(\mx^{(1)\top}\mx^{(1)})^{-1}\mx^{(1)\top}\mathbf{\Xi}^{(k,1)}\mx^{(1)}(\mx^{(1)\top}\mx^{(1)})^{-1}\mw^{(1,2)}\big)
\gtrsim n^{-1}.
$$
Weyl's inequality then implies
$
\lambda_{\min}(\mathbf{\Gamma}^{(k)})
\gtrsim n^{-1}.
$
The main reason for requiring a lower bound for the eigenvalues of $\mathbf{\Gamma}^{(k)}$ is that we do not require $\mathbf{\Gamma}^{(k)}$ to converge to any fixed matrix as $n \to \infty$, and thus  we cannot directly use $\mathbf{\Gamma}^{(k)}$ in our limiting normal approximation. Rather, we need to scale $\mw^\top \hat{\mathbf{y}}_k-{\mathbf{y}}_k$ by $(\mathbf{\Gamma}^{(k)})^{-1/2}$, and to ensure that this scaling is well-behaved, we need to control the smallest eigenvalue of $\mathbf{\Gamma}^{(k)}$. 
\end{remark1}

\subsection{Testing for shifted vertices} \label{sec:multi}


We next address the problem of determining whether vertices are shifted or unshifted. 
For a given vertex $k\in[n]$, we test the null hypothesis $\mathbb{H}_0: \mathbf{y}_k = \mathbf{0}$ against the alternative $\mathbb{H}_a: \mathbf{y}_k \neq \mathbf{0}$. 
We now transform the normal approximations for $\hat\my$ established in Theorem~\ref{thm:CLT}, into a test statistic for this hypothesis test.
Suppose the null hypothesis $\mathbb{H}_0:\mathbf{y}_k=\mathbf{0}$ is true. Then by Theorem~\ref{thm:CLT} we have
\begin{equation}\label{eq:test1}
\hat{\mathbf{y}}_k^\top \mw
(\mathbf{\Gamma}^{(k)})^{-1}
\mw^\top \hat{\mathbf{y}}_k 
\rightsquigarrow \chi_{d}^2
\end{equation}
as $n\to\infty$.
Our objective is to convert Eq.~\eqref{eq:test1} into a test statistic that depends only on estimates. We first define $\hat{\mathbf{\Gamma}}^{(k)}$  as
\begin{equation}\label{eq:hatGamma}
\begin{aligned}
	\hat{\mathbf{\Gamma}}^{(k)}
:&=(\hat\mx^{(2)\top}\hat\mx^{(2)})^{-1}\hat\mx^{(2)\top}\hat{\mathbf{\Xi}}^{(k,2)}\hat\mx^{(2)}(\hat\mx^{(2)\top}\hat\mx^{(2)})^{-1}
\\&+\hat\mw^{(1,2)\top}(\hat\mx^{(1)\top}\hat\mx^{(1)})^{-1}\hat\mx^{(1)\top}\hat{\mathbf{\Xi}}^{(k,1)}\hat\mx^{(1)}(\hat\mx^{(1)\top}\hat\mx^{(1)})^{-1}\hat\mw^{(1,2)},
\end{aligned}
\end{equation}
where $\hat{\mathbf{\Xi}}^{(k,i)}$ be an $n\times n$ diagonal matrix with diagonal entries defined as
$
\hat{\mathbf{\Xi}}^{(k,i)}_{\ell,\ell}
=\hat\mpp^{(i)}_{k\ell}(1-\hat\mpp^{(i)}_{k\ell})
$; here $\hat\mpp^{(i)}_{k\ell}$ represents the estimated connection probability between vertex $k$ and vertex $\ell$ in the network $i$ and we set $\hat\mpp^{(i)}=\hat\mx^{(i)}\hat\mx^{(i)\top}$.
The following lemma shows that 
$(\hat{\mathbf{\Gamma}}^{(k)})^{-1}$ is a consistent estimate of $\mw
(\mathbf{\Gamma}^{(k)})^{-1}
\mw^\top$.

\begin{lemma1}\label{lemma:sigma}
	Consider the setting of Theorem~\ref{thm:CLT}. We then have
	$$n^{-1}\|\mw
(\mathbf{\Gamma}^{(k)})^{-1}
\mw^\top-(\hat{\mathbf{\Gamma}}^{(k)})^{-1}\|\lesssim
(n\rho_n)^{-1/2}\log^{1/2} n$$
with high probability.
\end{lemma1}

Given Lemma~\ref{lemma:sigma}, the following result provides a test statistic for $\mathbb{H}_0: \mathbf{y}_k = \mathbf{0}$ that converges to a central (resp. non-central) $\chi^2$ under the null (resp. local alternative) hypothesis.

\begin{theorem1}\label{thm:HT}
	Consider the setting in Theorem~\ref{thm:CLT}.
	Define the test statistic
	\begin{equation}\label{eq:T_k}
		T_k=\hat{\mathbf{y}}_k^\top (\hat{\mathbf{\Gamma}}^{(k)})^{-1} \hat{\mathbf{y}}_k,
	\end{equation}
	where $\hat{\mathbf{\Gamma}}^{(k)}$ is given in Eq.~\eqref{eq:hatGamma}.
	Then under $\mathbb{H}_0:\mathbf{y}_k=\mathbf{0}$, we have $T_k \rightsquigarrow \chi^2_d$ as $n\to\infty$. 
	
	Next suppose that $\eta_k>0$ is a finite constant and that $\mathbf{y}_k\neq \mathbf{0}$ satisfies a local alternative hypothesis where $\mathbf{y}_k^\top ({\mathbf{\Gamma}}^{(k)})^{-1} \mathbf{y}_k\to\eta_k.$
	We then have $T_k \rightsquigarrow \chi^2_d(\eta_k)$ as $n\to\infty$, where $\chi^2_d(\eta)$ is the non-central chi-square distribution with $d$ degrees of freedom and noncentrality parameter $\eta_k$. 
	
	Furthermore, for any pair of indices $k\neq k'$, $T_k$ and $T_{k'}$ are asymptotically independent.
\end{theorem1}

Theorem~\ref{thm:HT} indicates that for each vertex $k$, for a chosen significance level $\alpha$, we reject $\mathbb{H}_0: \mathbf{y}_k = \mathbf{0}$ if $T_k > c_{\alpha}$, where $c_{\alpha}$ is the upper $\alpha$-quantile of the $\chi^2_d$ distribution.
We can also compute the corresponding p-value as
$p_k = \mathbb{P}(\chi^2_d > T_k) = 1 - F_{\chi^2_d}(T_k)$,
where $F_{\chi^2_d}(\cdot)$ is the cumulative distribution function of the $\chi^2_d$ distribution. This result 
enables the following multiple testing correction.


After obtaining the orthogonal transformation between the latent positions of the two networks based on a given seed set, we conduct hypothesis tests across all vertices to identify those exhibiting significant positional shifts.
Since we are performing $n$ simultaneous hypothesis tests, we employ the Benjamini-Hochberg procedure to control the false discovery rate (FDR) in this multiple testing scenario.
In our shift detection setting, the false discovery proportion (FDP)
$
= \frac{|\mathcal{U} \cap \hat{\mathcal{U}}^{\mathcal{C}}|}{|\hat{\mathcal{U}}^{\mathcal{C}}|},
$
where $\mathcal{U}$ and $\hat{\mathcal{U}}$ denote the true and estimated sets of invariant vertices, respectively (we define FDP$= 0$ when $|\hat{\mathcal{U}}^{\mathcal{C}}| = 0$).

\begin{algorithm}[htbp!]
\small
\begingroup
\setlength{\baselineskip}{1\baselineskip}
\caption{Benjamini-Hochberg procedure for vertex-wise shift detection.}\label{alg:BH}
\begin{algorithmic}[1]
\REQUIRE{P-values $\{p_k\}_{k=1}^n$, FDR level $\alpha$.}
\begin{enumerate}
    \item Sort the p-values in ascending order
$
p_{(1)} \leq p_{(2)} \leq \cdots \leq p_{(n)}.
$
Let $k_{(1)}, k_{(2)}, \ldots, k_{(n)}$ denote the corresponding vertex indices.

    \item Find the largest index $j^\star$ such that
$
j^\star = \max\left\{j \in [n] : p_{(j)} \leq \frac{j \alpha}{n}\right\}.
$
If no such $j$ exists, set $j^\star = 0$.

    \item Reject the null hypotheses $\mathbb{H}_0: \mathbf{y}_{k_{(i)}} = \mathbf{0}$ for all $i \in [j^\star]$.

    \item Obtain the estimated set of shifted vertices as
$
\hat{\mathcal{V}} = \{k_{(i)} : i \in [j^\star]\}.
$
The estimated set of invariant vertices is $\hat{\mathcal{U}} = [n] \setminus \hat{\mathcal{V}}$.
\end{enumerate}

\ENSURE{The estimated set of invariant vertices $\hat{\mathcal{U}}$.}

\end{algorithmic}

\endgroup
\end{algorithm}

The Benjamini-Hochberg procedure controls the FDR at level $\alpha$ under the assumption that p-values corresponding to true null hypotheses are independent. In our setting, this condition is satisfied since the test statistics $\{T_k\}_{k\in[n]}$ are asymptotically mutually independent. Therefore, the Benjamini-Hochberg procedure asymptotically guarantees that $\text{FDR}=\mathbb{E}[\text{FDP}]\leq p \alpha \leq \alpha$, where  $p$ is the proportion of unshifted vertices. Given that our framework generally assumes a considerable number of vertices remain unshifted between networks, $p$ is typically not too small, and thus the Benjamini-Hochberg procedure is not overly conservative.

\subsection{Seed set candidate filtering step of Algorithm~\ref{alg:without seeds}}\label{sec:filtering}

We include a filtering step for the randomly generated seed set candidates in Step~4 of Algorithm~\ref{alg:without seeds} to further reduce computational complexity. We now provide the theoretical justification for this step.

The key insight is that the underlying probability submatrices associated with the unshifted vertex set $\mathcal{U}$ are identical across both networks, i.e., $\mathbf{P}^{(1)}_{\mathcal{U},\mathcal{U}} = \mx_{\mathcal{U}}^{(1)}\mx_{\mathcal{U}}^{(1)\top}= (\mx_{\mathcal{U}}^{(1)}\mw^{(1,2)})(\mw^{(1,2)\top}\mx_{\mathcal{U}}^{(1)\top})=\mx_{\mathcal{U}}^{(2)}\mx_{\mathcal{U}}^{(2)\top}= \mathbf{P}^{(2)}_{\mathcal{U},\mathcal{U}}$, which implies $\Delta_{\mathcal{U},\mathcal{U}}=\mathbf{0}$, where we define $\Delta := {\mathbf{P}}^{(1)} - {\mathbf{P}}^{(2)}$. Then for any correct seed set $\mathcal{S} \subset \mathcal{U}$, we have $\Delta_{k,\ell} = 0$ for any vertex pair $(k,\ell)\in\mathcal{S}\times \mathcal{S}\subset \mathcal{U}\times \mathcal{U}$.

Let $\hat{\Delta} := \hat{\mathbf{P}}^{(1)} - \hat{\mathbf{P}}^{(2)}$ (recall that $\hat\mpp^{(i)}=\hat\mx^{(i)}\hat\mx^{(i)\top}$). The following Theorem~\ref{thm:max} shows that for any $(k, \ell)\in[n]\times [n]$, $\hat{\Delta}_{k,\ell}$ is asymptotically normal with mean $\Delta_{k,\ell}$ and variance of order $O(n^{-1}\rho_n)$. Based on this result, we establish a test statistic $\tilde{T}_{k,\ell}$ to test the null hypothesis $\mathbb{H}_0: \Delta_{k,\ell}=0$ against the alternative $\mathbb{H}_a: \Delta_{k,\ell}\neq 0$. Theorem~\ref{thm:max_HT} shows that $\tilde{T}_{k,\ell}$ converges to a standard normal distribution with mean zero under the null hypothesis and nonzero mean under the local alternative hypothesis.

Using this result, we filter candidate seed sets via the $n\times n$ test statistic matrix $\tilde{\mathbf{T}}=(\tilde{T}_{k,\ell})$, retaining only those candidates that satisfy $\|\tilde{\mathbf{T}}_{\mathcal{S},\mathcal{S}}\|_{\max} \leq \tilde{c}_{\tilde{\alpha}_B/2}$. Here, $\tilde{c}_{\tilde{\alpha}_B/2}$ is the upper $\tilde{\alpha}_B/2$-quantile of the standard normal distribution, where $\tilde{\alpha}_B=\frac{\tilde{\alpha}}{|\mathcal{S}|(|\mathcal{S}|+1)/2}$ applies the Bonferroni correction to the specified significance level $\tilde{\alpha}$. The denominator $|\mathcal{S}|(|\mathcal{S}|+1)/2$ accounts for the total number of hypothesis tests performed on the symmetric submatrix $\tilde{\mathbf{T}}_{\mathcal{S},\mathcal{S}}$.


\begin{theorem1}\label{thm:max}
Consider the setting in Theorem~\ref{thm:CLT} with the unshifted vertex set $\mathcal{U}$.
Let
\begin{equation}\label{eq:Pi}
	\begin{aligned}
	&\bm\Pi^{(i)}:=\mx^{(i)}(\mx^{(i)\top}\mx^{(i)})^{-1}\mx^{(i)\top}\text{ for } i\in \{1,2\},\\
	&\mathbf{\Psi}^{(k)}:=\bm\Pi^{(1)}\mathbf{\Xi}^{(k,1)}\bm\Pi^{(1)}+\bm\Pi^{(2)}\mathbf{\Xi}^{(k,2)}\bm\Pi^{(2)} \text{ for }k\in[n]
\end{aligned}
\end{equation}
where $\mathbf{\Xi}^{(k,i)}$ is defined in Eq.~\eqref{eq:Xi}.
We define $ {\mathbf{\Upsilon}}$ as a symmetric $n \times n$ matrix given by
\begin{equation}\label{eq;tilde Upsilon}
	 {\mathbf{\Upsilon}}_{k,\ell} =
\begin{cases} 
    \mathbf{\Psi}^{(k)}_{\ell,\ell} + \mathbf{\Psi}^{(\ell)}_{k,k}
    +2\bm\Pi^{(1)}_{k,k}\bm\Pi^{(1)}_{\ell,\ell}\mathbf{\Xi}^{(k,1)}_{\ell,\ell}
    +2\bm\Pi^{(2)}_{k,k}\bm\Pi^{(2)}_{\ell,\ell}\mathbf{\Xi}^{(k,2)}_{\ell,\ell} , & \text{for } k \neq \ell, \\
    4 \mathbf{\Psi}^{(k)}_{k,k}, & \text{for } k = \ell.
\end{cases}
\end{equation}
Note that for any vertex pair $(k,\ell)\in[n]\times [n]$, we have $| {\mathbf{\Upsilon}}_{k,\ell}|\lesssim n^{-2}(n\rho_n)$. Further suppose $| {\mathbf{\Upsilon}}_{k,\ell}|\gtrsim n^{-2}(n\rho_n)$.
We then have
$$
	( {\mathbf{\Upsilon}}_{k,\ell})^{-1/2}(\hat\Delta _{k,\ell}-\Delta _{k,\ell})\rightsquigarrow \mathcal{N}(0, 1)
$$
as $n\to\infty$.

\end{theorem1}

\begin{remark1}
Notice that $\hat{\mathbf{P}}^{(1)}$ and $\hat{\mathbf{P}}^{(2)}$ depend only on their respective adjacency matrices $\mathbf{A}^{(1)}$ and $\mathbf{A}^{(2)}$, without any dependence on seed set information. Consequently, while Theorem~\ref{thm:max} references the setting of Theorem~\ref{thm:CLT} for notational convenience, Theorem~\ref{thm:max} itself involves no seed sets and imposes no seed-set-related conditions.
\end{remark1}

We define $\hat{\mathbf{\Upsilon}}$ as a consistent estimator of ${\mathbf{\Upsilon}}$ in Eq.~\eqref{eq;tilde Upsilon} by
\begin{equation}\label{eq:hatUpsilon}
	 \hat{\mathbf{\Upsilon}}_{k,\ell} =
\begin{cases} 
    \hat{\mathbf{\Psi}}^{(k)}_{\ell,\ell} + \hat{\mathbf{\Psi}}^{(\ell)}_{k,k}
    +2\hat{\bm\Pi}^{(1)}_{k,k}\hat{\bm\Pi}^{(1)}_{\ell,\ell}\hat{\mathbf{\Xi}}^{(k,1)}_{\ell,\ell}
    +2\hat{\bm\Pi}^{(2)}_{k,k}\hat{\bm\Pi}^{(2)}_{\ell,\ell}\hat{\mathbf{\Xi}}^{(k,2)}_{\ell,\ell} , & \text{if } k \neq \ell, \\
    4 \hat{\mathbf{\Psi}}^{(k)}_{k,k}, & \text{if } k = \ell,
\end{cases}
\end{equation}
where $\hat{\bm\Pi}^{(i)}:=\hat{\mathbf{X}}^{(i)}(\hat{\mathbf{X}}^{(i)\top}\hat{\mathbf{X}}^{(i)})^{-1}\hat{\mathbf{X}}^{(i)\top},
	\hat{\mathbf{\Psi}}^{(k)}:=\hat{\bm\Pi}^{(1)}\hat{\mathbf{\Xi}}^{(k,1)}\hat{\bm\Pi}^{(1)}+\hat{\bm\Pi}^{(2)}\hat{\mathbf{\Xi}}^{(k,2)}\hat{\bm\Pi}^{(2)},$
and $\hat{\mathbf{\Xi}}^{(k,i)}$ is defined after Eq.~\eqref{eq:hatGamma}.
Notice that to compute $\hat{\mathbf{\Upsilon}}$, we only need the diagonal elements of $\hat{\mathbf{\Psi}}^{(k)}$. Therefore, in practice, computing $\hat{\mathbf{\Psi}}^{(k)}$ only requires calculating its diagonal elements, which significantly reduces both computational cost and storage requirements.

\begin{theorem1}\label{thm:max_HT}
	Consider the setting in Theorem~\ref{thm:max}.
	We define the test statistic
	\begin{equation}\label{eq:tildeT_kl}
		\tilde T_{k,\ell}=(\hat{\mathbf{\Upsilon}}_{k,\ell})^{-1/2} \hat\Delta _{k,\ell}.
	\end{equation}
	Then under $\mathbb{H}_0:\Delta_{k,\ell}= 0$, we have $\tilde T_{k,\ell} \rightsquigarrow \mathcal{N}(0, 1)$ as $n\to\infty$.
	
	Next suppose that $\eta>0$ is a finite constant and that $\Delta_{k,\ell}\neq 0$ satisfies a local alternative hypothesis where $({\mathbf{\Upsilon}}_{k,\ell})^{-1/2} \Delta_{k,\ell}\to\tilde\eta_{k,\ell}.$
	We then have $\tilde T_{k,\ell} \rightsquigarrow \mathcal{N}(\tilde\eta_{k,\ell}, 1)$ as $n\to\infty$.
\end{theorem1}

Notice that $\Delta_{\mathcal{S},\mathcal{S}}=\mathbf{0}$ is a necessary but not sufficient condition for $\mathcal{S}\subset\mathcal{U}$. The condition $\Delta_{\mathcal{S},\mathcal{S}}=\mathbf{0}$ actually implies that the latent positions of vertices in $\mathcal{S}$ across the two networks can be aligned by a common orthogonal transformation, i.e., there exists some $\mathbf{O}\in\mathcal{O}_d$ such that $\mathbf{X}^{(2)}_\mathcal{S}=\mathbf{X}^{(1)}_\mathcal{S}\mathbf{O}$. 
The unshifted vertices in $\mathcal{U}$ satisfy $\mathbf{X}^{(2)}_\mathcal{U}=\mathbf{X}^{(1)}_\mathcal{U}\mathbf{W}^{(1,2)}$, meaning their latent positions across the two networks can all be aligned by $\mathbf{W}^{(1,2)}\in\mathcal{O}_d$, so they satisfy this condition. And if the vertices in some $\mathcal{S}\not\subset\mathcal{U}$ are shifted but happen to have latent positions that can be aligned by another orthogonal transformation $\mathbf{O}\neq\mathbf{W}^{(1,2)}$, they would also satisfy the condition $\Delta_{\mathcal{S},\mathcal{S}}=\mathbf{0}$. (Nevertheless, such seed set candidates $\mathcal{S}$ corresponding to $\mathbf{O}\neq\mathbf{W}^{(1,2)}$ tend to yield fewer estimated unshifted vertices compared to correct seed sets $\mathcal{S}\subset\mathcal{U}$ corresponding to $\mathbf{W}^{(1,2)}$; see the discussion at the end of this section).
This filtering step can ensure, with high probability, that the latent positions of all vertices in each surviving seed set candidate can be aligned via a common orthogonal transformation, under mild conditions such that the non-zero entries ${\Delta}_{k,\ell}$ satisfy $|{\Delta}_{k,\ell}| = \omega(n^{-1/2}\rho_n^{1/2})$, as the variance of $\hat{\Delta}_{k,\ell}$ is of order $n^{-1}\rho_n$ as shown in Theorem~\ref{thm:max}.
Note that $n^{-1/2}\rho_n^{1/2} = (n\rho_n)^{-1/2}\rho_n$, and for $\mpp^{(i)}$ with homogeneous entries, $\mpp^{(i)}_{k,\ell} \asymp \rho_n$. Thus such condition is very mild and allows the differences between corresponding entries relative to the entries themselves to approach zero at a rate close to $(n\rho_n)^{-1/2}$.

Note that as discussed in Section~\ref{sec:alg_no seed}, although $d$ is bounded under our low-rank assumption, when $d$ is relatively large, the number of required random seed set candidates can still be substantial.
In such cases, rather than randomly generating a large number of candidates and then filtering out most of them using the above filtering step, we provide an alternative algorithm that directly generates random seed set candidates within the feasible region such that $|\tilde{T}_{k,\ell}| \leq \tilde{c}_{\tilde{\alpha}_B/2}$ for all pairs in each candidate set. See Algorithm~\ref{alg:feasible_sampling} for details.

\begin{algorithm}[htbp!]
\small
\begingroup
\setlength{\baselineskip}{1\baselineskip}
\caption{Sampling of feasible seed set candidates}\label{alg:feasible_sampling}
\begin{algorithmic}[1]
\REQUIRE{Test statistic matrix $\tilde{\mathbf{T}}$, threshold $\tilde{c}_{\tilde{\alpha}_B/2}$, seed set size $L\geq d$, number of candidate seed sets $M$.}
\begin{enumerate}
    \item Construct feasibility matrix $\mathbf{F} \in \{0,1\}^{n \times n}$ where
$$\mathbf{F}_{k,\ell} = \begin{cases} 
1, & \text{if } |\tilde{T}_{k,\ell}| \leq \tilde{c}_{\tilde{\alpha}_B/2} \\
0, & \text{otherwise}
\end{cases}$$

    \item Initialize the candidate set collection $\mathcal{C} = \varnothing$.

    \item \textbf{While} $|\mathcal{C}| < M$:
    \begin{enumerate}
        \item[(a)] Randomly select an initial vertex $v_1 \in [n]$ with $\mathbf{F}_{v_1,v_1}=1$.
        \item[(b)] Initialize candidate seed set $\mathcal{S}_{\text{temp}} = \{v_1\}$.
        \item[(c)] \textbf{For} $i = 2$ to $L$:
        \begin{enumerate}
            \item[i.] Compute feasible vertices: 
            $$\mathcal{V}_{\text{feasible}} = \{v \in [n] \setminus \mathcal{S}_{\text{temp}} : \mathbf{F}_{v,v}=1 \text{ and } \mathbf{F}_{u,v} = 1 \text{ for all } u \in \mathcal{S}_{\text{temp}}\}$$
            \item[ii.] \textbf{If} $\mathcal{V}_{\text{feasible}} = \varnothing$, break and restart with new initial vertex.
            \item[iii.] Randomly select $v_i \in \mathcal{V}_{\text{feasible}}$ and update $\mathcal{S}_{\text{temp}} = \mathcal{S}_{\text{temp}} \cup \{v_i\}$.
        \end{enumerate}
        \item[(d)] \textbf{If} $|\mathcal{S}_{\text{temp}}| = L$, set $\mathcal{C} = \mathcal{C} \cup \{\mathcal{S}_{\text{temp}}\}$.
    \end{enumerate}

\end{enumerate}

\ENSURE{Collection of feasible seed set candidates $\mathcal{C}$.}

\end{algorithmic}

\endgroup
\end{algorithm}

An important advantage of this direct sampling approach is that it significantly reduces the required number of candidates $M$ when $d$ is relatively large. For example, consider a scenario where all vertices outside the unshifted vertex set $\mathcal{U}$ can be partitioned into small groups such that vertices within \textit{each} group have latent positions across the two networks that are alignable by a common orthogonal transformation (different from $\mw^{(1,2)}$).
Since the direct sampling approach, with high probability, only generates seed set candidates where the latent positions of all vertices in each candidate can be aligned by a common orthogonal transformation (as discussed above), whether a candidate entirely belongs to $\mathcal{U}$ depends on whether the first randomly generated vertex belongs to $\mathcal{U}$.
Recall that $p$ is the proportion of unshifted vertices. Therefore, each time we generate a candidate by the direct sampling approach, there is a probability $p$ of obtaining a candidate that belongs to $\mathcal{U}$.
Thus, to ensure that at least one candidate seed set is correct with probability at least $q$ among the total $M$ candidates, it suffices to have 
$M \geq \frac{\ln(1-q)}{\ln(1-p)}.$
This bound does not depend on $d$, so it is not influenced by the relatively large $d$. 
Notice that this analysis considers a challenging scenario, whereas in the problem setting we consider, as shown in our motivating example in Figure~\ref{fig:food_country 15}, the shifts for shifted vertices usually do not have the same direction, so only a few shifted vertices would coincidentally be aligned by the common orthogonal transformation. Therefore, in practice, the required number of candidates can be even smaller.

Finally, we discuss the theoretical guarantee of Algorithm~\ref{alg:without seeds} for selecting a correct seed set that belongs to $\mathcal{U}$. Recall that the problem we are concerned with is that between two networks, a substantial number of unshifted vertices have latent positions that can be aligned by the common orthogonal transformation $\mathbf{W}^{(1,2)}$, while other vertices exhibit a shift after being rotated by $\mathbf{W}^{(1,2)}$.
Now we classify all vertices into vertices whose latent positions across the two networks cannot be aligned by any orthogonal transformation and the vertices that can be aligned by some orthogonal transformation.
Furthermore, among the vertices that can be aligned by some orthogonal transformation, we can define vertex sets according to which specific orthogonal transformation aligns them. 
Our problem actually assumes that among the alignable vertices, there exists a unique largest set, namely $\mathcal{U}$ corresponding to $\mathbf{W}^{(1,2)}\in\mathcal{O}_d$, while other sets, denoted by $\mathcal{V}_1,\ldots,\mathcal{V}_f$ corresponding to $\mathbf{O}_1,\ldots,\mathbf{O}_f\in\mathcal{O}_d$, are all relatively much smaller (recall our motivating example in Figure~\ref{fig:food_country 15}, where for shifted vertices, these shifts usually do not have the same direction). In our analysis here, we allow a few shifted vertices to coincidentally be aligned by a common orthogonal transformation, giving rise to $\mathcal{V}_1,\ldots,\mathcal{V}_f$. 
Let $p:=|\mathcal{U}|/n$ and $p_i:=|\mathcal{V}_i|/n$. In our setting, we assume that each $p_i$ is much smaller than $p$.

Our filtering step ensures that each surviving seed set candidate is a subset of one of $\mathcal{U},\mathcal{V}_1,\ldots,\mathcal{V}_f$ with high probability (as discussed above). 
1) For $\mathcal{S}\subset\mathcal{U}$ and the corresponding estimated unshifted vertex set $\hat{\mathcal{U}}$ obtained with $\mathcal{S}$, according to Corollary~3 in \cite{izmirlian2020strong} for the Benjamini-Hochberg procedure, we have $\frac{|\hat{\mathcal{U}}|}{|{\mathcal{U}}|}\geq 1-\alpha$ almost surely as $n\to\infty$, and thus $|\hat{\mathcal{U}}|\geq (1-\alpha)pn$ almost surely.
2) For $\mathcal{S}\subset\mathcal{V}_i$ and the corresponding estimated unshifted vertex set $\hat{\mathcal{U}}$, according to Theorem~2 in \cite{izmirlian2020strong}, we have $|\hat{\mathcal{U}}|\leq p_i n+(1-\pi_{\text{pi}})(1-p_i)n$ almost surely as $n\to\infty$, where $\pi_{\text{pi}}$ is the test power (see Corollary~4 in \cite{izmirlian2020strong} for details), and $\pi_{\text{pi}}$ goes to $1$ for a given $\alpha$, provided mild conditions such that for vertices not in $\mathcal{V}_i$, the shifts remaining after rotation with $\mathbf{O}_i$ have $\ell_2$ norm of order $\omega(n^{-1/2})$, as the covariance matrices of the estimated shifts have order $n^{-1}$ as shown in Theorem~\ref{thm:CLT}. Note that the latent positions of vertices have $\ell_2$ norm of order $n^{-1/2}(n\rho_n)^{1/2}$; thus this condition is very mild and allows the shifts relative to the latent positions to approach zero at a rate close to $(n\rho_n)^{1/2}$.
Therefore, if $p > p_i/(1-\alpha)$ for all $i$, i.e., the gap between $p$ and $\max_i\{p_i\}$ is large enough (this condition holds in our problem setting as mentioned above), then the seed set candidate with the largest corresponding $|\hat{\mathcal{U}}|$ is a subset of $\mathcal{U}$ with high probability. That is, Algorithm~\ref{alg:without seeds} selects a correct seed set that belongs to $\mathcal{U}$ with high probability. See Section~\ref{sec:simu_noseeds} for simulation results demonstrating the high empirical accuracy of Algorithm~\ref{alg:without seeds}.

\section{Simulation Experiments}\label{sec:sim}

We now present simulation experiments to complement our theoretical results and showcase the performance of our algorithms.

\subsection{Estimation error of shifts}
\label{sec:estimation error}

We consider two RDPGs on $n = 2000$ vertices, where the latent positions of half the vertices are shifted between them, and the dimension of their latent positions is $d = 3$.
The two networks share the latent positions for the first $1000$ vertices, and the latent positions in $\mx^{(1)}_\mathcal{U}=\mx^{(2)}_\mathcal{U}$, where $\mathcal{U}=\{1,2,\dots,1000\}$, are independently and randomly generated. Specifically, we first uniformly generate values between $0$ and $1$, then take their square root and divide by $\sqrt{d}$, which ensures that the values in the resulting probability matrix fall between $0$ and $1$.
We use the same procedure to separately generate the latent positions of their distinct last $1000$ vertices, $\mx^{(1)}_{\mathcal{U}^\mathcal{C}}$ and $\mx^{(2)}_{\mathcal{U}^\mathcal{C}}$, where $\mathcal{U}^\mathcal{C}=\{1001,\dots,2000\}$. In this setting, $\my = \mx^{(2)} - \mx^{(1)}$ with the shift $\my_{\mathcal{U}^\mathcal{C}} = \mx^{(2)}_{\mathcal{U}^\mathcal{C}} - \mx^{(1)}_{\mathcal{U}^\mathcal{C}}$ and $\my_{\mathcal{U}} = \mathbf{0}$.
We then simulate adjacency matrices $\ma^{(1)}$ and $\ma^{(2)}$ using the edge probability matrices $\mpp^{(1)}=\mx^{(1)}\mx^{(1)\top}$ and $\mpp^{(2)}=\mx^{(2)}\mx^{(2)\top}$.
Given $\ma^{(1)}$ and $\ma^{(2)}$, we estimate $\my$ with the seed set $\mathcal{S}=\{1,2,\dots,100\}$ by Algorithm~\ref{alg:with seeds}.

\begin{figure}[htbp!]
\centering
\subfigure
{\includegraphics[height=4cm]{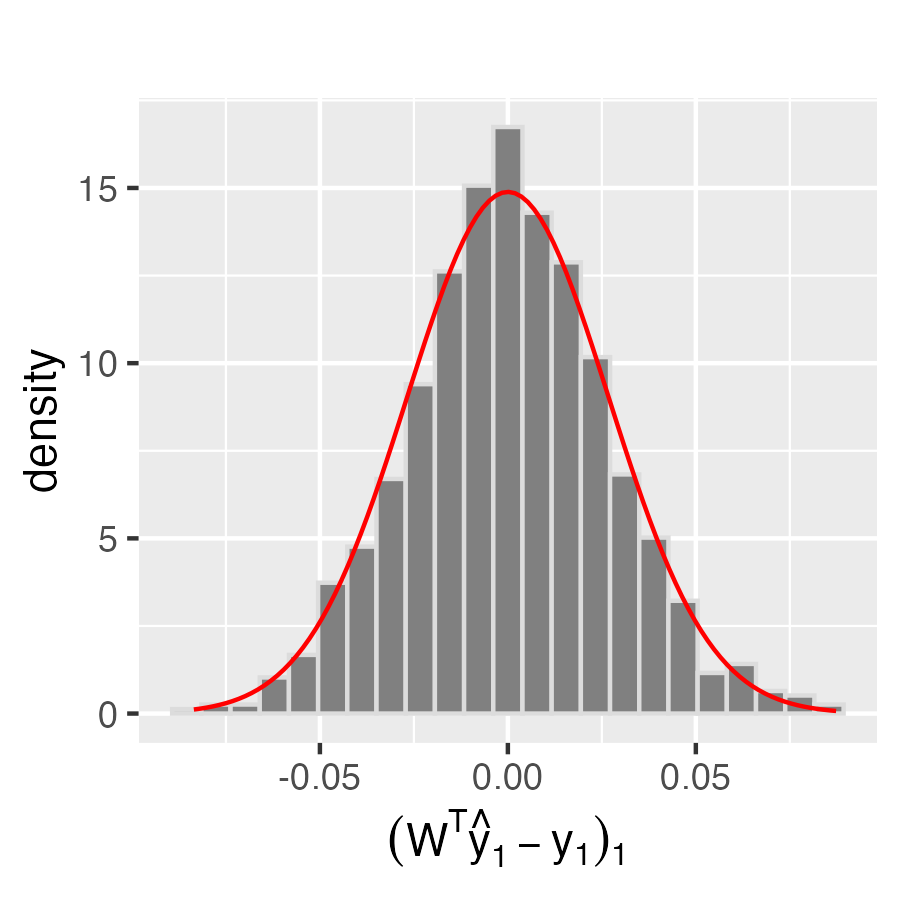}}
\subfigure
{\includegraphics[height=4cm]{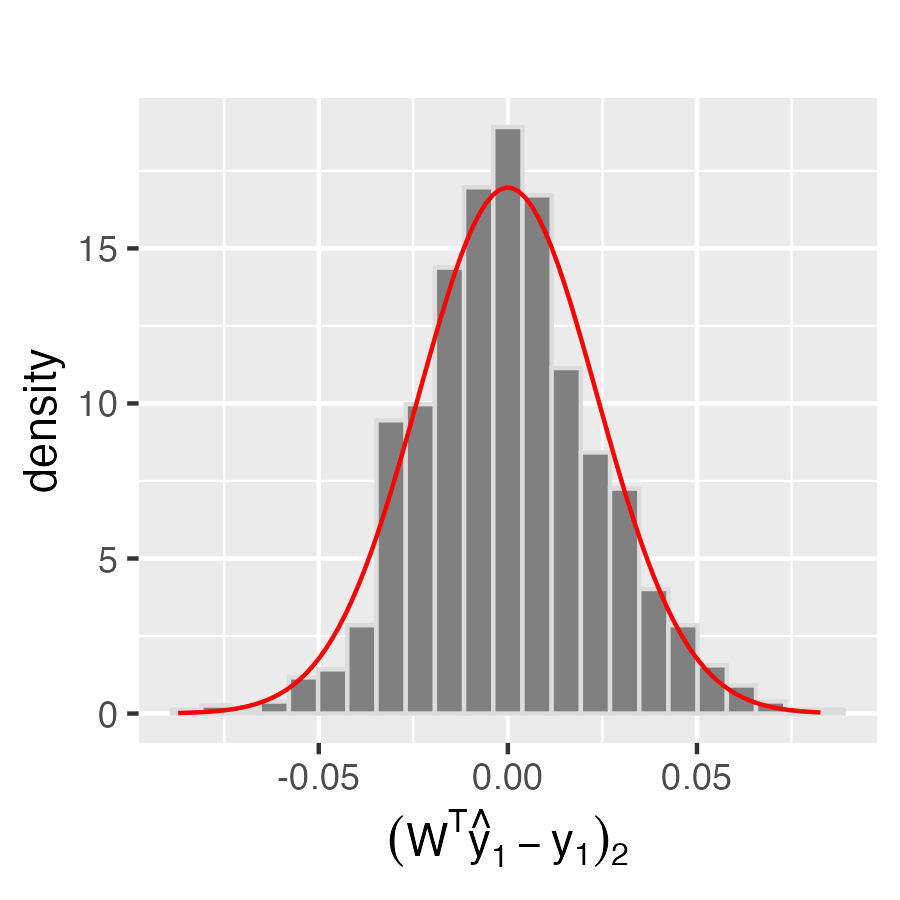}}
\subfigure
{\includegraphics[height=4cm]{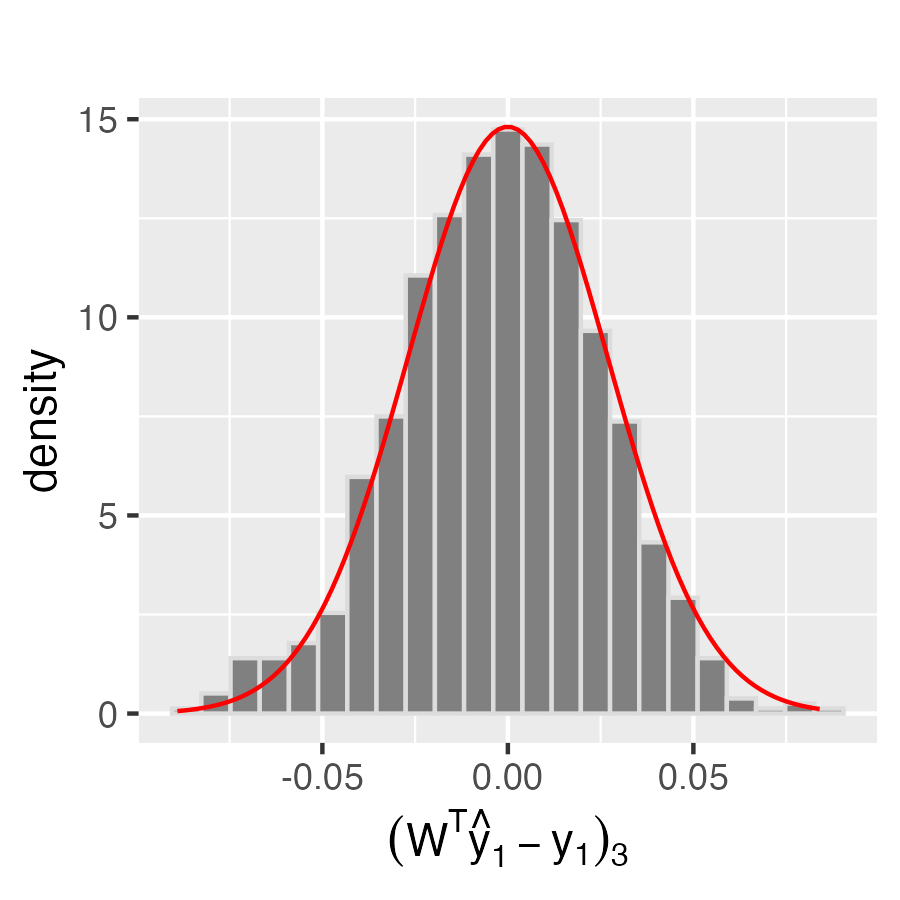}}

\caption{\footnotesize Histograms of the empirical distributions of the entries of the estimation error $\mw^\top\hat{\mathbf{y}}_k - {\mathbf{y}}_k$ for $k = 1$. 
These histograms are based on $1000$ independent Monte Carlo replicates of two RDPGs on $n = 2000$ vertices with $d = 3$ dimensional latent positions, where half of the vertices are shifted.
The red lines represent the probability density functions of the normal distributions with parameters specified in Theorem~\ref{thm:CLT}.
}
\label{fig:simulation_CLT1}
\end{figure}

\begin{figure}[htbp!]
\centering

\subfigure
{\includegraphics[height=4.2cm]{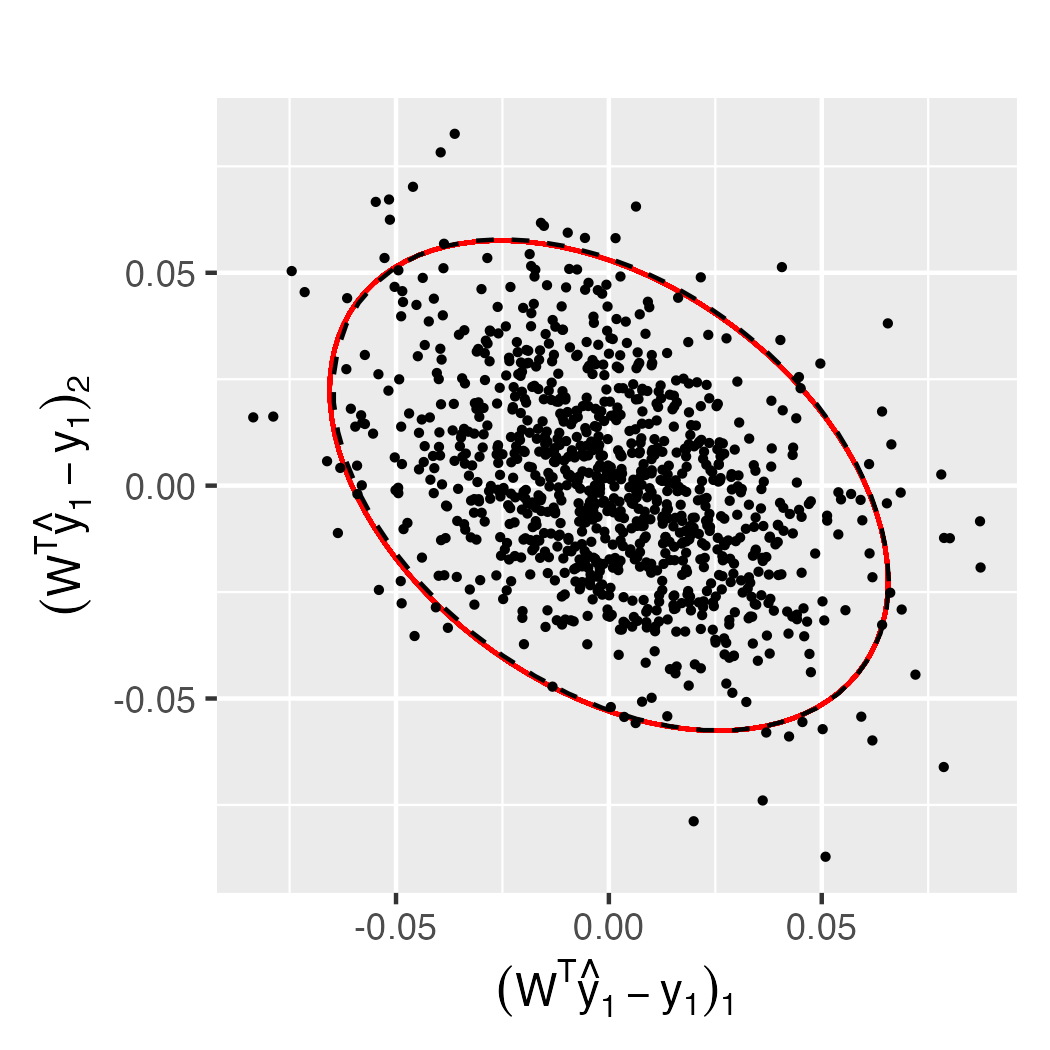}}
\subfigure
{\includegraphics[height=4.2cm]{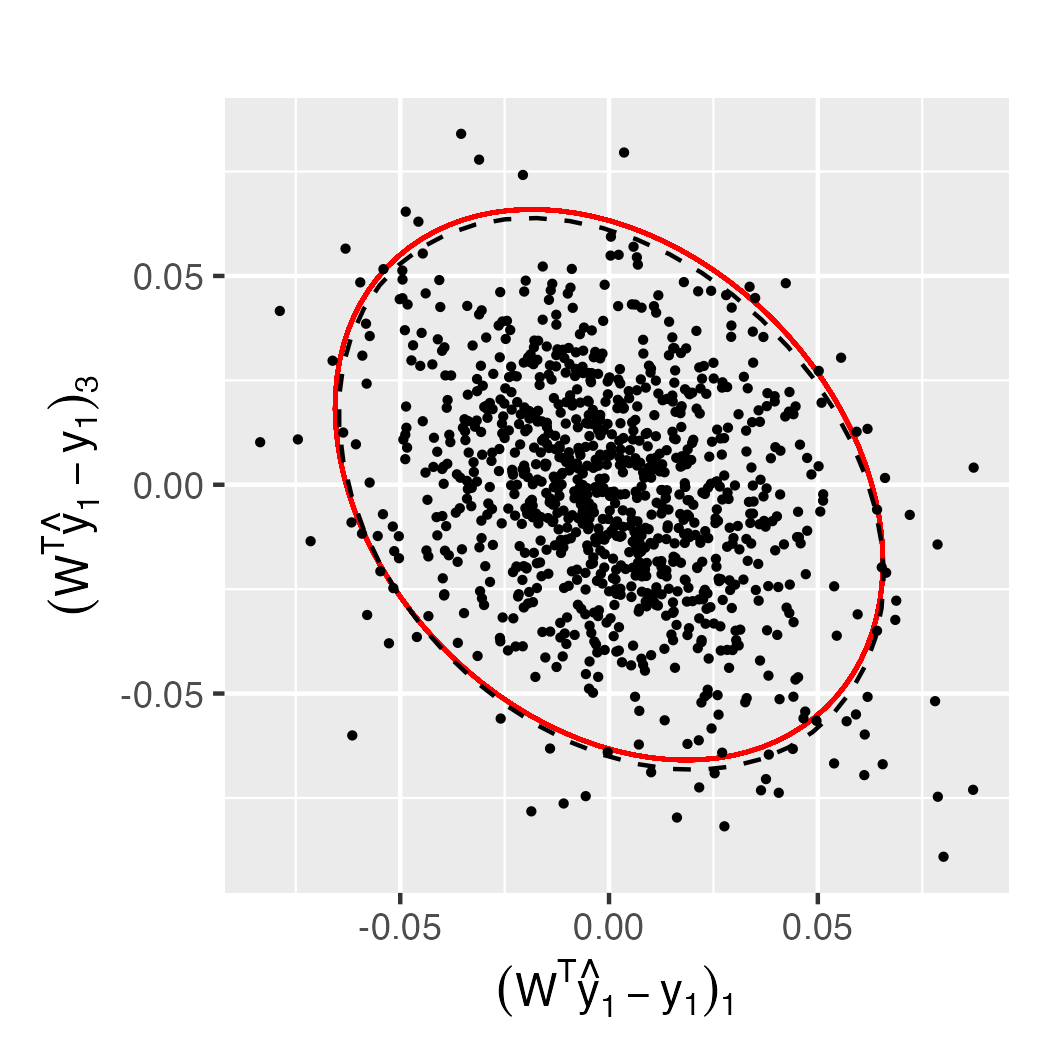}}
\subfigure
{\includegraphics[height=4.2cm]{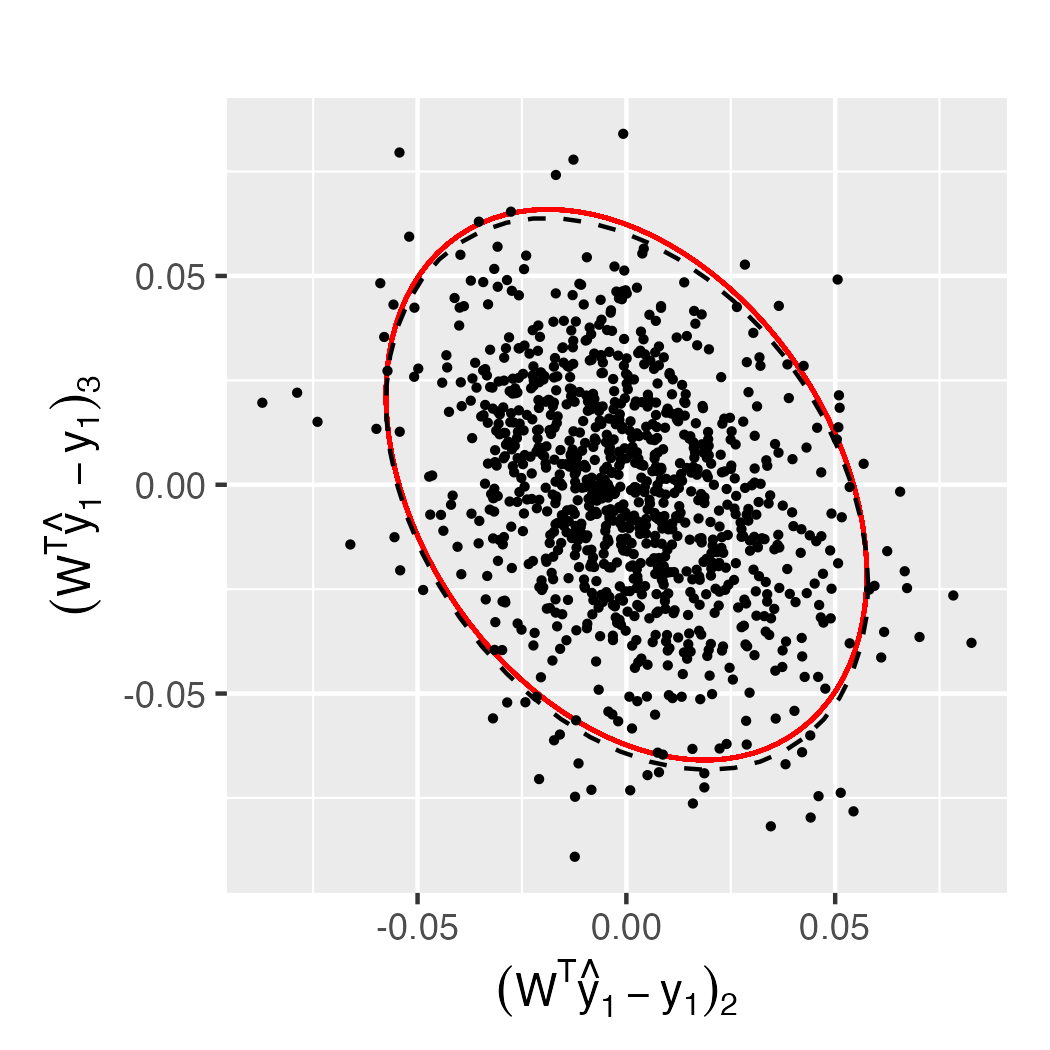}}

\caption{\footnotesize Bivariate plots for the empirical distributions between the entries of the estimation error $\mw^\top\hat{\mathbf{y}}_k - {\mathbf{y}}_k$ for $k = 1$ based on $1000$ Monte Carlo replicates of the same setting with Figure~\ref{fig:simulation_CLT1}. Dashed black ellipses represent 95\% level curves for the empirical distributions while solid red ellipses represent 95\% level curves for the theoretical distributions as specified in Theorem~\ref{thm:CLT}.
}
\label{fig:simulation_CLT2}
\end{figure}

We repeat the above steps for $1000$ Monte Carlo iterations to obtain an empirical distribution of the estimation error $\mw^\top\hat{\mathbf{y}}_k-{\mathbf{y}}_k$ for $k=1$ which we then compare against the limiting distribution given in Theorem~\ref{thm:CLT}. The results are summarized in Figure~\ref{fig:simulation_CLT1} and Figure~\ref{fig:simulation_CLT2}. Henze-Zirkler's normality test indicates that the empirical distribution of $\mw^\top\hat{\mathbf{y}}_k-{\mathbf{y}}_k$ is well-approximated by a multivariate normal distribution and furthermore the empirical covariances for $\mw^\top\hat{\mathbf{y}}_k-{\mathbf{y}}_k$ are very close to the theoretical covariances.

\subsection{Test statistic for $\mathbb{H}_0:\mathbf{y}_k=\mathbf{0}$}

\begin{figure}[htbp!]
\centering

\subfigure
{\includegraphics[height=4cm]{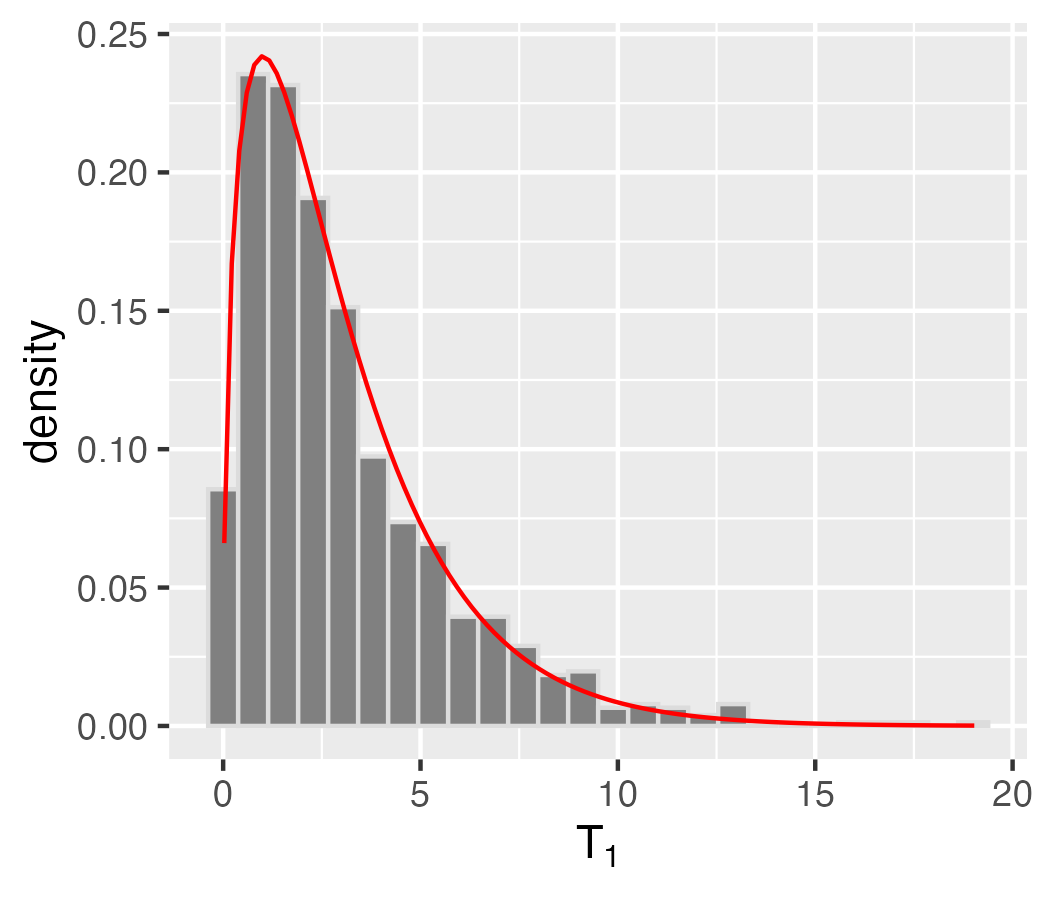}}
\subfigure
{\includegraphics[height=4cm]{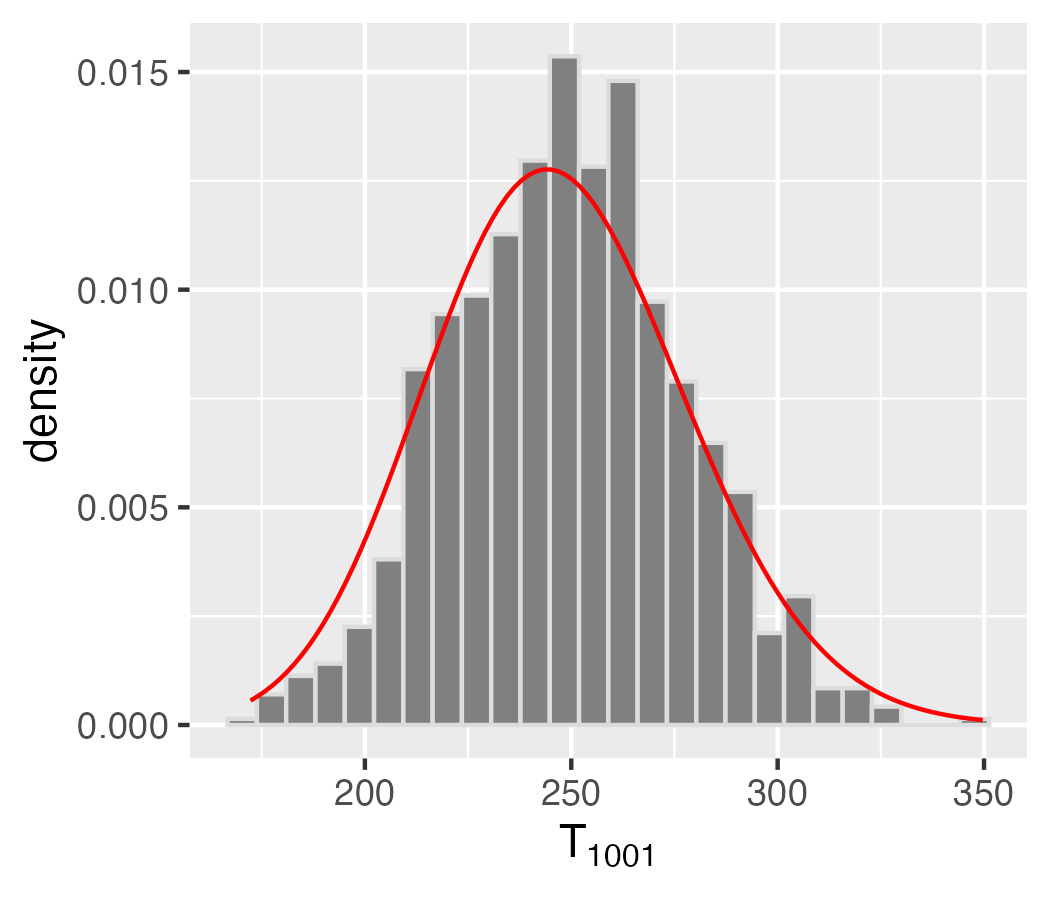}}

     \caption{\footnotesize Histograms of the empirical distributions of $T_k$ for an unshifted vertex ($k = 1$) and a shifted vertex ($k = 1001$), corresponding to the null and local alternative hypotheses, respectively.
     The histograms are based on $1000$ independent Monte Carlo replicates of two RDPGs on $n = 2000$ vertices with $d = 3$ dimensional latent positions, where half of the vertices are shifted.
     The red lines are the probability density functions for the central and non-central chi-square distributions with $d=3$ degrees of freedom and non-centrality parameters given in Theorem~\ref{thm:HT}.
}
\label{fig:simulation_HT}
\end{figure}

Under the setting described in Section~\ref{sec:estimation error}, we test $\mathbb{H}_0: \mathbf{y}_k = \mathbf{0}$ against $\mathbb{H}_a: \mathbf{y}_k \neq \mathbf{0}$ for an unshifted vertex ($k = 1$) and a shifted vertex ($k = 1001$), respectively. The vertex $k = 1$, which retains the same latent position in both networks, corresponds to the null hypothesis, while the vertex $k = 1001$, whose latent position shifts between the two networks, corresponds to the local alternative hypothesis.
For each Monte Carlo replicate, we compute the test statistic $T_k$ from Theorem~\ref{thm:HT} and compare its empirical distributions under the null and alternative hypotheses to the central and non-central $\chi^2$ distributions with $d = 3$ degrees of freedom. The results, summarized in Figure~\ref{fig:simulation_HT}, show that the empirical distributions of $T_k$ are well approximated by the theoretical distributions.

\subsection{Performance with limited seeds}

\label{sec:simu_small seeds}

\begin{figure}[htbp!]
\centering

\subfigure
{\includegraphics[height=5cm]{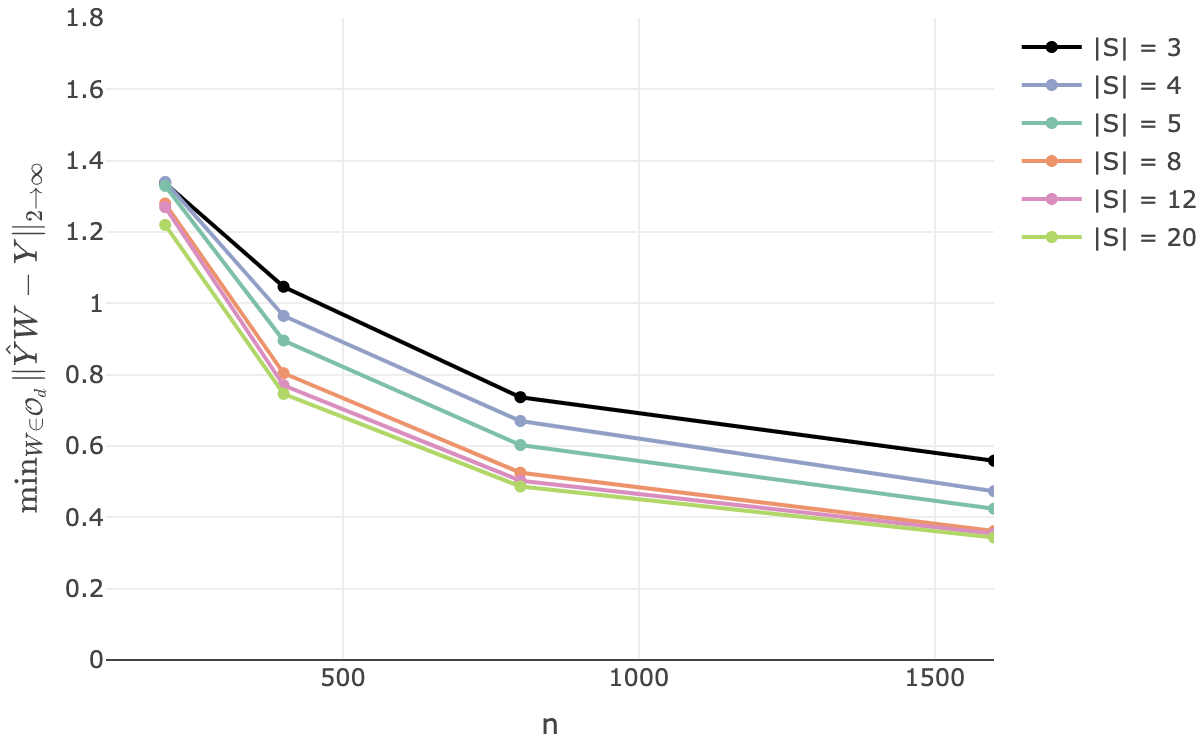}}

     \caption{\footnotesize 
     The empirical error rate of $\min_{\mw\in\mathcal{O}_d}\|\hat\my \mw-\my\|_{2\to\infty}$ as we vary the number of vertices $n \in \{200,400,800,1600\}$ for the seed set size $|\mathcal{S}|=\{3,4,5,8,12,20\}$.
The errors are averaged over $100$ independent Monte Carlo replicates of two RDPGs on $n$ vertices with $d=3$ dimensional latent positions, where half of the vertices are shifted.
}
\label{fig:simulation_minimalS}
\end{figure}

Remark~\ref{rem:small S} states that, according to Theorem~\ref{thm:X1W-X2}, our algorithms remain effective even when the seed set size $|\mathcal{S}|$ is small, as the estimation error of the shift $\my$ in terms of the $2\to\infty$ norm decreases at a rate of $n^{-1/2} \log^{1/2} n$, regardless of $|\mathcal{S}|$.
To validate the results, we conduct simulations using the same settings of two RDPGs as in Section~\ref{sec:estimation error}, but with the number of vertices $n$ varying. The first half of the vertices consistently exhibit no shift, i.e., $\mathcal{U} = \{1, 2, \dots, n/2\}$, while the second half of the vertices are shifted, i.e., $\mathcal{U}^\mathcal{C} = \{n/2+1, n/2+2, \dots, n\}$.
We compute the estimation error $\min_{\mw \in \mathcal{O}_d} \|\hat{\my} \mw - \my\|_{2\to\infty}$ as we vary the number of vertices $n$ for different values of the seed set size $|\mathcal{S}|$, including the minimal possible seed set size $|\mathcal{S}| = d = 3$.
The results are summarized in Figure~\ref{fig:simulation_minimalS}.
The error rate of each line in Figure~\ref{fig:simulation_minimalS} approximately matches the theoretical error rate of $n^{-1/2} \log^{1/2}{n}$ in Theorem~\ref{thm:X1W-X2}.
Figure~\ref{fig:simulation_minimalS} shows that, as mentioned in Remark~\ref{rem:small S}, the estimation errors obtained for different $|\mathcal{S}|$ are of the same order, and the increase in $|\mathcal{S}|$ has a limited effect on reducing the estimate error.
For example, the errors for $|\mathcal{S}| = 8$ and $|\mathcal{S}| = 20$ are almost the same.

\subsection{Test statistics for $\mathbb{H}_0:\Delta_{k,\ell}= 0$}\label{sec:simu_delta}

\begin{figure}[htbp!]
\centering

\subfigure
{\includegraphics[height=3.8cm]{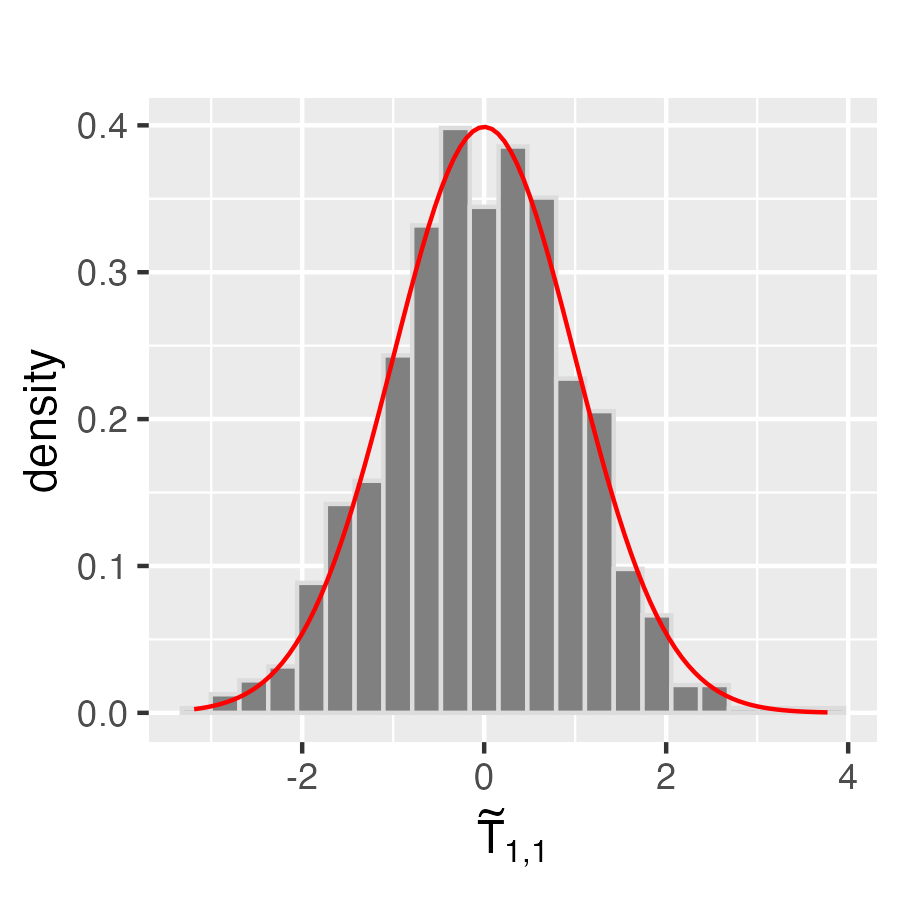}}
\subfigure
{\includegraphics[height=3.8cm]{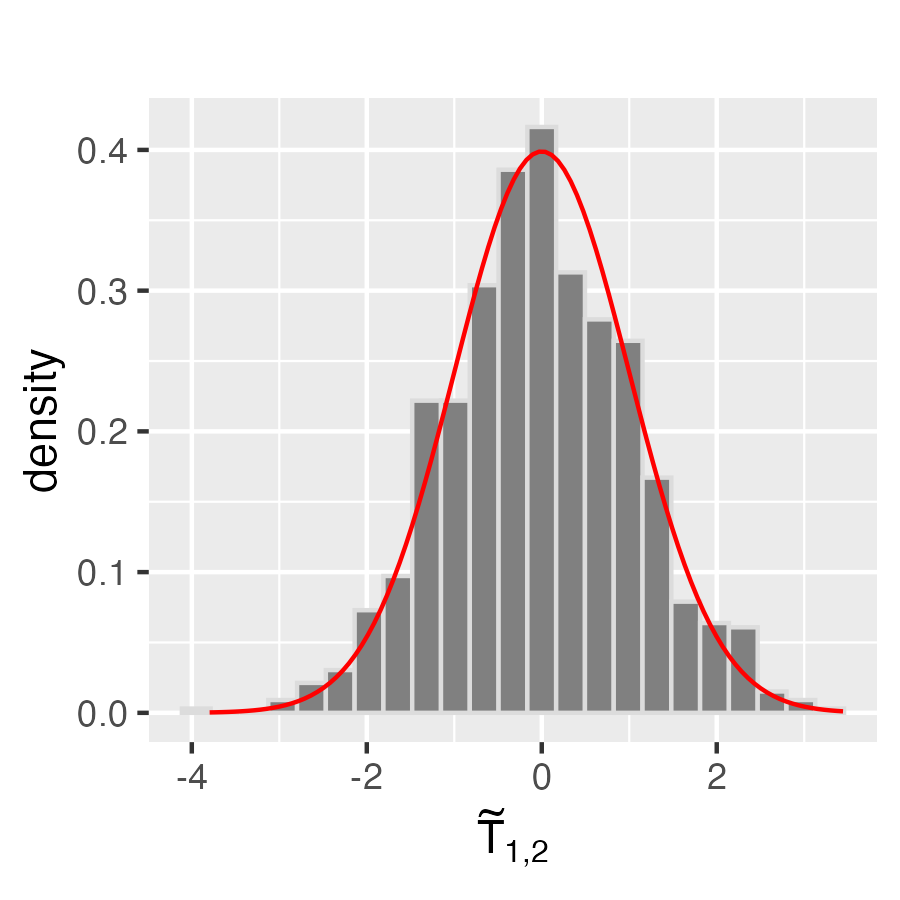}}
\subfigure
{\includegraphics[height=3.8cm]{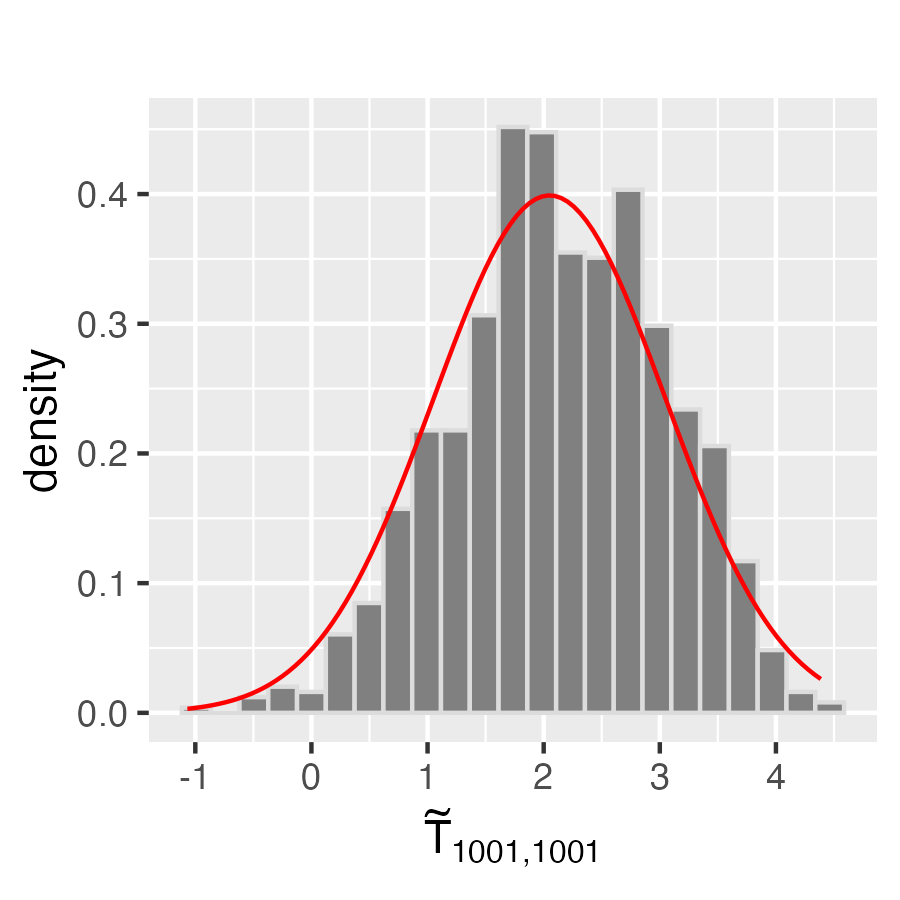}}
\subfigure
{\includegraphics[height=3.8cm]{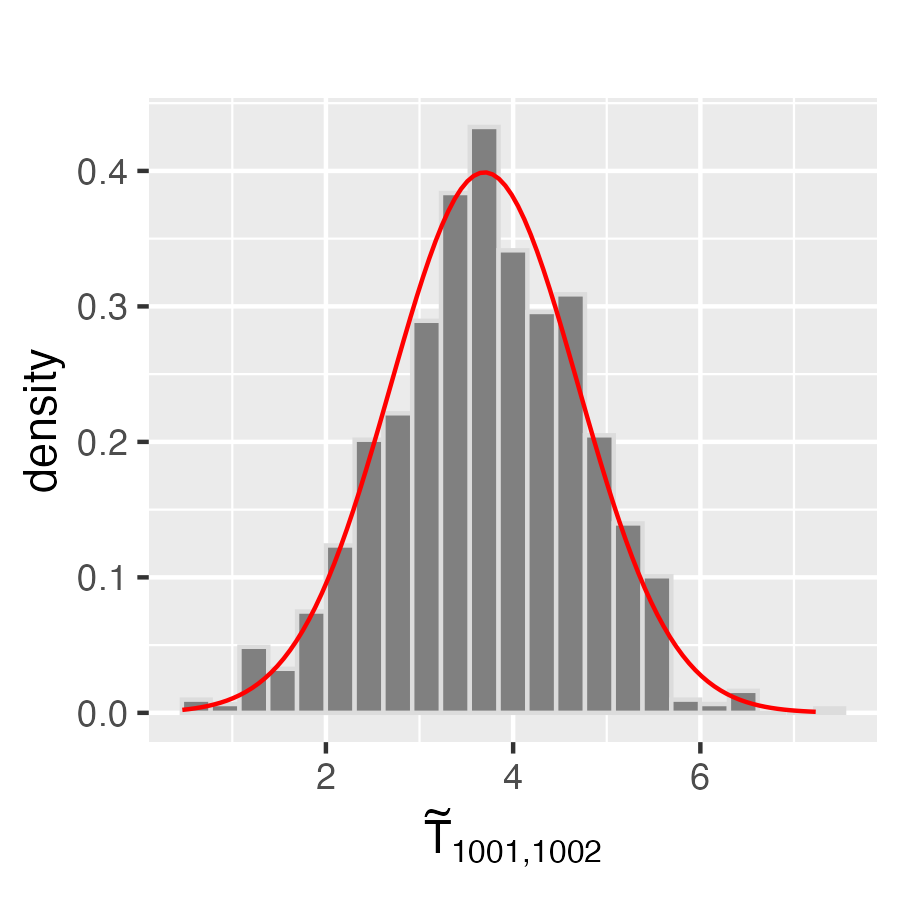}}

     \caption{\footnotesize Histograms of the empirical distributions of $\tilde{T}_{k,\ell}$ for two pairs from unshifted vertices $(1,1)$ and $(1,2)$, corresponding to the null hypothesis, and for two pairs from shifted vertices $(1001,1001)$ and $(1001,1002)$, corresponding to the local alternative hypotheses.
The histograms are based on $1000$ independent Monte Carlo replicates of two RDPGs on $n = 2000$ vertices with $d = 3$ dimensional latent positions, where half of the vertices are shifted.
The red lines are the probability density functions of the standard normal distributions with zero mean and nonzero mean given in Theorem~\ref{thm:max_HT}.
}
\label{fig:simulation_max_HT}
\end{figure}

Under the setting described in Section~\ref{sec:estimation error}, we test $\mathbb{H}_0:\Delta_{k,\ell}= 0$ against $\mathbb{H}_a:\Delta_{k,\ell}\neq 0$ for two pairs from unshifted vertices $(1,1)$ and $(1,2)$ and two pairs from shifted vertices $(1001,1001)$ and $(1001,1002)$.
Note that for the two pairs from unshifted vertices $(1,1)$ and $(1,2)$, the corresponding entries across the two probability matrices are the same, satisfying the null hypothesis $\Delta_{k,\ell}=0$. For the two pairs from shifted vertices $(1001,1001)$ and $(1001,1002)$, the corresponding entries across the two probability matrices are different, satisfying the alternative hypothesis $\Delta_{k,\ell}\neq 0$.
For each Monte Carlo replicate, we compute the test statistic $\tilde{T}_{k,\ell}$ for each pair $(k,\ell)$ from Theorem~\ref{thm:max_HT} and compare its empirical distribution under the null and alternative hypotheses to the standard normal distributions with zero mean and nonzero mean, respectively. The results, summarized in Figure~\ref{fig:simulation_max_HT}, show that the empirical distributions of $\tilde{T}_{k,\ell}$ are well approximated by the theoretical distributions.

\subsection{Shift detection without seeds}

\label{sec:simu_noseeds}

We now explore the performance of Algorithm~\ref{alg:without seeds} in the case where no seeds are available.
Recall that in a stochastic block model (SBM) \cite{holland1983stochastic} with $n$ vertices and $d$ blocks, the probability matrix is given by $\mathbf{P} = \mathbf{Z}\mathbf{B}\mathbf{Z}^\top$, where $\mathbf{Z} \in \mathbb{R}^{n \times d}$, with entries in $\{0, 1\}$ satisfying $\sum_{k=1}^d \mathbf{Z}_{s,k} = 1$ for all $s \in [n]$, represents the block assignments of the vertices, and $\mathbf{B} \in \mathbb{R}^{d \times d}$, with entries in $[0,1]$, represents the edge probabilities between blocks.
We consider two SBMs in which the block assignments of half of vertices are shifted.
We let the block-wise probability matrix 
$$\mb= 
\begin{bmatrix}
0.7 & 0.1 & 0.1 \\
0.1 & 0.65 & 0.1 \\
0.1 & 0.1 & 0.6
\end{bmatrix},
$$
and the block assignments of the vertices are randomly generated.
More specifically, the block assignments of the vertices for the first SBM are randomly generated, with each vertex independently assigned to one of the three blocks with equal probability of $1/3$.
For the second SBM, the first half of the vertices retain the same block assignments as in the first SBM, i.e., $\mathbf{Z}_{\mathcal{U}}^{(1)} = \mathbf{Z}_{\mathcal{U}}^{(2)}$, where $\mathcal{U} = \{1, 2, \dots, n/2\}$, and the block assignments of the second half of the vertices are changed, with each shifted vertex randomly assigned to one of the two remaining blocks (other than its original block in the first SBM) with equal probability of $1/2$, which means $\mathbf{Z}_{\mathcal{U}^\mathcal{C}}^{(1)} = \mathbf{Z}_{\mathcal{U}^\mathcal{C}}^{(2)}$, where ${\mathcal{U}^\mathcal{C}} = \{n/2+1, n/2+2, \dots, n\}$.

\begin{figure}[htbp!]
\centering

\subfigure[\footnotesize Accuracy of shift detection]
{\includegraphics[height=4.5cm]{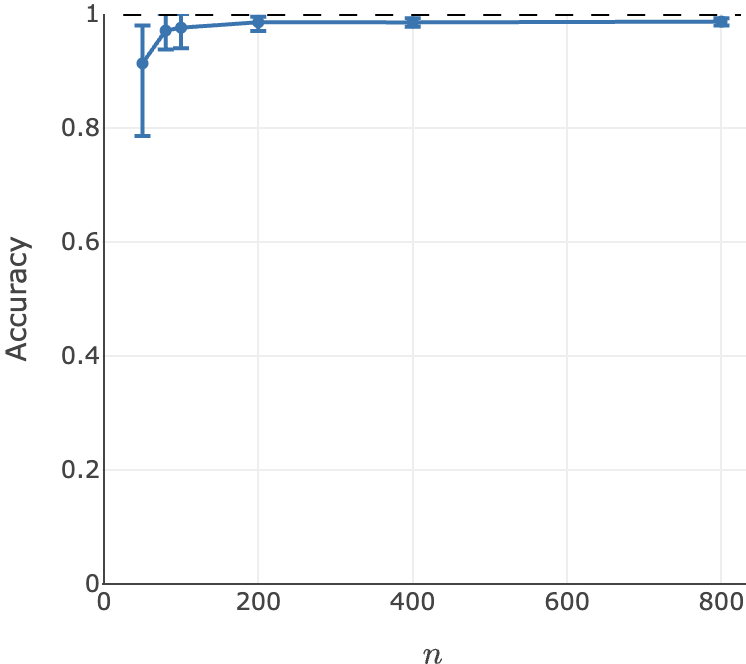}}
\subfigure[\footnotesize Running time]
{\includegraphics[height=4.5cm]{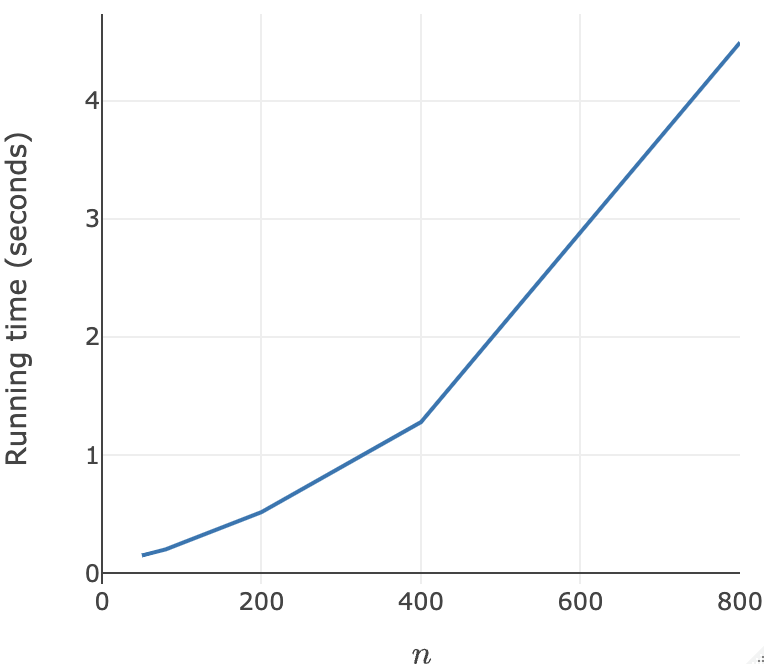}}

        \caption{\footnotesize Panel~(a) illustrates the accuracy of detecting shifted and unshifted vertices without seeds, as we vary the number of vertices $n \in \{50, 80, 100, 200, 400, 800\}$ while setting $L=3$, $M=1000$, $\alpha=0.05$, $\tilde\alpha=0.3$.
     Panel (a) reports the mean, along with the $0.05$ and $0.95$-quantile points, based on $100$ independent Monte Carlo replicates of two SBMs on $n$ vertices with $d=3$ blocks, where the block assignments of half of the vertices are shifted.
Panel~(b) reports the average running time on a standard computer across all replicates.
}
\label{fig:simulation_noseeds}
\end{figure}

Assuming no prior knowledge of the unshifted vertices, we apply Algorithm~\ref{alg:without seeds} to detect which vertices are shifted and which are unshifted.
The performance is then evaluated using accuracy, which is defined as the proportion of vertices correctly classified as shifted or unshifted out of the total number of vertices.
Figure~\ref{fig:simulation_noseeds} presents the accuracy results and running time as $n$ varies in $\{50, 80, 100, 200, 400, 800\}$, while the parameters of Algorithm~\ref{alg:without seeds} are set to $L = 3$, $M = 1000$, $\alpha = 0.05$, and $\tilde\alpha = 0.3$.
It demonstrates that Algorithm~\ref{alg:without seeds} is both highly accurate, achieving high detection accuracy with as few as just $n = 80$ vertices, and computationally efficient, requiring only about $4.5$ seconds per Monte Carlo replicate on a standard computer even when $n = 800$.

\section{Real Data Experiments}\label{sec:real}

In this section, we apply our algorithms to analyze real datasets, including comparing the brain networks of ADHD patients and controls to detect key brain regions associated with ADHD, as well as analyzing the dynamic chocolate trading network to identify countries with changes in trading patterns over time.

\subsection{Brain networks comparison study for attention deficit hyperactivity disorder}\label{sec:brain}

Attention Deficit Hyperactivity Disorder (ADHD) affects at least 5-10\% of school-age children \citep{polanczyk2007worldwide} and is associated with substantial lifelong impairment.
To understand the neurobiological mechanisms underlying ADHD, it is essential to investigate the key brain regions associated with ADHD. 
In this experiment, we perform our comparison study on the brain networks of ADHD patients and healthy controls. By identifying vertices that exhibit significant embedding shifts, we aim to pinpoint the corresponding brain regions that are likely to play critical roles in ADHD, offering insights into its neurobiological basis. 

ADHD-200 data is shared through the Intenational Neuroimaging Datasharing Initiative \citep{mennes2013making}, and includes rs-fMRI, structural MRI, and basic information for individuals: some typically-developing controls and patients diagnosed with ADHD \citep{adhd2012adhd}.
In this experiment, we use the frontal2D dataset from the NBR package \citep{nbr}, which was derived from the ADHD200 dataset, with the variables manipulated to differ from the original data.
Functional connectivity was measured as the Pearson correlation between the average fMRI signal from the brain regions of interest, i.e., $n=28$ anatomical areas of the frontal lobe.
We consider the total of $17$ female individuals in the frontal2D dataset, including $5$ ADHD patients and $12$ healthy controls.

For any pair of individuals, we can use Algorithm~\ref{alg:without seeds} to compare their brain networks and estimate the latent position shifts for all brain regions.
To determine the embedding dimension $d$ for Algorithm~\ref{alg:without seeds}, we apply the automatic dimensionality selection procedure described in \cite{zhu2006automatic}, and it selects dimensions of $1$, $2$, and $3$ for $4$, $9$, and $4$ networks, respectively, among the total $17$ brain networks. Based on this, we choose $d = 3$, as it suffices to capture the complexity of all networks (more details are provided in Section~\ref{sec:different rank}).
We set the FDR level to detect shifts as $\alpha = 0.01$. The other parameters of Algorithm~\ref{alg:without seeds} mainly affect computational complexity, and we choose $L = d = 3$, $M = 5000$, and $\tilde\alpha = 0.01$.

\begin{figure}[htbp!]
\centering

\subfigure
{\includegraphics[height=4.5cm]{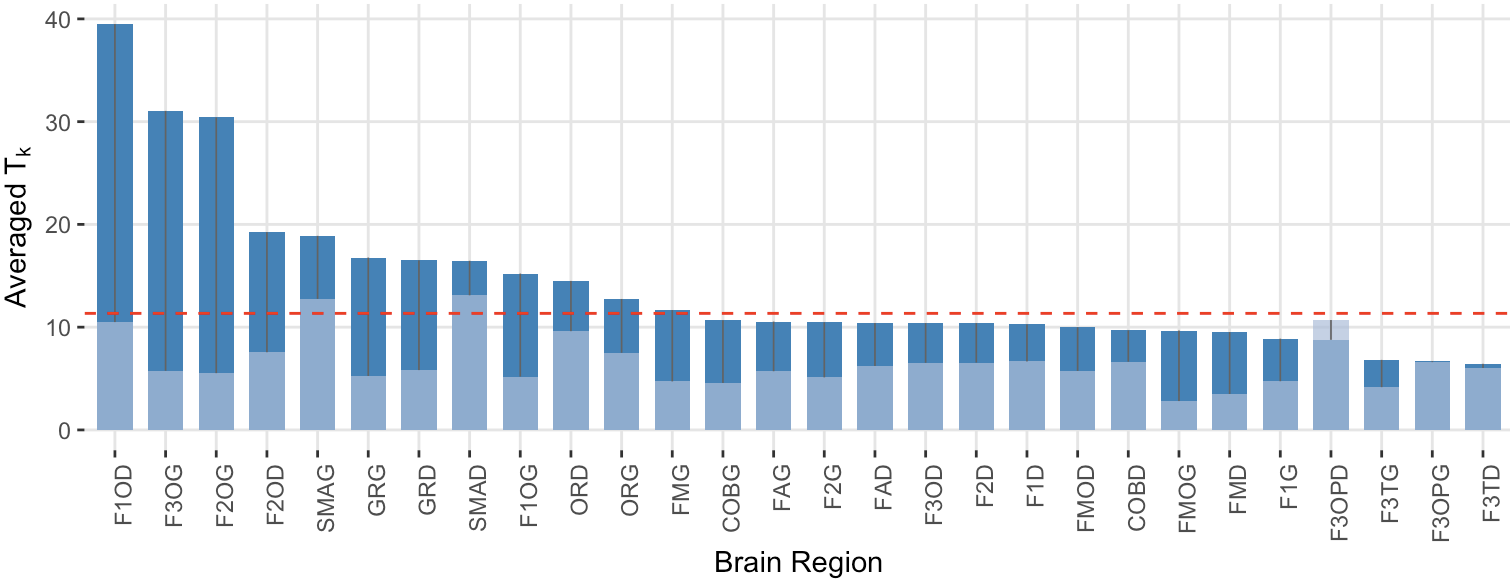}}

\caption{\footnotesize Comparison of brain region latent position shifts between ADHD patients and controls, and within controls using Algorithm~\ref{alg:without seeds}. 
The blue bars represent the average test statistic $T_k$ (as defined in Theorem~\ref{thm:HT}) by comparing the brain networks across patients and controls. 
The lighter-colored bars represent the average $T_k$ by comparing brain networks within the control group. 
The horizontal red dashed line represents the critical value for detecting shifts, as specified in Theorem~\ref{thm:HT}.
}
\label{fig:brain}
\end{figure}

Figure~\ref{fig:brain} presents a bar chart comparing the differences in the averaged test statistic $T_k$ as defined in Theorem~\ref{thm:HT} for detecting shifts across all $n=28$ brain regions using Algorithm~\ref{alg:without seeds}, with a focus on identifying regions where ADHD patients deviate the most from controls. For each brain region, the blue bar represents the test statistic $T_k$ obtained by comparing the brain networks of each ADHD patient with each control across all combinations, with the final value being the average of these $T_k$ values. The lighter-colored bar represents $T_k$ obtained by comparing brain networks within the control group, considering all possible pairs, with the final value being the average of these $T_k$ values. Note that while we use multiple testing correction in Algorithm~\ref{alg:without seeds} to align each pair of networks with more unshifted vertices, here for summarizing results across all network pairs, we consider individual $T_k$ for each brain region $k$ and compute the average across all network pairs.

From Figure~\ref{fig:brain}, we can see that the averaged test statistics $T_k$ between controls, computed for all brain regions, are generally on the same scale and are almost always smaller than the critical value for detecting shifts, as specified in Theorem~\ref{thm:HT} and represented by the horizontal red dashed line in Figure~\ref{fig:brain}.
The averaged $T_k$ across patients and controls for all brain regions are typically larger than those between controls alone. Brain regions with larger differences in these $T_k$ values are likely to be more significant for understanding ADHD.
Notably, the differences in the leftmost $7$ brain regions in Figure~\ref{fig:brain} are larger than those in other regions and significantly exceed the critical value. In addition, the regions SWAG and SMAD also exhibit $T_k$ values across patients and controls that are notably greater than the critical value.
Finally, these brain regions -- F1OD, F2OG, F3OG, F2OD, SMAG, GRG, GRD, SMAD, F1OG,  and ORD  -- are likely to be important for understanding ADHD.

\subsection{Dynamic chocolate trading network}\label{sec:chocolatenetwork}

We consider the dynamic trading network between countries for chocolate during the year from $2010$ to $2022$. The data is
collected by the Food and Agriculture Organization of the United
Nations and is available at
\url{https://www.fao.org/faostat/en/#data/TM}.
We construct a dynamic trading network of chocolate for $T=2022-2020+1=13$ time points,
where vertices represent countries
and edges in each network represent trade relationships between countries.
For each time point $t$, we obtain the adjacency
matrix $\mathbf{A}^{(t)}$ by (1) we set $(\ma^{(t)})_{r,s}=(\ma^{(t)})_{s,r}=1$ if there is trade between countries $r$ and $s$; (2) we ignore the links between
countries $r$ and $s$ in $\mathbf{A}^{(t)}$ if their
total trade amount at the time point $t$ is less than two hundred thousands US
dollars; (3) finally we extract the {
 intersection} of the 
  largest connected components of $\{\mathbf{A}^{(t)}\}$ to get the networks for the common involved countries.
  The resulting adjacency matrices $\{\ma^{(t)}\}_{t\in[T]}$ corresponding to $T=13$ time points on a set of $n=169$ vertices.
  
\begin{figure}[htbp!]
\centering

\subfigure
{\includegraphics[height=3.9cm]{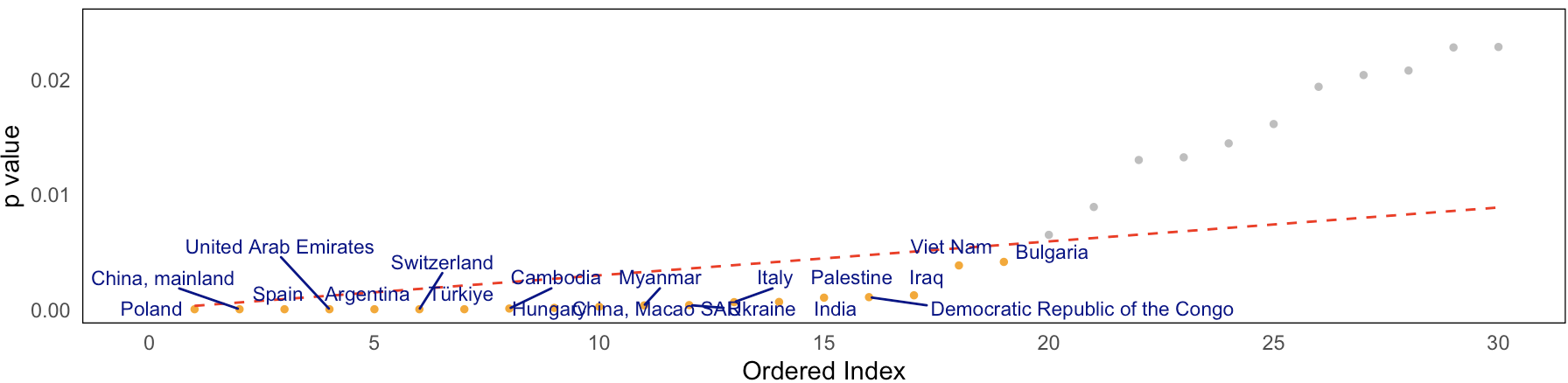}}

\caption{\footnotesize The ordered p-values $\{p_{(j)}\}_j$ for the first 30 countries ranked by significance. The red dashed line indicates the Benjamini-Hochberg threshold $\frac{j\alpha}{n}$ with $\alpha=0.05$, used to control the false discovery rate. Countries with significant shifts in embeddings (those below the threshold line) are labeled by their names.
}
\label{fig:food_T_k}
\end{figure}

\begin{figure}[htbp!]
\centering

\subfigure[\footnotesize Pie chart for $\hat{\mathcal{U}}$]
{\includegraphics[height=2.2cm]{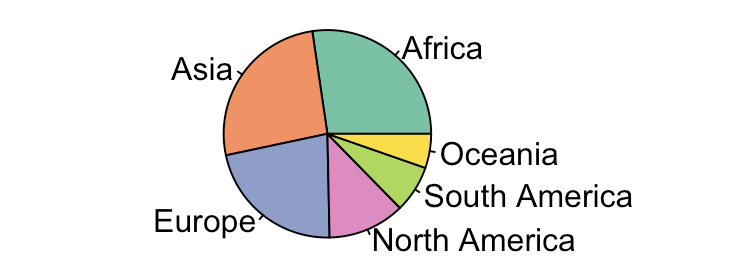}}
\subfigure[\footnotesize Pie chart for $\hat{\mathcal{U}}^{\mathcal{C}}$]
{\includegraphics[height=2.2cm]{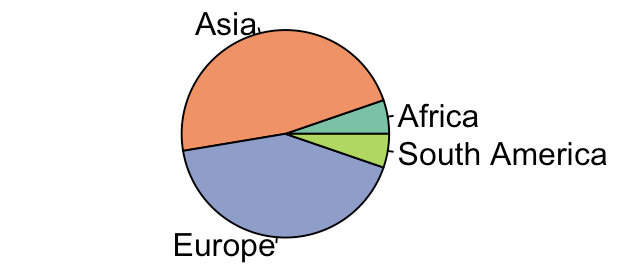}}

\caption{\footnotesize Proportions of countries with unshifted and shifted embeddings.}
\label{fig:food_country distribution}
\end{figure}

\begin{figure}[htbp!]
\centering

\subfigure
{\includegraphics[height=6cm]{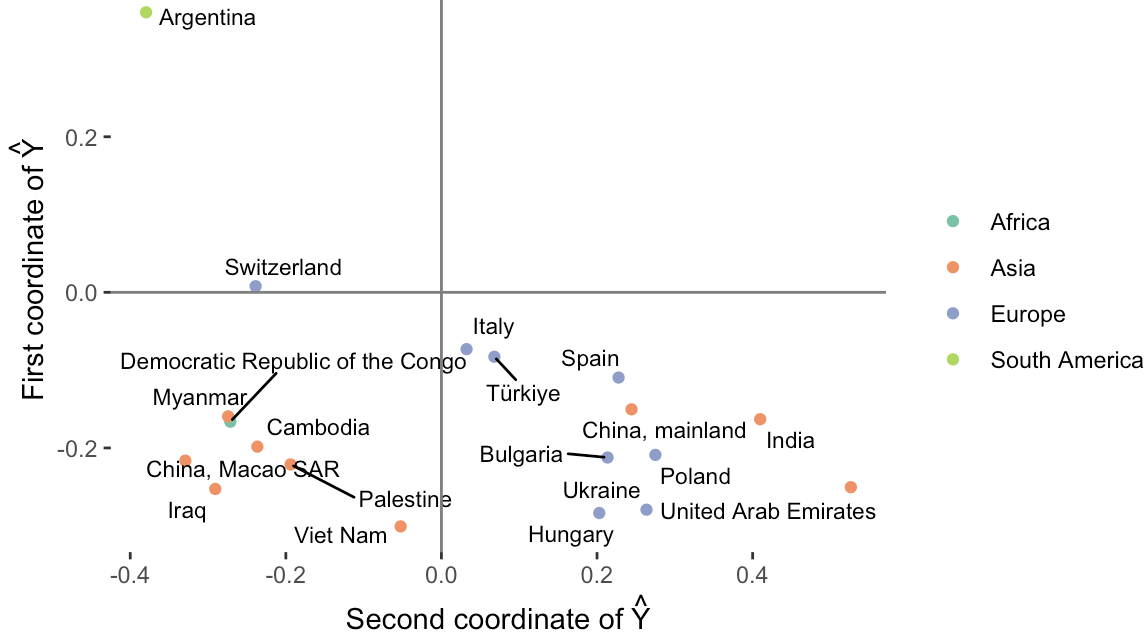}}

\caption{\footnotesize Embedding shifts for countries in $\hat{\mathcal{U}}^{\mathcal{C}}$, corresponding to the rows of $\hat\my_{\hat{\mathcal{U}}^{\mathcal{C}}}$.}
\label{fig:food_Y}
\end{figure}

We now consider the first year, 2010, and the last year, 2022, to compare the trading networks $\ma^{(1)}$ and $\ma^{(13)}$, estimating the sets of countries with unshifted embeddings and shifted embeddings, $\mathcal{U}$ and $\mathcal{U}^\mathcal{C}$, as well as the shifts $\my$ by Algorithm~\ref{alg:without seeds}.
We set the embedding dimension to $d = 2$, the significance level for detecting shifts to $\alpha = 0.05$, and the other parameters of Algorithm~\ref{alg:without seeds} to $L= 3$, $M = 5000$, and $\tilde\alpha = 0.01$.
We identify $|\hat{\mathcal{U}}|=150$ countries with unshifted embeddings $\hat{\mathcal{U}}$, leaving $19$ countries with shifted embeddings $\hat{\mathcal{U}}^{\mathcal{C}}$.
Figure~\ref{fig:food_T_k} shows the test statistics ${T_k}$ for all countries. In Figure~\ref{fig:food_T_k}, countries with significant $T_k$ values often align with real-world changes in chocolate trade over the past decade. For instance, Poland, which has a smallest $p_k$, is a major chocolate producer and exporter in Europe with growing export volumes. Similarly, China has experienced a significant surge in chocolate consumption and imports over the past decade, which explains its significant $p_k$.
Figure~\ref{fig:food_country distribution} shows the proportions of countries with unshifted and shifted embeddings across different regions. It can be observed that a higher proportion of Asian and European countries experience shifts in their embeddings for chocolate trade between these two time points.
Figure~\ref{fig:food_Y} provides the estimates of embedding changes for countries with shifted embeddings, i.e., rows of $\hat\my_{\hat{\mathcal{U}}^{\mathcal{C}}}$.
Figure~\ref{fig:food_Y} shows that countries on the same continent are generally placed close together, indicating a strong correlation between the estimated embedding shifts and the true underlying geographic proximities. 
Here, we just compare the networks for two specific time points, while in Section~\ref{sec:mirror} we extend the analysis to the entire dynamic chocolate trading network by considering all pairs of time points.

\newpage

\bibliographystyle{unsrtnat}
\bibliography{ref}

\newpage

\appendix

\begin{center}%
    {\LARGE Supplementary Material for ``Detection and estimation of vertex-wise latent position shifts across networks"\par}%
  \end{center}

\counterwithin{figure}{section}
\counterwithin{theorem}{section}
\counterwithin{assumption}{section}

\section{Extensions}

\subsection{Related Euclidean mirrors for dynamic networks}\label{sec:mirror}

For a dynamic network, \cite{athreya2024discovering} proposes the idea of \textit{mirror} to visualize the network evolution pattern with a curve in low-dimensional Euclidean space.
More specifically, for a dynamic network $\{\ma^{(i)}\}_{i \in [T]}$, where $T$ is the number of time points and $\ma^{(t)}$ is the adjacency matrix of the network at time point $t$, we measure the difference $\hat{\mathbf{D}}_{i,j}$ between each pair of networks $(\ma^{(i)}, \ma^{(j)})$ for all observed time points $i, j \in [T]$, and then classical multidimensional scaling (CMDS) \citep{borg2007modern,wickelmaier2003introduction,li2020central} is applied to obtain a configuration in a low-dimensional space that approximately preserves the dissimilarity in $\hat{\mathbf{D}}\in\mathbb{R}^{T\times T}$. This results in a curve whose coordinates are the rows of $\hat\mm \in \mathbb{R}^{T \times r}$ in $r$-dimensional Euclidean space, representing the evolution of the network dynamics over the $T$ time points, referred to as the mirror of the dynamic network.
After the mirror is obtained, using the ISOMAP technique \citep{tenenbaum2000global}, the mirror in $r$-dimensional space can be further reduced to a $1$-dimensional curve, called the \textit{iso-mirror}, where the Euclidean distances approximate the geodesic distances along the mirror, and thus larger changes along the y-axis of the iso-mirror correspond to significant changes in the network.

In the initial mirror algorithm proposed by \cite{athreya2024discovering}, the difference $\hat{\mathbf{D}}_{i,j}$ between each pair of networks $(\ma^{(i)}, \ma^{(j)})$ is defined as
$$
\hat{\mathbf{D}}_{i,j}=\min_{\mo\in\mathcal{O}_d}\|\hat\mx^{(i)}\mo-\hat\mx^{(j)}\|/\sqrt{n},
$$
which estimates the pairwise distance between the two estimated latent position matrices $\hat\mx^{(i)},\hat\mx^{(j)}\in\mathbb{R}^{n\times d}$.
In this paper, under the model where the latent positions of some vertices remain unchanged while others are shifted, the distance between two networks can alternatively be defined, such as by the proportion of shifted vertices, given by
\begin{equation}\label{eq:Dij}
	\hat{\mathbf{D}}_{i,j}=|(\hat{\mathcal{U}}^{(i,j)})^{\mathcal{C}}|/ n,
\end{equation}
where $(\hat{\mathcal{U}}^{(i,j)})^{\mathcal{C}}$ denotes the set of shifted vertices between networks $i$ and $j$, and can be obtained by Algorithm~\ref{alg:without seeds}.

\begin{figure}[htbp!]
\centering

\subfigure[\footnotesize Mirror]
{\includegraphics[height=3.3cm]{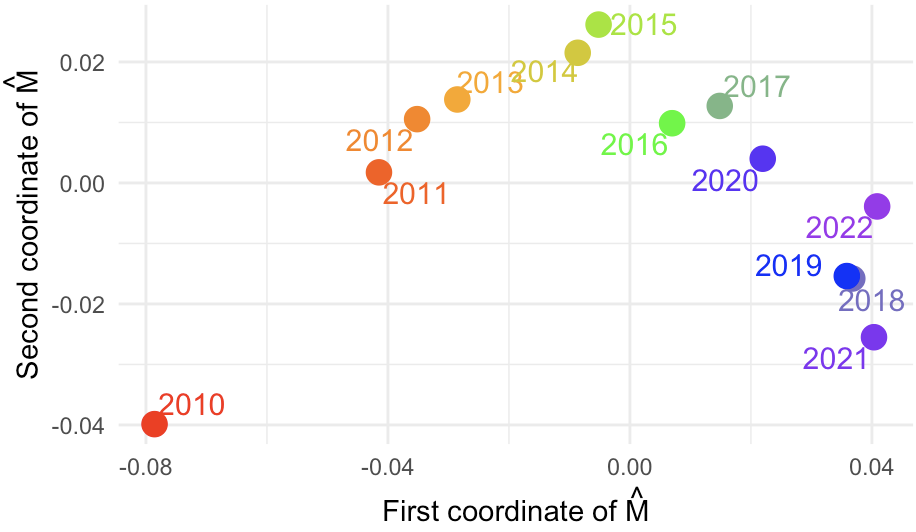}}
\subfigure[\footnotesize Iso-mirror]
{\includegraphics[height=3.3cm]{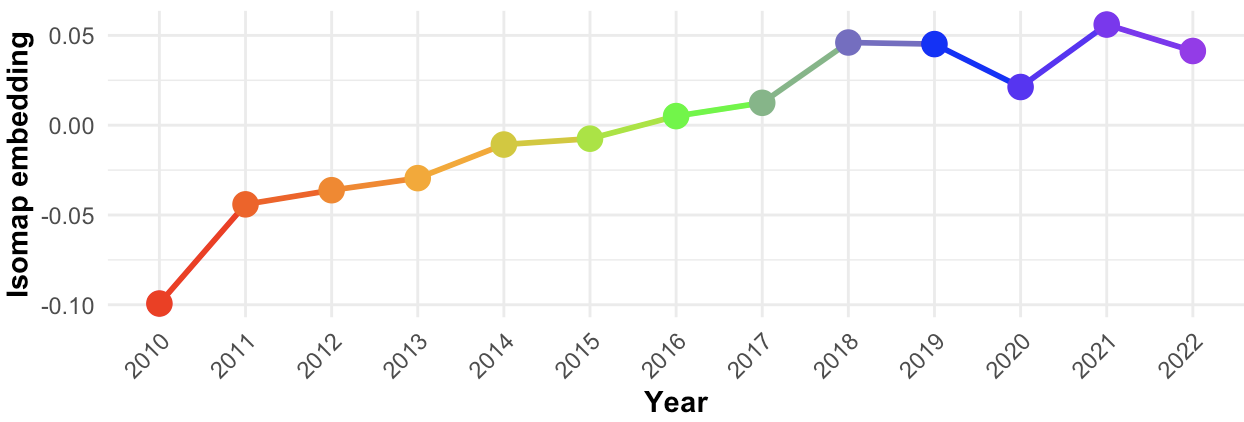}}

\caption{\footnotesize Mirror and iso-mirror for the entire chocolate dynamic trading network.}
\label{fig:food_mirror}
\end{figure}

We now consider the entire dynamic trading network $\{\ma^{(t)}\}_{t \in [13]}$ for chocolate trading, as described in Section~\ref{sec:chocolatenetwork}, and construct its mirror to approximately preserve the distances defined in Eq.~\eqref{eq:Dij} with the embedding dimension $r$ set to $2$. 
The mirror and corresponding iso-mirror are shown in Figure~\ref{fig:food_mirror}, and they show that there are more countries with changes in global chocolate trading patterns between 2010 and 2011, between 2017 and 2018, and between 2020 and 2021.

The above idea focuses on constructing a low-dimensional mirror to represent the evolution pattern of the \textit{entire} dynamic network. Similarly, we can also focus on \textit{a specific vertex} and construct a mirror to describe the changes in its latent position over time. For a specific vertex $k \in [n]$, we define the distance between each pair of time points $(i, j)$ as
\begin{equation}\label{eq:Dkij}
	\hat{\mathbf{D}}^{[k]}_{i,j}=
\begin{cases}
0 & \text{if }k\in\hat{\mathcal{U}}^{(i,j)},\\
\|\hat{\mathbf{y}}_k\| & \text{if }k\in(\hat{\mathcal{U}}^{(i,j)})^{\mathcal{C}}.
\end{cases}
\end{equation}
This means that if vertex $k$ is estimated to be unshifted between time points $i$ and $j$, the distance is defined as $0$, and if vertex $k$ is estimated to be shifted, the distance is defined as the norm of its estimated shift $\hat{\mathbf{y}}_k$.

\begin{figure}[htbp!]
\centering

\subfigure[\footnotesize Mirror]
{\includegraphics[height=2.82cm]{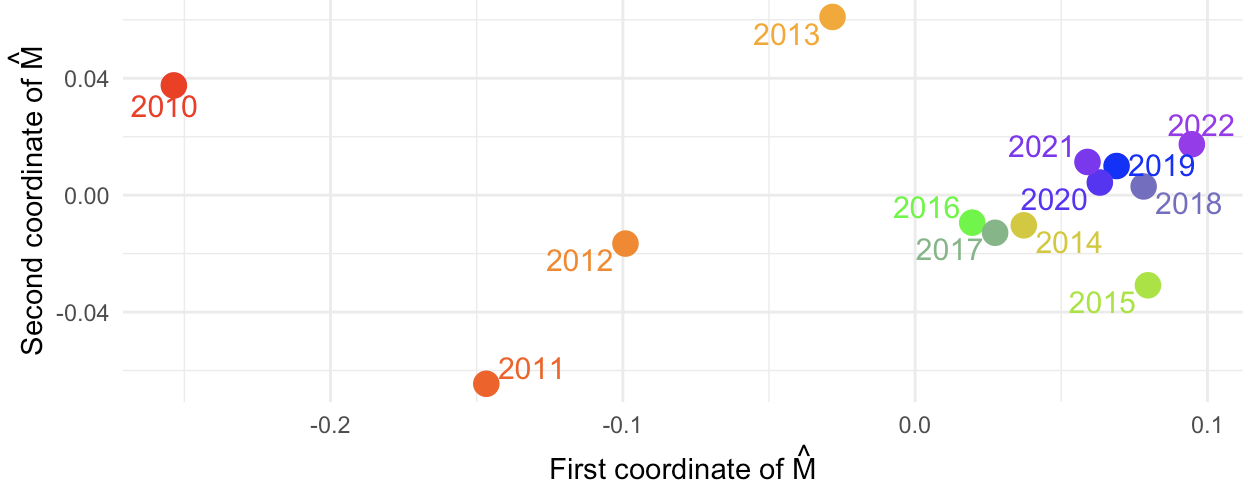}}
\subfigure[\footnotesize Iso-mirror]
{\includegraphics[height=2.82cm]{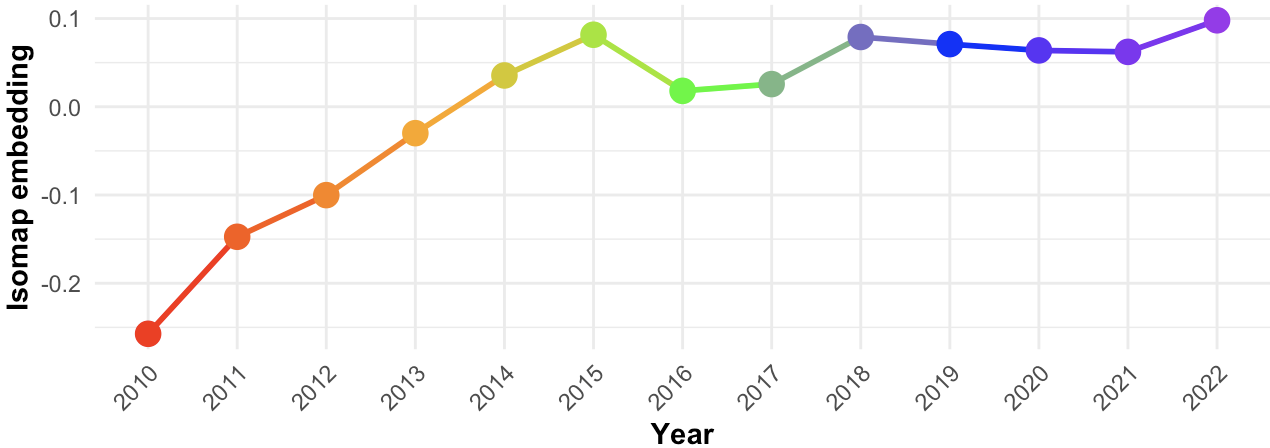}}

\caption{\footnotesize Mirror and iso-mirror for Poland in the chocolate dynamic trading network.}
\label{fig:food_mirror_poland}
\end{figure}

Using the distance defined in Eq.~\eqref{eq:Dkij}, we generate the mirror and corresponding iso-mirror for Poland as a vertex in the dynamic chocolate trading network, as shown in Figure~\ref{fig:food_mirror_poland}. Figure~\ref{fig:food_mirror_poland} indicates that Poland's chocolate trading pattern experienced significant changes before 2016 and became relatively stable thereafter.

\subsection{Extension to generalized random dot product graphs}

We extend the model to pairs of random networks based on the GRDPG model \citep{rubin2017statistical,young2007random}, which generalizes the RDPG model by allowing for indefinite inner products.

\begin{definition}[Generalized random dot product graph (GRDPG)]
Let $d\geq 1$ be given and let $\mathcal{X}$ be a subset of $\mathbb{R}^d$ such that $x^\top \mi_{d_+,d_-} y\in[0,1]$ for any $x,y\in\mathcal{X}$.
Here, $\mi_{d_+,d_-}$ is a $d \times d$ diagonal matrix with $d_+$ diagonal entries equal to $+1$ and $d_-$ diagonal entries equal to $-1$, where $d_+$ and $d_-$ are non-negative integers satisfying $d_++d_-=d$.
For a given $n\geq 1$, let $\mx=[\mathbf{x}_1,\mathbf{x}_2,\dots,\mathbf{x}_n]^\top$ be a $n\times d$ matrix with $\mathbf{x}_k\in\mathcal{X}$ for all $k\in[n]$. A random network $G$ is said to be a generalized random dot product graph with latent positions of the vertices in $\mx$, where each row $\mathbf{x}_k\in\mathbb{R}^d$ denotes the latent position for the $k$th vertex, if the adjacency matrix $\ma$ of $G$ is a symmetric matrix whose upper triangular entries $\{\ma_{s,t}\}_{s<t}$ are independent Bernoulli random variables with
$$
\ma_{s,t}\sim \operatorname{Bernoulli}(\mathbf{x}_s^\top \mi_{d_+,d_-}\mathbf{x}_t).
$$
We define $\mpp:=\mx\mi_{d_+,d_-}\mx^\top$ as the connection probability matrix of $G$ and denote such a graph by GRDPG$(\mx\mi_{d_+,d_-}\mx^\top)$.
In this case, the success probabilities of $\{\ma_{s,t}\}_{s<t}$ are given by  $\mpp_{s,t}=\mathbf{x}_s^\top \mi_{d_+,d_-}\mathbf{x}_t$.
\end{definition}

\begin{remark}[Indefinite orthogonal nonidentifiability in GRDPGs]\label{rm:noni GRDPG}
Note that if $G \sim \operatorname{GRDPG}(\mx\mi_{d_+,d_-}\mx^\top)$ with the latent position matrix $\mx \in \mathbb{R}^{n \times d}$, then for any orthogonal matrix $\mw \in \mathcal{O}_{d_+,d_-}$, $\mx\mw$ also gives rise to a GRDPG with the same probability distribution.
Here $\mathcal{O}_{d_+,d_-}:=\{\mo\in\mathbb{R}^{d \times d}\mid \mo\mi_{d_+,d_-}\mo^\top=\mi_{d_+,d_-}\}$ is the indefinite orthogonal group.
\end{remark}

Assume that $G^{(1)}$ and $G^{(2)}$ are independent GRDPGs with latent position matrices $\mx^{(1)}$ and $\mx^{(2)}$, respectively, i.e., $G^{(i)} \sim \operatorname{GRDPG}(\mx^{(i)}\mi_{d_+, d_-}\mx^{(i)\top})$ for $i \in \{1, 2\}$. For the set of unshifted vertices $\mathcal{U} \subseteq [n]$, there exists an indefinite orthogonal matrix $\mw^{(1,2)} \in \mathcal{O}_{d+, d_-}$ such that
$
\mx^{(2)}_{\mathcal{U}} = \mx^{(1)}_{\mathcal{U}}\mw^{(1,2)}.
$
And for the latent position matrix $\my \in \mathbb{R}^{n \times d}$ with $\my_{\mathcal{U}} = \mathbf{0}$, we have
$
\mx^{(2)} = \mx^{(1)}\mw^{(1,2)} + \my.
$

Given each observed adjacency matrix $\ma^{(i)}$ of the GRDPG, the estimated latent position matrix is computed as
$
\hat\mx^{(i)} = \hat\muu^{(i)}|\hat{\mathbf{\Lambda}}^{(i)}|^{1/2},
$
where $\hat{\mathbf{\Lambda}}^{(i)} = \operatorname{diag}(\hat{\mathbf{\Lambda}}+^{(i)}, \hat{\mathbf{\Lambda}}-^{(i)})$ is a $d \times d$ diagonal matrix, and $\hat{\mathbf{\Lambda}}+^{(i)}$ and $\hat{\mathbf{\Lambda}}-^{(i)}$ contain the $d_+$ largest positive and $d_-$ largest (in magnitude) negative eigenvalues of $\ma^{(i)}$, respectively. $\hat\muu^{(i)}=[\hat\muu_+^{(i)},\hat\muu_-^{(i)}]$ contains the corresponding eigenvectors.

In this case, for a given seed set $ \mathcal{S} \subset \mathcal{U} $ with $ |\mathcal{S}| \geq d $, we aim to obtain $\hat\mw^{(1,2)}$ by solving the indefinite orthogonal Procrustes problem
$$\hat\mw^{(1,2)}=\argmin\limits_{\mo \in \mathcal{O}_{d_+,d_-}} \|\hat\mx^{(1)}_{\mathcal{S}}\mo-\hat\mx^{(2)}_{\mathcal{S}}\|_F
=\argmin\limits_{\mo \in \mathcal{O}_{d_+,d_-}} \|\hat\mx^{(1)}_{\mathcal{S}}\mi_{d_+,d_-}-\hat\mx^{(2)}_{\mathcal{S}}\mi_{d_+,d_-}\mo^\top\|_F,$$
but there is no longer an analytical solution due to the indefinite orthogonality constraint, so we approximate the solution by relaxing the problem into two unconstrained least squares problems
\begin{equation}\label{eq:indefinite1}
	\hat\mw^{(1,2)}_L
	=\underset{\mo\in \mathbb{R}^{d\times d}}{\operatorname{argmin}} \|\hat\mx^{(1)}_{\mathcal{S}}\mo-\hat\mx^{(2)}_{\mathcal{S}}\|_F,
\end{equation}
\begin{equation}\label{eq:indefinite2}
	\hat\mw^{(1,2)}_R
	=\underset{\mo\in \mathbb{R}^{d\times d}}{\operatorname{argmin}} \|\hat\mx^{(1)}_{\mathcal{S}}\mi_{d_+,d_-}-\hat\mx^{(2)}_{\mathcal{S}}\mi_{d_+,d_-}\mo^\top\|_F,
\end{equation}
The solution to Eq.~\eqref{eq:indefinite1} is given by 
$\hat\mw^{(1,2)}_L 
= (\mx^{(1)\top}_{\mathcal{S}} \mx^{(1)}_{\mathcal{S}})^{-1}
\mx^{(1)\top}_{\mathcal{S}} \mx^{(2)}_{\mathcal{S}}
=(\mx^{(1)}_{\mathcal{S}})^{\dagger}\mx^{(2)}_{\mathcal{S}}$,
where $(\cdot)^{\dagger}$ denotes the Moore-Penrose pseudoinverse. 
Similarly, the solution to Eq.~\eqref{eq:indefinite2} is $\hat\mw^{(1,2)}_R
= [(\hat\mx^{(2)}_{\mathcal{S}}\mi_{d_+,d_-})^{\dagger}\hat\mx^{(1)}_{\mathcal{S}}\mi_{d_+,d_-}]^\top$.
We combine the two solutions by setting $\hat\mw^{(1,2)}=\frac{1}{2}(\hat\mw^{(1,2)}_L+\hat\mw^{(1,2)}_R)$, and subsequently compute the estimated shift matrix as $\hat\my=\hat\mx^{(2)}-\hat\mx^{(1)}\hat\mw^{(1,2)}$.
Based on Theorem~3.2 of \cite{xie2024entrywise} and the analysis in Lemma~D.13 of \cite{zheng2024chain}, we can derive the test statistic $T_k$ with an almost identical formula as in Eq.~\eqref{eq:T_k} with the only difference being the definition of the estimated covariance matrix, which is given by
$$
\begin{aligned}
	\hat{\mathbf{\Gamma}}^{(k)}
:&=\mi_{d_+,d_-}(\hat\mx^{(2)\top}\hat\mx^{(2)})^{-1}\hat\mx^{(2)\top}\hat{\mathbf{\Xi}}^{(k,2)}\hat\mx^{(2)}(\hat\mx^{(2)\top}\hat\mx^{(2)})^{-1}\mi_{d_+,d_-}
\\&+\hat\mw^{(1,2)\top}\mi_{d_+,d_-}(\hat\mx^{(1)\top}\hat\mx^{(1)})^{-1}\hat\mx^{(1)\top}\hat{\mathbf{\Xi}}^{(k,1)}\hat\mx^{(1)}(\hat\mx^{(1)\top}\hat\mx^{(1)})^{-1}\mi_{d_+,d_-}\hat\mw^{(1,2)}.
\end{aligned}
$$
With the above test statistic $T_k$, we have the corresponding Algorithm~\ref{alg:with seeds} and Algorithm~\ref{alg:without seeds} for the GRDPG framework. 

\begin{figure}[htbp!]
\centering
\subfigure[\footnotesize Accuracy of shift detection]
{\includegraphics[height=4.5cm]{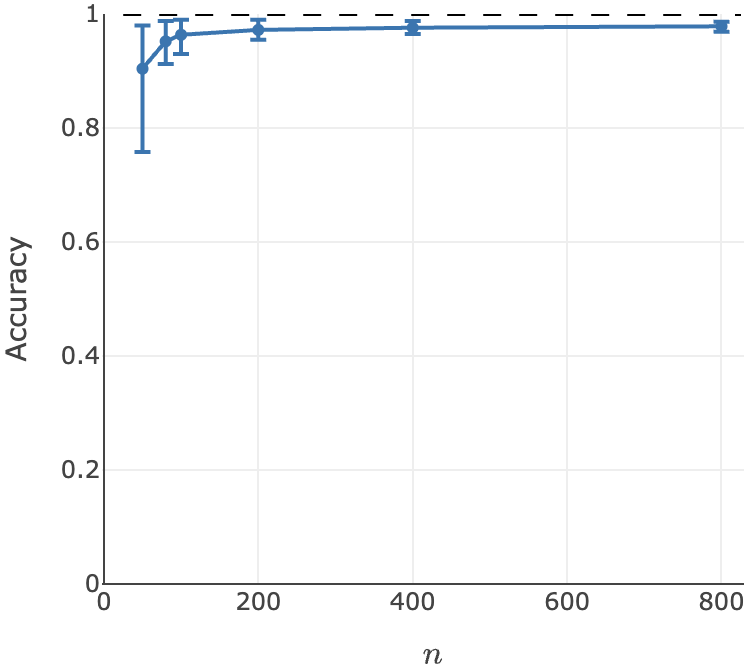}}
\subfigure[\footnotesize Running time]
{\includegraphics[height=4.5cm]{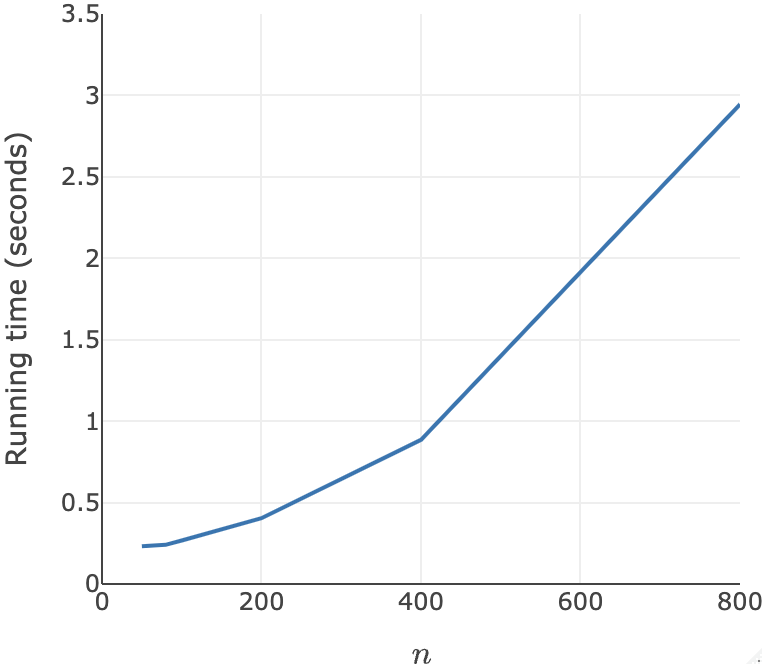}}

     \caption{\footnotesize Panel~(a) illustrates the accuracy of detecting shifted and unshifted vertices without seeds, as we vary the number of vertices $n \in \{50, 80, 100, 200, 400, 800\}$ while setting $L=3$, $M=1000$, $\alpha=0.05$, $\tilde\alpha=0.04$.
     Panel (a) reports the mean, along with the $0.05$ and $0.95$-quantile points, based on $100$ independent Monte Carlo replicates of two SBMs on $n$ vertices with $d=3$ blocks consisting of $d_+=2$ positive and $d_-=1$ negative eigenvalues, where the block assignments of half of the vertices are shifted.
Panel~(b) reports the running time on a standard computer for $100$ Monte Carlo replicates.
}
\label{fig:simulation_noseeds_GRDPG}
\end{figure}

We explore the performance of the corresponding Algorithm~\ref{alg:without seeds} in the no-seed scenario under the GRDPG framework.
Consider the setting of Section~\ref{sec:simu_noseeds}, but
with the block-wise probability matrix 
$$\mb= 
\begin{bmatrix}
0.7 & 0.1 & 0.1 \\
0.1 & 0.3 & 0.8 \\
0.1 & 0.8 & 0.5
\end{bmatrix},
$$
and thus $d_+=2,d_-=1$. 
Figure~\ref{fig:simulation_noseeds_GRDPG} presents the accuracy results and running time of the algorithm, and it demonstrate that the algorithm achieves high accuracy in identifying unshifted vertices while maintaining computational efficiency.

\subsection{Extension to networks of different ranks}
\label{sec:different rank}

In this paper, motivated by the RDPG model, we primarily consider two networks whose vertices have latent positions in the same space of dimension $d$, and assume that probability matrices $\mpp^{(1)}$ and $\mpp^{(2)}$ share the same rank $d$. Such an assumption is commonly made in multiple network analysis; see e.g., \cite{tang2017nonparametric,tang2017semiparametric,paul2020spectral, lei2023bias, lei2024computational,nielsen2018multiple,arroyo2021inference}.

We now consider the extension to two networks on $n$ vertices with probability matrices $\mpp^{(1)}$ and $\mpp^{(2)}$ of ranks $d_1$ and $d_2$, respectively, where we assume $d_1 < d_2$ without loss of generality.
Such a model can encompass scenarios such as an SBM where one community splits, resulting in a dimension increase of $1$.
Suppose there exist latent position matrices $\mx^{(1)}\in\mathbb{R}^{n\times d_1}$ and $\mx^{(2)}\in\mathbb{R}^{n\times d_2}$ such that $\mpp^{(1)}=\mx^{(1)}\mx^{(1)\top}$ and $\mpp^{(2)}=\mx^{(2)}\mx^{(2)\top}$, and suppose there exists an unshifted vertex set $\mathcal{U}\subseteq[n]$ satisfying $\mpp^{(1)}_{\mathcal{U},\mathcal{U}}=\mpp^{(2)}_{\mathcal{U},\mathcal{U}}$ with rank $d_1$. 
Then there exists a unique matrix $\mw^{(1,2)} \in \mathbb{R}^{d_1 \times d_2}$ with full row rank such that 
$$
\mx^{(2)}_{\mathcal{U}} = \mx_{\mathcal{U}}^{(1)} \mw^{(1,2)}.
$$
We denote the vertex-wise shifts in latent positions by $\my \in \mathbb{R}^{n \times d_2}$, satisfying
$
\mx^{(2)} = \mx^{(1)}\mw^{(1,2)} + \my,
$
where $\my_{\mathcal{U}} = \mathbf{0}$ for the unshifted vertices, and $\my_{\mathcal{U}^\mathcal{C}}$ captures the shifts for the shifted vertices.

Suppose $\ma^{(1)}$ and $\ma^{(2)}$ are observed adjacency matrices for $\mpp^{(1)}$ and $\mpp^{(2)}$, and let $\hat\mx^{(1)}$ and $\hat\mx^{(2)}$ be the corresponding estimates of $\mx^{(1)}$ and $\mx^{(2)}$, respectively.
For a given seed set $\mathcal{S}\subseteq\mathcal{U}$, we can obtain $\hat\mw^{(1,2)}$ by aligning $\hat\mx^{(1)}_{\mathcal{S}}$ and $\hat\mx^{(2)}_{\mathcal{S}}$ with
$$
	\hat\mw^{(1,2)}=\underset{\mo\in \mathbb{R}^{d_1\times d_2}}{\operatorname{argmin}} \|\hat\mx^{(1)}_{\mathcal{S}}\mo-\hat\mx^{(2)}_{\mathcal{S}}\|_F,
$$
and the solution is given by
$\hat\mw^{(1,2)} = (\hat\mx^{(1)}_{\mathcal{S}})^{\dagger}\hat\mx^{(2)}_{\mathcal{S}}.$
Then the estimated shifts can be obtained by $\hat\my=\hat\mx^{(2)}- \hat\mx^{(1)}\hat\mw^{(1,2)}$.
Following analysis similar to that for networks of the same rank, we can derive test statistics using the same formulas as those for the algorithms in Section~\ref{sec:alg}.

\begin{figure}[htbp!]
\centering
\subfigure[\footnotesize Accuracy of shift detection]
{\includegraphics[height=4.5cm]{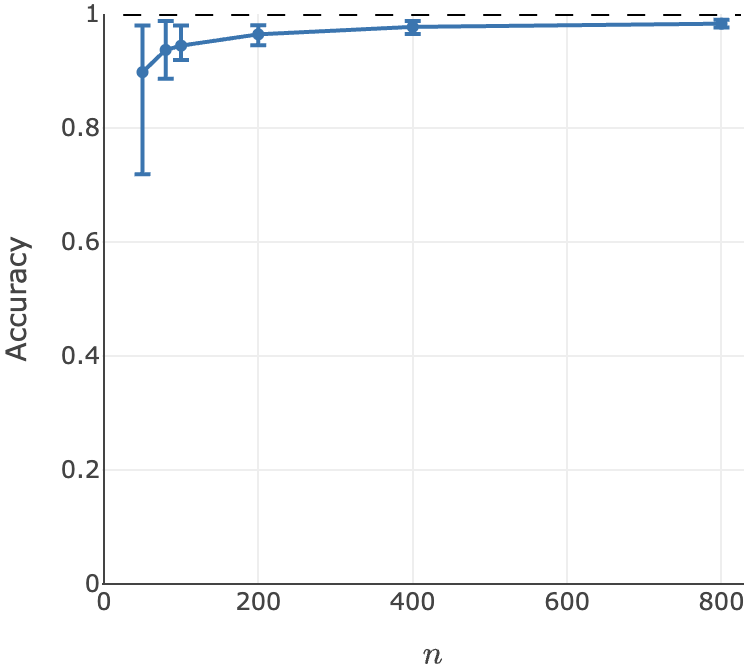}}
\subfigure[\footnotesize Relative estimation error of shifts]
{\includegraphics[height=4.5cm]{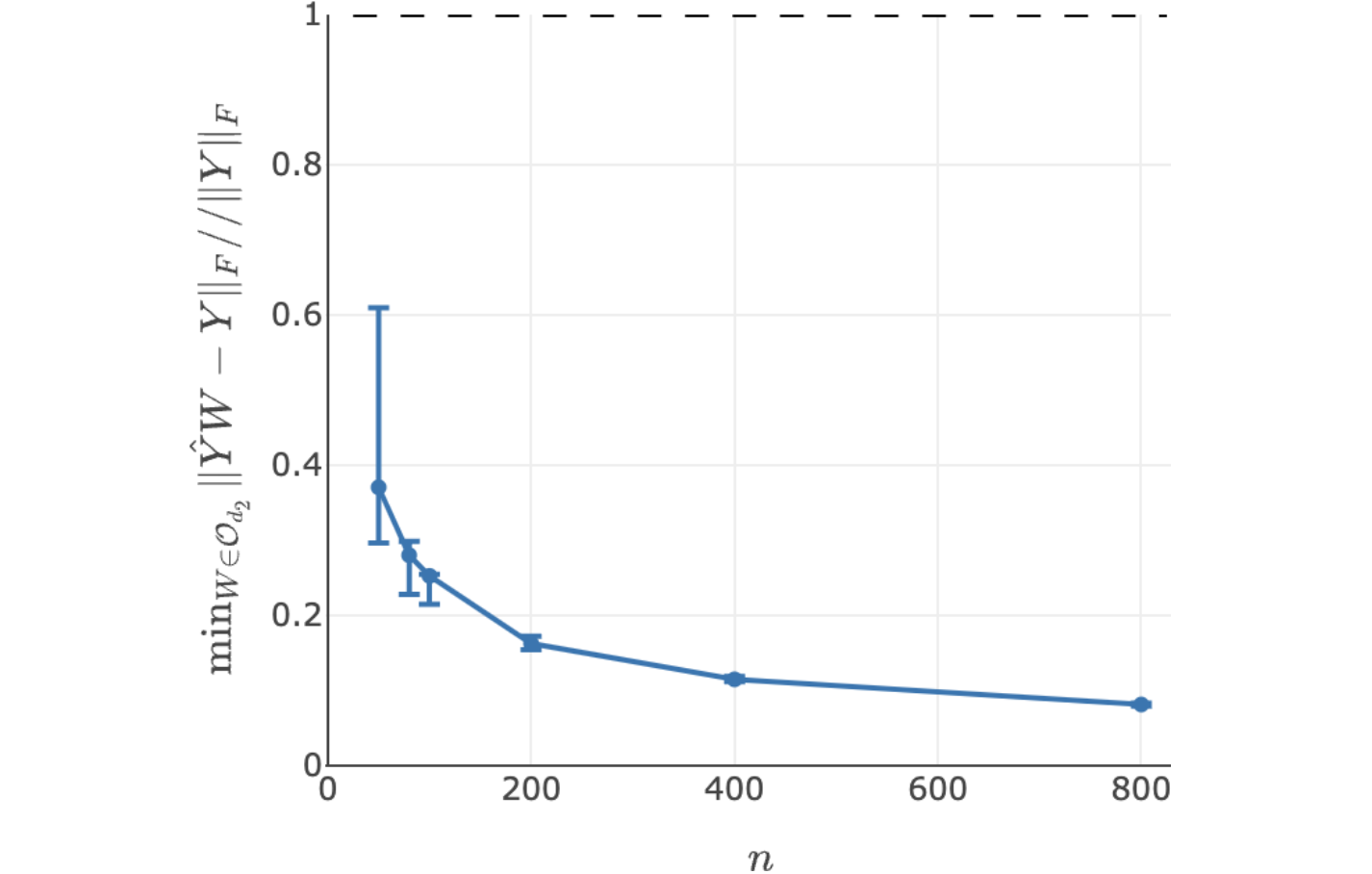}}

     \caption{\footnotesize Panel~(a) illustrates the accuracy of detecting shifted and unshifted vertices without seeds, as we vary the number of vertices $n \in \{50, 80, 100, 200, 400, 800\}$ while setting $L=3$, $M=1000$, $\alpha=0.05$, $\tilde\alpha=0.3$.
     Panel (a) reports the mean, along with the $0.05$ and $0.95$-quantile points, based on $100$ independent Monte Carlo replicates of two SBMs on $n$ vertices with $d_1=2$ and $d_2=3$ blocks, respectively, where half of the vertices have shifted block assignments.
Panel~(b) reports the relative Frobenius norm error for shift matrix estimation, $\min_{\mw\in\mathcal{O}_{d_2}}\|\hat\my\mw-\my\|_F/\|\my\|_F$.
}
\label{fig:simulation_noseeds_d1d2}
\end{figure}

We explore the performance of the corresponding Algorithm~\ref{alg:without seeds} when applied to networks of different ranks.
We consider the setting in Section~\ref{sec:simu_noseeds} involving two SBM networks with 3 blocks, where half of the vertices undergo shifts, but now in network 1, vertices are randomly assigned to only the first 2 blocks, whereas in network 2, vertices are distributed across all 3 blocks, which results in $d_1=2$ and $d_2=3$.
Figure~\ref{fig:simulation_noseeds_d1d2} shows the shift detection accuracy and relative estimation error of $\my$.
Note that $\my$ still is unique up to rotations in $d_2$-dimensional space, so we measure estimation error via $\min_{\mw\in\mathcal{O}_{d_2}}\|\hat\my\mw-\my\|_F$.
The results demonstrate high accuracy in both identifying unshifted vertices and estimating shifts.

\begin{figure}[htbp!]
\centering
\subfigure
{\includegraphics[height=1.7cm]{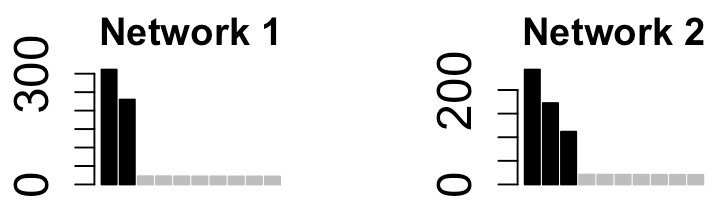}}

     \caption{\footnotesize Scree plots for two networks on $n=800$ with probability matrix ranks $d_1=2$ and $d_2=3$. Black bins indicate the dimensions selected by the automatic dimensionality selection procedure in \cite{zhu2006automatic}.
}
\label{fig:simulation_noseeds_d1d2_scree}
\end{figure}

We now discuss the selection of embedding dimensions in more detail. Theoretically, choosing a dimension smaller than the true dimension can result in information loss, while choosing a dimension larger than the true dimension may retain more noise, so it is better for the selected dimension to match the true model dimension.
In practice, when analyzing real data, we can examine the scree plots for networks to determine the embedding dimensions.
If we observe very clear distinctions in eigenvalue magnitudes for each network, and the two networks exhibit different individual optimal dimensions, then the algorithms for networks of different ranks $d_1\neq d_2$ would be more appropriate.
For example, Figure~\ref{fig:simulation_noseeds_d1d2_scree} shows the scree plots from adjacency matrices generated under the above simulation setting for two networks of ranks $d_1=2$ and $d_2=3$, where we can see the numbers of significantly large eigenvalues correspond to the true ranks, while the remaining eigenvalues are comparatively very small. 
The algorithms for networks of the same rank are appropriate when the selected optimal dimensions are the same, or when adopting a uniform dimension across networks is also reasonable.
For example, for the 17 female brain networks studied in Section~\ref{sec:brain}, Figure~\ref{fig:realdata_brain_scree} in Section~\ref{sec:add_brain} shows the corresponding scree plots. Although the automatic dimensionality selection procedure in \cite{zhu2006automatic} selects different optimal dimensions for some individual networks, choosing a uniform $d=3$ is still reasonable since the first three eigenvalues of each network are always relatively large compared to most of the remaining eigenvalues.

Finally, we emphasize that algorithms for networks of different ranks should not be uniformly applied when networks share a common rank.
The key difference lies in the alignment procedures: the algorithms for common rank $d$ use
$\hat{\mathbf{W}}^{(1,2)}= \underset{\mo\in \mathcal{O}_d}{\operatorname{argmin}} \|\hat{\mathbf{X}}^{(1)}_{\mathcal{S}}\mathbf{O}-\hat{\mathbf{X}}^{(2)}_{\mathcal{S}}\|_F,$
while the algorithms for $d_1\neq d_2$ use 
$\hat\mw^{(1,2)}=\underset{\mo\in \mathbb{R}^{d_1\times d_2}}{\operatorname{argmin}} \|\hat\mx^{(1)}_{\mathcal{S}}\mo-\hat\mx^{(2)}_{\mathcal{S}}\|_F.$
For networks of the same rank, we know that aligning the true latent positions of unshifted vertices should use an orthogonal transformation, so when aligning the estimated latent positions of the seed set, we incorporate this prior knowledge to obtain more accurate estimates, while for networks with different ranks, we do not have such prior knowledge.


\subsection{Discussion}

We now mention several potential directions for future research.
Firstly, in this paper we focus on the setting where the unshifted vertices between the two networks share the same latent positions (up to an orthogonal transformation), i.e., $\mx^{(2)}_{\mathcal{U}}=\mx^{(1)}_{\mathcal{U}}\mw^{(1,2)}$. A natural extension would be to consider more general scenarios where the latent positions of the unshifted vertices undergo scaling or diagonal transformations. Such flexible settings have been explored in the context of other two-sample network inference problems in the existing literature \citep{tang2017semiparametric, tang2017nonparametric, du2023hypothesis}. Our framework can also be extended to address the shift detection and estimation problem under these generalized models, such as considering $\mx^{(2)}_{\mathcal{U}}=c_n\mx^{(1)}_{\mathcal{U}}\mw^{(1,2)}$ for some scaling factor $c_n$ depending on $n$, or $\mx^{(2)}_{\mathcal{U}}=\mathbf{D}_n\mx^{(1)}_{\mathcal{U}}\mw^{(1,2)}$ for some diagonal $\mathbf{D}_n$.

Secondly, our current model focuses on undirected networks, which can be extended to directed networks \citep{leicht2008community,malliaros2013clustering} by assuming $\mpp^{(i)} = \mx^{(i)}\my^{(i)\top}$, where $\mx^{(i)}$ represents the latent positions of the source vertices and $\my^{(i)}$ represents the latent positions of the target vertices in network $i$. Denoting the set of unshifted source vertices by $\mathcal{U}$ and the set of unshifted target vertices by $\mathcal{V}$, our method can be extended to identify the sets $\mathcal{U}$ and $\mathcal{V}$ and estimate the corresponding shifts.

Finally, the concept of detecting latent position shifts can be extended beyond networks to various latent position models in machine learning applications. 
For instance, latent factor models, such as matrix factorization \citep{ma2008sorec,he2016fast}, extract latent features from user-item interaction data and are widely used in recommendation systems, and detecting shifts in these latent representations can provide insights into evolving user preferences and changes in item popularity over time.

\section{Additional Experiment Results}

\subsection{Multiple testing correction for shift detection}\label{sec:simu_multiple}
\begin{figure}[htbp!]
\centering
\subfigure
{\includegraphics[height=3.5cm]{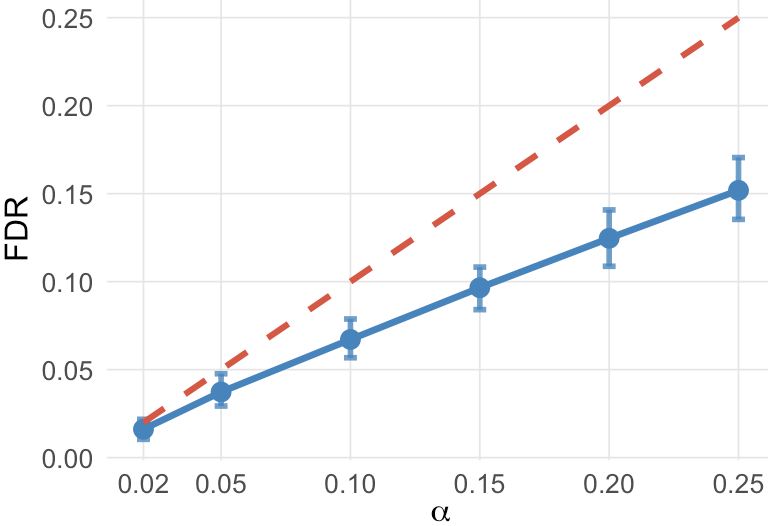}}

 \caption{\footnotesize Empirical FDR versus nominal FDR level for $\alpha \in \{0.02, 0.05, 0.1, 0.15, 0.2, 0.25\}$. The blue solid line represents the empirical FDR (mean FDP across replicates) based on $100$ independent Monte Carlo replicates of two RDPGs with $n = 2000$ vertices and $d = 3$ dimensional latent positions, where half of the vertices are shifted. Error bars indicate the $5$th and $95$th percentiles of the FDP distribution across replicates. The red dashed line shows the nominal FDR level $\alpha$.
}
\label{fig:simulation_BH}
\end{figure}

Under the setting described in Section~\ref{sec:estimation error}, we validate the multiple testing correction procedure described in Section~\ref{sec:multi} to control the FDR at the nominal level $\alpha$. We vary the nominal FDR level $\alpha$, and for each Monte Carlo replicate, we compute the estimated set of invariant vertices $\hat{\mathcal{U}}$ using the Benjamini-Hochberg procedure (Algorithm~\ref{alg:BH}) and calculate the FDP. The results summarized in Figure~\ref{fig:simulation_BH} demonstrate that the empirical FDR is effectively controlled at the nominal level $\alpha$.

\subsection{Additional results for Section~\ref{sec:estimation error}}

\begin{figure}[htbp!]
\centering

\centering
\subfigure
{\includegraphics[height=5cm]{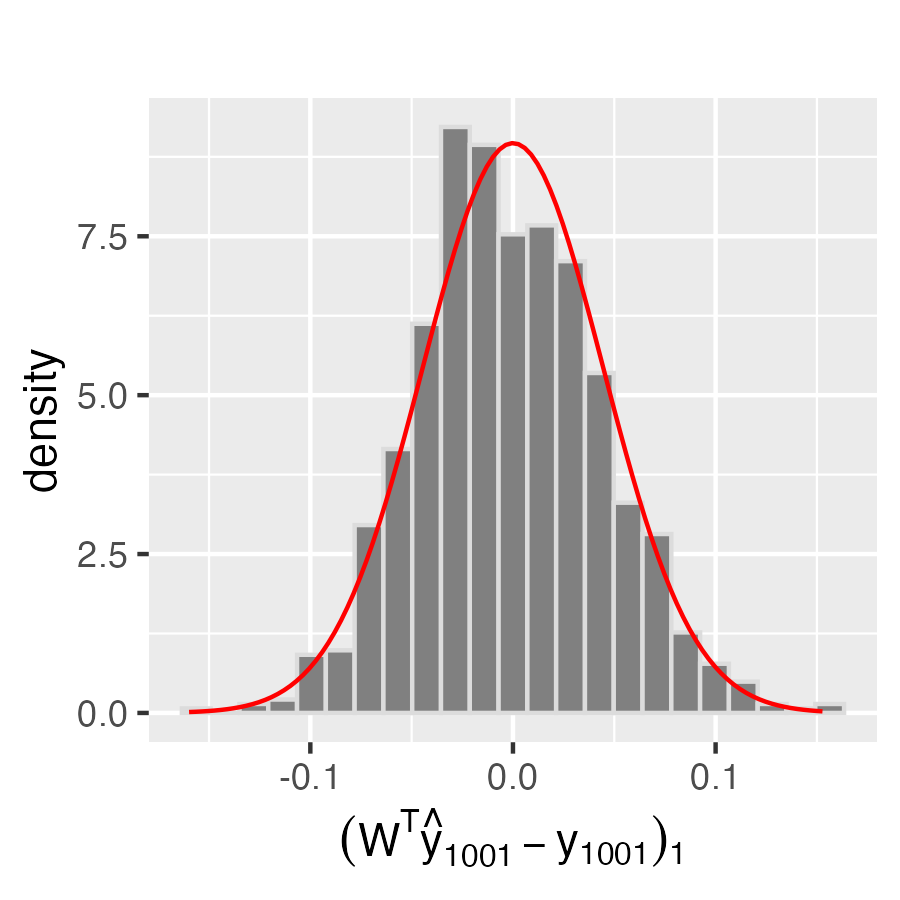}}
\subfigure
{\includegraphics[height=5cm]{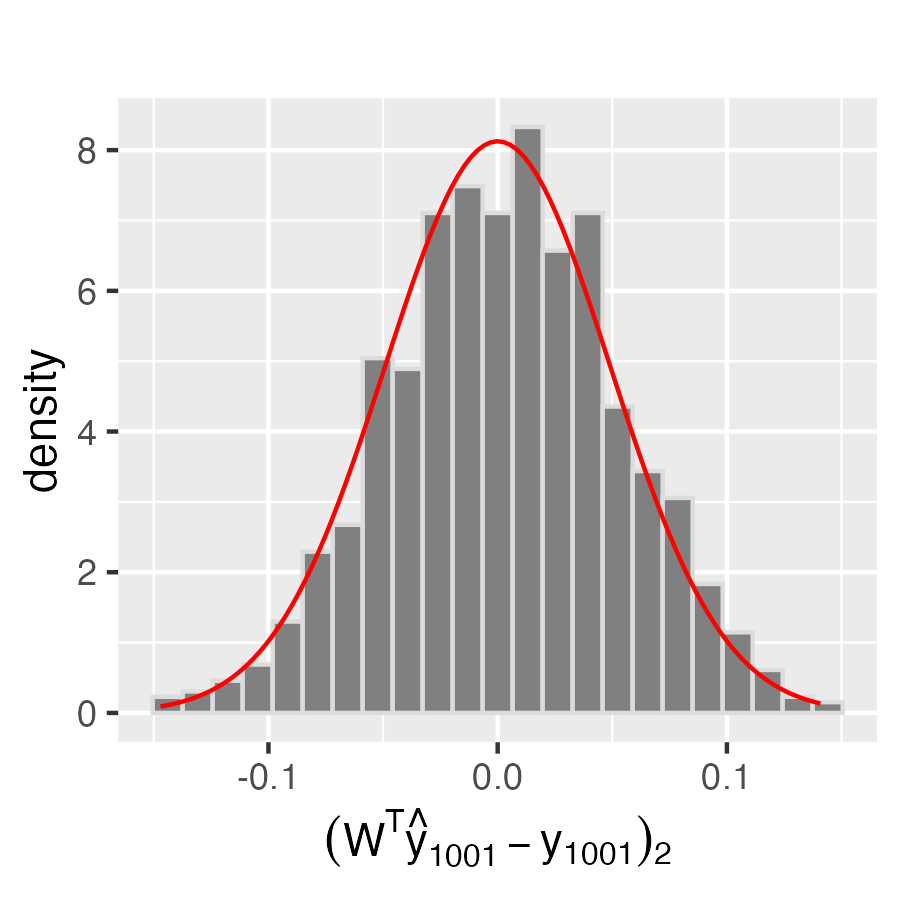}}
\subfigure
{\includegraphics[height=5cm]{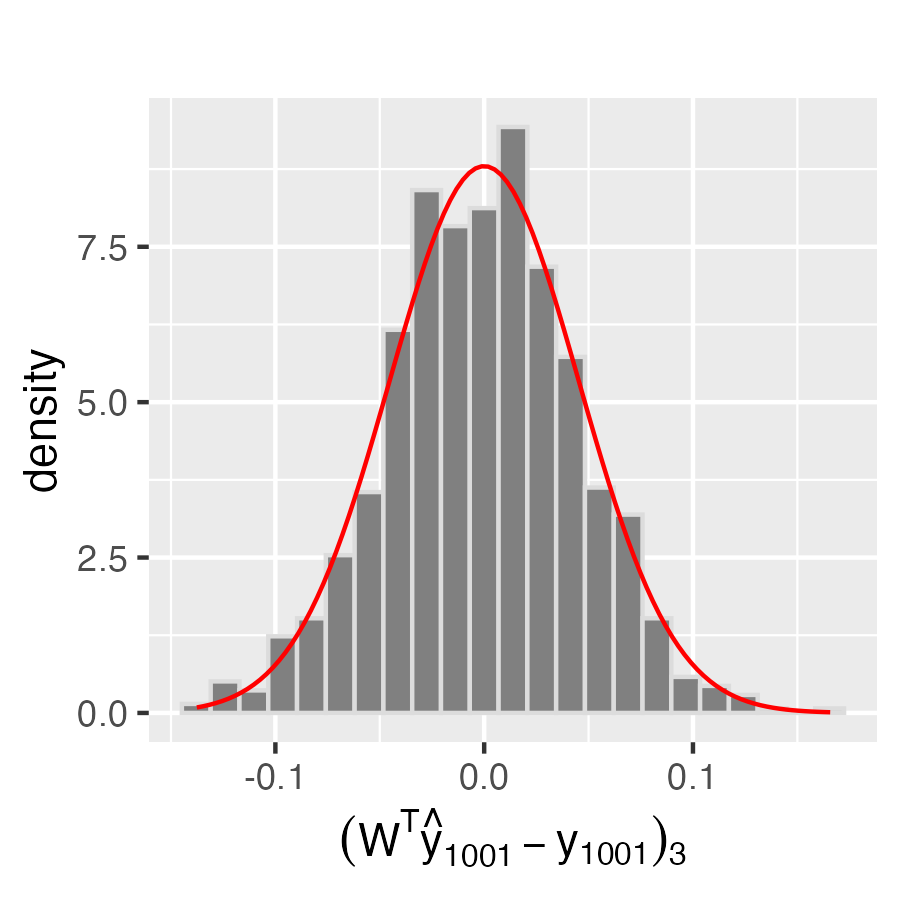}}

\caption{\footnotesize Histograms of the empirical distributions of the entries of the estimation error $\mw^\top\hat{\mathbf{y}}_k - {\mathbf{y}}_k$ for $k = 1001$. 
These histograms are based on $1000$ independent Monte Carlo replicates of two RDPGs on $n = 2000$ vertices with $d = 3$ dimensional latent positions, where half of the vertices are shifted.
The red lines represent the probability density functions of the normal distributions with parameters specified in Theorem~\ref{thm:CLT}.
}
\label{fig:simulation_CLT1_v}
\end{figure}

\begin{figure}[htbp!]
\centering

\subfigure
{\includegraphics[height=5cm]{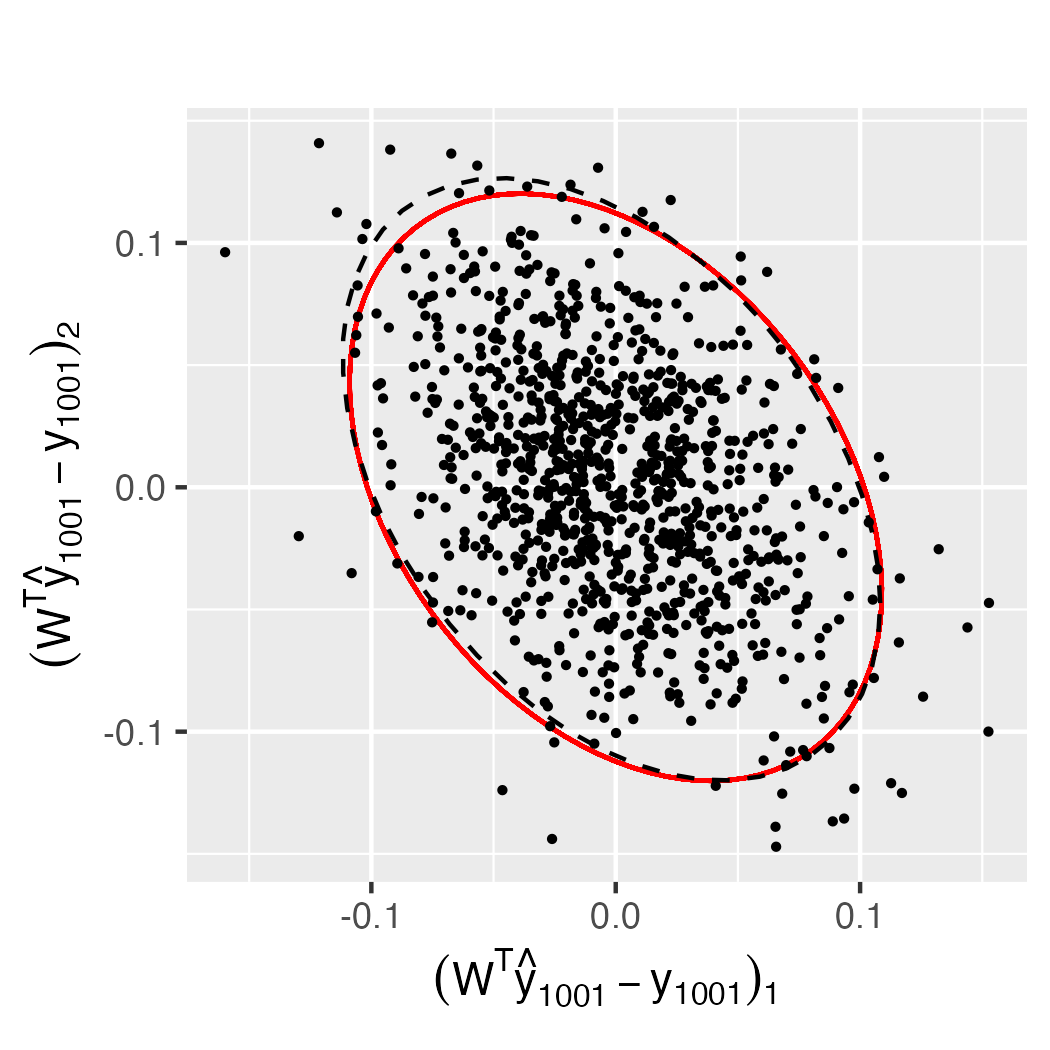}}
\subfigure
{\includegraphics[height=5cm]{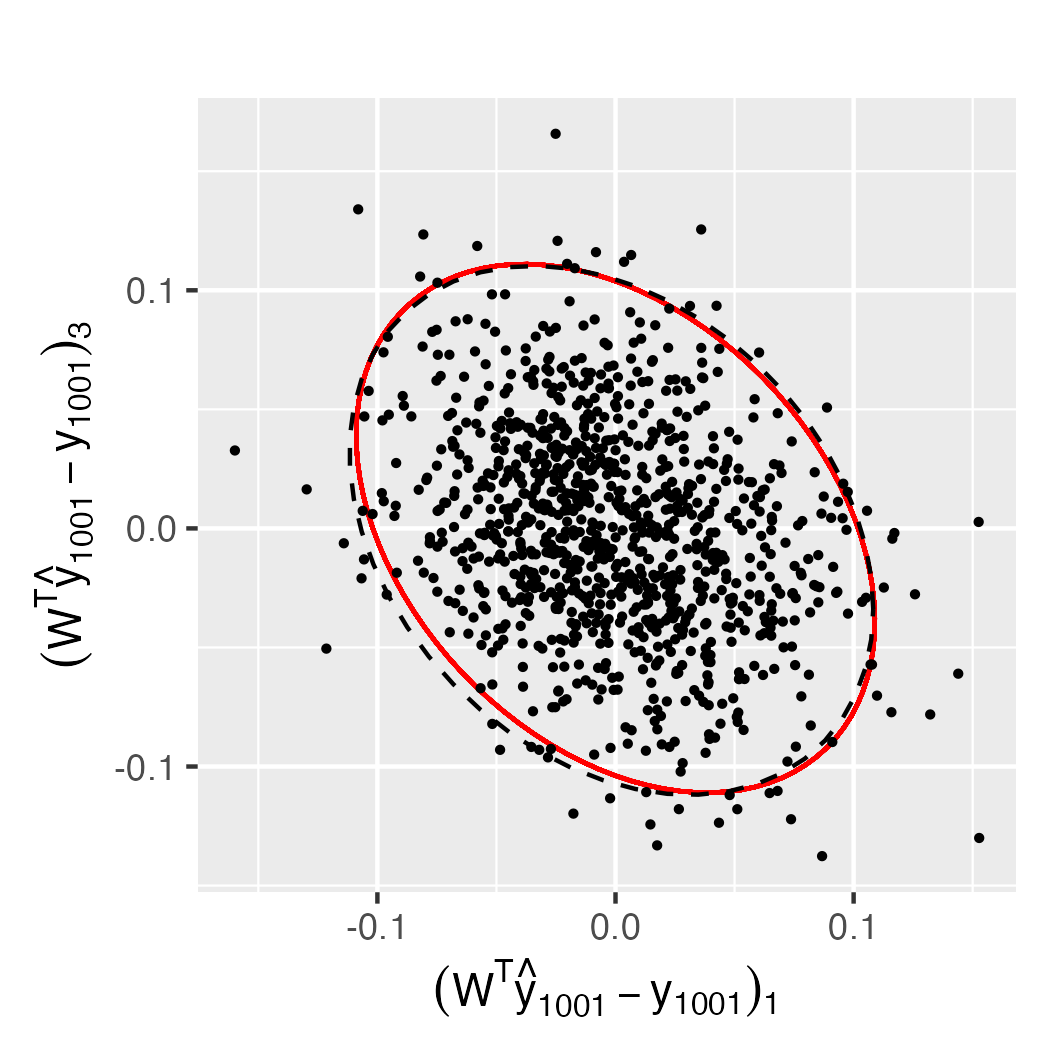}}
\subfigure
{\includegraphics[height=5cm]{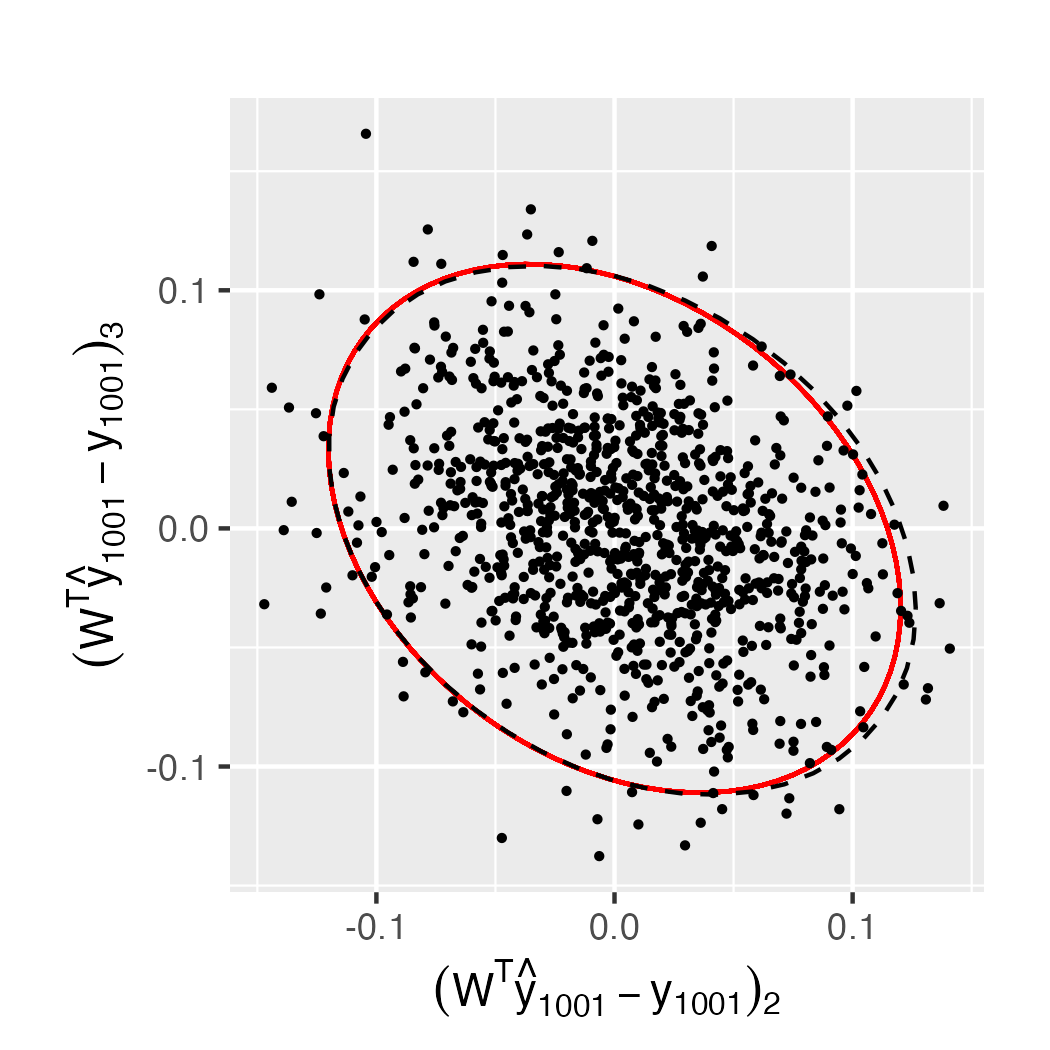}}

\caption{\footnotesize Bivariate plots for the empirical distributions between the entries of the estimation error $\mw^\top\hat{\mathbf{y}}_k - {\mathbf{y}}_k$ for $k = 1001$ based on $1000$ Monte Carlo replicates of the same setting with Figure~\ref{fig:simulation_CLT1_v}. Dashed black ellipses represent 95\% level curves for the empirical distributions while solid red ellipses represent 95\% level curves for the theoretical distributions as specified in Theorem~\ref{thm:CLT}.
}
\label{fig:simulation_CLT2_v}
\end{figure}

In Section~\ref{sec:estimation error}, we present the error results for an unshifted vertex ($k=1$) for the simulation experiment. Here, we further present the estimation error results for a shifted vertex ($k = 1001$) in Figure~\ref{fig:simulation_CLT1_v} and Figure~\ref{fig:simulation_CLT2_v}.
Henze-Zirkler's normality test indicates that the empirical distribution of $\mw^\top\hat{\mathbf{y}}_k - {\mathbf{y}}_k$ for $k = 1001$ is well-approximated by a multivariate normal distribution, and Figure~\ref{fig:simulation_CLT1_v} and Figure~\ref{fig:simulation_CLT2_v} illustrate that the empirical distribution closely matches the theoretical distribution provided in Theorem~\ref{thm:CLT}.

\newpage

\subsection{Additional results for Section~\ref{sec:brain}}
\label{sec:add_brain}

We now present additional results and analysis to complement Section~\ref{sec:brain}. 

\begin{figure}[htbp!]
\centering

\subfigure
{\includegraphics[height=7cm]{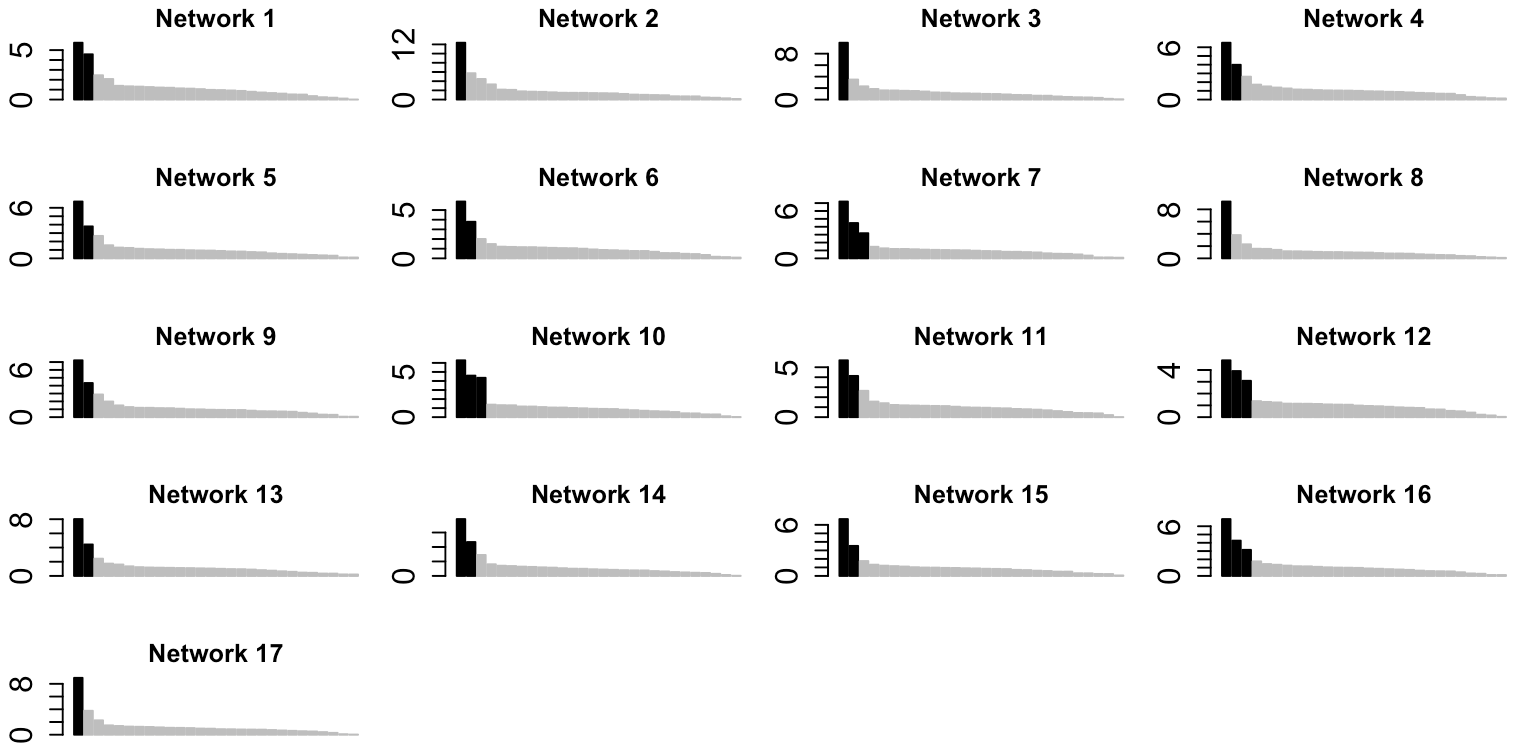}}
\caption{\footnotesize Scree plots for $17$ female brain networks. Black bins indicate the dimensions selected by the automatic dimensionality selection procedure in \cite{zhu2006automatic}.}
\label{fig:realdata_brain_scree}
\end{figure}

Figure~\ref{fig:realdata_brain_scree} shows the scree plots for the $17$ female brain networks, and the automatic dimensionality selection procedure in \cite{zhu2006automatic} selects dimensions of $1$, $2$, and $3$ for $4$, $9$, and $4$ networks, respectively.
In Section~\ref{sec:brain}, we select $d=3$ as it suffices to capture the complexity of all networks, and from the scree plots, choosing $d=3$ does not introduce excessive noise since the first three eigenvalues remain reasonably large even for networks where the elbow occurs at dimensions $1$ or $2$.

\begin{figure}[htbp!]
\centering

\subfigure
{\includegraphics[height=4.5cm]{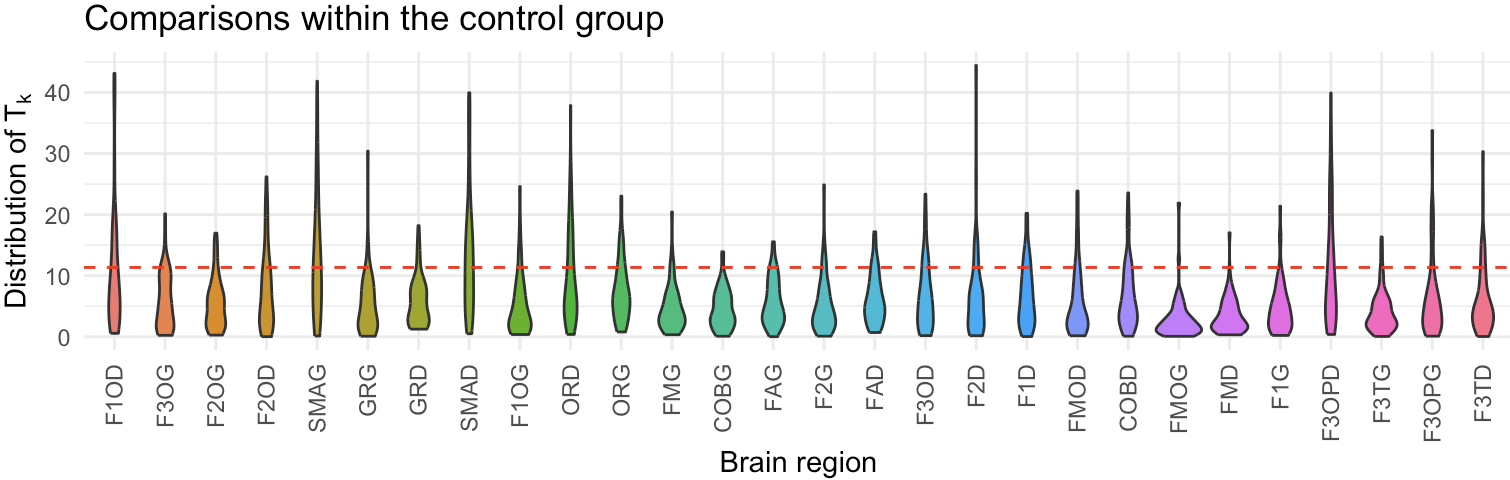}}
\subfigure
{\includegraphics[height=4.5cm]{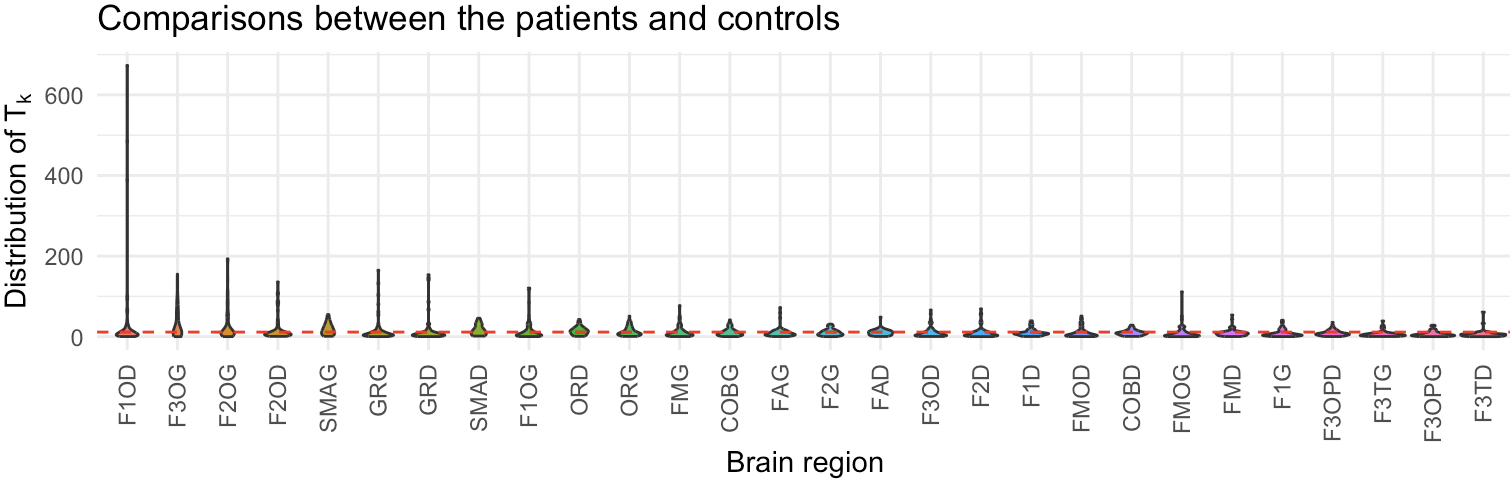}}
\subfigure
{\includegraphics[height=4.5cm]{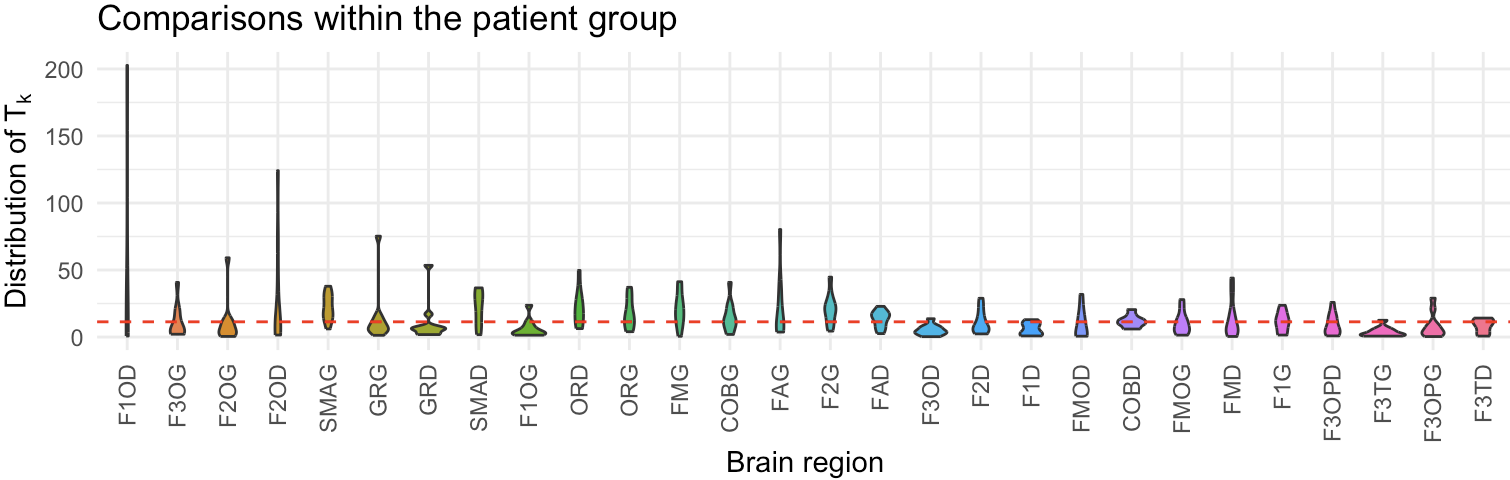}}

\caption{\footnotesize Analysis of brain region latent position shifts (1) within controls, (2) between ADHD patients and controls, and (3) within patients, using Algorithm~\ref{alg:without seeds}. The violin plots illustrate the distributions of the test statistic $T_k$ for these three scenarios. The horizontal red dashed lines represent the critical value for detecting shifts, as specified in Theorem~\ref{thm:HT}.
}
\label{fig:realdata_brain_violin}
\end{figure}

From Figure~\ref{fig:realdata_brain_violin}, we observe that within the control group, the test statistic $T_k$ for all brain regions are generally smaller than the critical value for detecting shifts. This indicates that there are no significant shifts in any brain regions between controls.
Furthermore, Figure~\ref{fig:realdata_brain_violin} shows that the number of brain regions with shifts is considerably large between patients and controls, as well as within patients.

In Section~\ref{sec:brain}, we concluded that the brain regions -- F1OD, F2OG, F3OG, F2OD, SMAG, GRG, GRD, SMAD, F1OG, and ORD -- are likely to be important for understanding ADHD. 
Among these regions, some, such as F1OG, F2OG, and F3OG, exhibit relatively small $T_k$ values within patients. This indicates a high degree of similarity in these brain regions among patients, suggesting that they warrant further focused investigation.

\newpage

\section{Proofs of Main Results}

\subsection{Proof of Theorem~\ref{thm:X1W-X2}}

We first state two important technical lemmas, one for the error of $\hat\mx^{(i)}$ as an estimate for the true $\mx^{(i)}$ for each $i=1,2$,
and another for the difference between $\mw^{(1)\top}\hat\mw^{(1,2)}\mw^{(2)}$ and $\mw^{(1,2)}$.
\begin{lemma}
  \label{lemma:Xhat-XW}
  Consider the setting in Theorem~\ref{thm:X1W-X2}.
  Define $\mw^{(i)}$ as a
  minimizer of $\|\hat{\muu}^{(i)} \mathbf{O} - \muu^{(i)}\|_{F}$ over all
  $d \times d$ orthogonal matrix $\mo$.
  We then have
  \begin{equation}\label{eq:hatXiWi-Xi}
  	 \hat\mx^{(i)}\mw^{(i)}-\mx^{(i)}=\me^{(i)}\mx^{(i)}(\mx^{(i)\top}\mx^{(i)})^{-1}
    	+\mr^{(i)},
  \end{equation}
  where $\mr^{(i)}$ is a $n \times d$ matrix satisfying
  \begin{equation*}
    \begin{aligned}
    \|\mr^{(i)}\|_{2\to\infty} \lesssim n^{-1/2}(n\rho_n)^{-1/2}\log n
    \end{aligned}
  \end{equation*}
  with high probability, and thus
  \begin{equation*}
  	\begin{aligned}
  	\|\hat\mx^{(i)}\mw^{(i)}-\mx^{(i)}\|_{2\to\infty}
  		\lesssim n^{-1/2}\log^{1/2}n
  	\end{aligned}
  \end{equation*}
  with high probability.
\end{lemma}

Lemma~\ref{lemma:Xhat-XW} is an application of Theorem~3.2 of \cite{xie2024entrywise}.

\begin{lemma}
  \label{lemma:WWW-W}
  Consider the setting in Theorem~\ref{thm:X1W-X2}.
  We then have
  $$
	\|\mw^{(1)\top}\hat\mw^{(1,2)}\mw^{(2)} -\mw^{(1,2)}\|
	\lesssim |\mathcal{S}|^{-1/2}(n\rho_n)^{-1/2}\log^{1/2} n
    + (n\rho_n)^{-1}\log n
$$
with high probability.  
\end{lemma}
  
  The proof of Lemma~\ref{lemma:WWW-W} is presented in Section~\ref{sec:proof of lemma:WWW-W}.

We now proceed with the proof of Theorem~\ref{thm:X1W-X2}.
Let $\xi_i := \hat\mx^{(i)} \mw^{(i)}-\mx^{(i)}$ for $i=1,2$.
We decompose $\hat\my\mw^{(2)}$ as
$$
\begin{aligned}
	\hat\my\mw^{(2)}
	&=(\hat\mx^{(2)}-\hat\mx^{(1)}\hat\mw^{(1,2)})\mw^{(2)}
	\\&
	=\xi_2
	+\mx^{(2)}
	-(\xi_1+\mx^{(1)})\mw^{(1)\top}\hat\mw^{(1,2)}\mw^{(2)}
	\\&
	=\xi_2
	+(\mx^{(1)}\mw^{(1,2)}+\my)
	-(\xi_1+\mx^{(1)})\mw^{(1,2)}
	-(\xi_1+\mx^{(1)})(\mw^{(1)\top}\hat\mw^{(1,2)}\mw^{(2)}-\mw^{(1,2)})
	\\&\quad \text{(recall Eq.~\eqref{eq:X2=X1W+Y})}
	\\&
	=\my
	+\xi_2
	-\xi_1\mw^{(1,2)}
	-(\xi_1+\mx^{(1)})(\mw^{(1)\top}\hat\mw^{(1,2)}\mw^{(2)}-\mw^{(1,2)})	
	\\&
	=\my
	+(\me^{(2)}\mx^{(2)}(\mx^{(2)\top}\mx^{(2)})^{-1}
    	+\mr^{(2)})
	-(\me^{(1)}\mx^{(1)}(\mx^{(1)\top}\mx^{(1)})^{-1}
    	+\mr^{(1)})\mw^{(1,2)}
	\\&-(\xi_1+\mx^{(1)})(\mw^{(1)\top}\hat\mw^{(1,2)}\mw^{(2)}-\mw^{(1,2)})	
	\quad \text{(recall Eq.~\eqref{eq:hatXiWi-Xi} in Lemma~\ref{lemma:Xhat-XW})}
	\\&
	=\my+\me^{(2)}\mx^{(2)}(\mx^{(2)\top}\mx^{(2)})^{-1}-\me^{(1)}\mx^{(1)}(\mx^{(1)\top}\mx^{(1)})^{-1}\mw^{(1,2)}+\mr,
\end{aligned}
$$
where
\begin{equation*}
	\begin{aligned}
		\mr
		:=\mr^{(2)}
	    -\mr^{(1)}\mw^{(1,2)}
	    -(\xi_1+\mx^{(1)})(\mw^{(1)\top}\hat\mw^{(1,2)}\mw^{(2)}-\mw^{(1,2)}).
	\end{aligned}
\end{equation*}
Note $$\|\mx^{(i)}\|_{2\to\infty}\leq \|\muu^{(1)}\|_{2\to\infty}\cdot \|(\mLambda^{(1)})^{1/2}\|\lesssim n^{-1/2}(n\rho_n)^{1/2},$$
and by Lemma~\ref{lemma:Xhat-XW}, Lemma~\ref{lemma:WWW-W} we have
\begin{equation}\label{eq:r 2toinfty}
	\begin{aligned}
	\|\mr\|_{2\to\infty}
	&\lesssim \|\mr^{(2)}\|_{2\to\infty}
	+\|\mr^{(1)}\|_{2\to\infty}
	+(\|\xi_1\|_{2\to\infty}+\|\mx^{(i)}\|_{2\to\infty})
	\cdot \|\mw^{(1)\top}\hat\mw^{(1,2)}\mw^{(2)}-\mw^{(1,2)}\|
	\\& 
	\lesssim n^{-1/2}(n\rho_n)^{-1/2}\log n
	\\&+(n^{-1/2}\log^{1/2}n
	+n^{-1/2}(n\rho_n)^{1/2})
	\cdot (|\mathcal{S}|^{-1/2}(n\rho_n)^{-1/2}\log^{1/2} n
    + (n\rho_n)^{-1}\log n)
	\\&
	\lesssim n^{-1/2}(n\rho_n)^{-1/2}\log n
	+
	|\mathcal{S}|^{-1/2}n^{-1/2}\log^{1/2} n
    + n^{-1/2}(n\rho_n)^{-1/2}\log n
    \\&
    \lesssim 
	|\mathcal{S}|^{-1/2}n^{-1/2}\log^{1/2} n
    + n^{-1/2}(n\rho_n)^{-1/2}\log n
\end{aligned}
\end{equation}
with high probability.

Finally, by Lemma~\ref{lemma|E|_2|UEV|F} we have
\begin{equation}\label{eq:m1 2toinfty}
\begin{aligned}
	\|\me^{(2)}\mx^{(2)}(\mx^{(2)\top}\mx^{(2)})^{-1}\|_{2\to\infty}
	&\leq \|\me^{(2)}\muu^{(2)}(\mLambda^{(2)})^{-1/2}\|_{2\to\infty}\\
	&\leq \|\me^{(2)}\muu^{(2)}\|_{2\to\infty}\cdot \|(\mLambda^{(2)})^{-1/2}\|_{2\to\infty}\\
	&\lesssim (\rho_n \log n)^{1/2}\cdot (n\rho_n)^{-1/2}
	\lesssim n^{-1/2}\log^{1/2} n
\end{aligned}
\end{equation}
and 
\begin{equation}\label{eq:m2 2toinfty}
\begin{aligned}
	\|\me^{(1)}\mx^{(1)}(\mx^{(1)\top}\mx^{(1)})^{-1}\mw^{(1,2)}\|_{2\to\infty}\lesssim n^{-1/2}\log^{1/2} n
\end{aligned}
\end{equation}
with high probability.
Then by combining Eq.~\eqref{eq:r 2toinfty}, Eq.~\eqref{eq:m1 2toinfty}, and Eq.~\eqref{eq:m2 2toinfty}, we have $\|\hat\my\mw-\my\|_{2\to\infty}\lesssim n^{-1/2}\log^{1/2}n$ with high probability.


\subsection{Proof of Theorem~\ref{thm:CLT}}

From Theorem~\ref{thm:X1W-X2}, for the $k$th rows $\hat{\mathbf{y}}_k$ of $\hat\my$ and ${\mathbf{y}}_k$ of $\my$, let $$\bm{\theta}^{(k)}:=(\mx^{(2)\top}\mx^{(2)})^{-1}\mx^{(2)\top}\mathbf{e}_k^{(2)}
		-\mw^{(1,2)\top}(\mx^{(1)\top}\mx^{(1)})^{-1}\mx^{(1)\top}\mathbf{e}_k^{(1)},$$
we then have
\begin{equation}\label{eq:clt1}
	\begin{aligned}
		\mw^\top \hat{\mathbf{y}}_k-{\mathbf{y}}_k
		=\bm{\theta}^{(k)}
		+\mathbf{r}_k,
	\end{aligned}
\end{equation}
where $\mathbf{e}_k^{(2)},\mathbf{e}_k^{(1)},\mathbf{r}_k$ are the $k$th rows of $\me^{(2)},\me^{(1)},\mr$, respectively.
Because $\operatorname{Var}[\me^{(i)}_{st}] = \mpp^{(i)}_{st}(1 - \mpp^{(i)}_{st})$, and $\me^{(1)},\me^{(2)}$ are independent symmetric matrices, where the upper triangular entries are independent, we have
\begin{equation*}
	\begin{aligned}
		\operatorname{Var}\big[\bm{\theta}^{(k)}\big]
		=\mathbf{\Gamma}^{(k)}.
	\end{aligned}
\end{equation*}

Denote $\mz^{(2)}:=(\mx^{(2)\top}\mx^{(2)})^{-1}\mx^{(2)\top}, \mz^{(1)}:=-\mw^{(1,2)\top}(\mx^{(1)\top}\mx^{(1)})^{-1}\mx^{(1)\top}$ and let $\mathbf{z}^{(i)}_\ell$ be the $\ell$th column of $\mz^{(i)}$ for any $\ell\in[n]$.
Furthermore, let
$$
\begin{aligned}
	\my^{(k,i)}_{\ell}:=\me^{(i)}_{k,\ell}\mathbf{z}^{(i)}_{\ell}
	\text{ for }i=1,2\text{ and any }\ell\in[n].
\end{aligned}
$$
We then have
$$
\begin{aligned}
\bm{\theta}^{(k)}=\sum_{i=1,2}\sum_{\ell\in[n]}\my^{(k,i)}_{\ell}.
\end{aligned}
$$
Note $\{\my^{(k,1)}_{\ell},\my^{(k,2)}_{\ell}\}_{\ell\in[n]}$ are mutually independent zero-mean random vectors.
Let
$$
\tilde\my^{(k,i)}_{\ell}:=(\mathbf{\Gamma}^{(k)})^{-1/2}\my^{(k,i)}_{\ell}
\text{ for }i=1,2\text{ and any }\ell\in[n].
$$
For any fixed $k$ and $\ell$, we can bound the spectral norm of $\tilde\my^{(k,1)}_{\ell}$ by
\begin{equation}\label{eq:Yk1l}
	\begin{aligned}
		\|\tilde\my^{(k,1)}_{\ell}\|
		&=\|(\mathbf{\Gamma}^{(k)})^{-1/2}\|
		\cdot|\me^{(1)}_{k,\ell}|
		\cdot\|\mw^{(1,2)}\|
		\cdot\|(\mx^{(1)\top}\mx^{(1)})^{-1}\|
		\cdot\|\mx^{(1)}\|_{2\to\infty}
		\\& \lesssim n^{1/2}\cdot 1\cdot 1\cdot (n\rho_n)^{-1}\cdot n^{-1/2}(n\rho_n)^{1/2}
		\lesssim (n\rho_n)^{-1/2}
	\end{aligned}
\end{equation}
almost surely.
For any fixed $\epsilon>0$, Eq.~\eqref{eq:Yk1l} implies that, for sufficiently large $n$, we have $\|\tilde\my^{(k,1)}_{\ell}\|\leq \epsilon$ almost surely; by similar analysis we also have $\|\tilde\my^{(k,2)}_{\ell}\|\leq \epsilon$ almost surely.
And thus
$$
\lim_{n \to \infty}
\sum_{i=1,2}\sum_{\ell\in[n]}\mathbb{E}\big[\|\tilde\my^{(k,i)}_{\ell}\|^2\cdot\mathbb{I}\{\|\tilde\my^{(k,i)}_{\ell}\|>\epsilon\}\big]
=0.
$$
As $\epsilon>0$ is fixed but arbitrary, the collection $\{\tilde\my^{(k,i)}_{\ell}\}$ satisfies the condition of the Lindeberg-Feller central limit theorem (see e.g.,
Proposition~2.27 in \cite{van2000asymptotic}) and hence
\begin{equation}\label{eq:clt2}
	(\mathbf{\Gamma}^{(k)})^{-1/2}
	\bm{\theta}^{(k)}
		\rightsquigarrow \mathcal{N}(\mathbf{0}, \mathbf{I})
\end{equation}
as $n\to\infty$.
By Theorem~\ref{thm:X1W-X2}, we also have
\begin{equation*}
\begin{aligned}
	\|(\mathbf{\Gamma}^{(k)})^{-1/2}\mathbf{r}_k\|
	&\lesssim \|(\mathbf{\Gamma}^{(k)})^{-1/2}\|\cdot \|\mr\|_{2\to\infty}
	\\&\lesssim n^{1/2}\cdot (|\mathcal{S}|^{-1/2}n^{-1/2}\log^{1/2} n
    + n^{-1/2}(n\rho_n)^{-1/2}\log n)
    \\&\lesssim |\mathcal{S}|^{-1/2}\log^{1/2} n
    + (n\rho_n)^{-1/2}\log n
\end{aligned}
\end{equation*}
with high probability. Thus we have
\begin{equation}\label{eq:clt3}
	(\mathbf{\Gamma}^{(k)})^{-1/2}\mathbf{r}_k\  \xrightarrow{p} \mathbf{0}
\end{equation}
as $n\to\infty$ provided that {\color{black}$|\mathcal{S}|=\omega(\log n)$ and $n\rho_n=\omega(\log^2 n)$}.
Combining Eq.~\eqref{eq:clt1}, Eq.~\eqref{eq:clt2}, Eq.~\eqref{eq:clt3}, and applying Slutsky's theorem, the desired result is obtained.

\subsection{Proof of Theorem~\ref{thm:HT}}

Define
$$
\zeta_k:=
\hat{\mathbf{y}}_k^\top \mw
(\mathbf{\Gamma}^{(k)})^{-1}
\mw^\top \hat{\mathbf{y}}_k.
$$
Under $\mathbb{H}_0$ we have $\zeta_k\rightsquigarrow \chi_{d}^2$ ; see Eq.~\eqref{eq:test1}. 
As $d$ is finite, we conclude that $\zeta_k$ is bounded in probability.
By the assumption $\lambda_\ell(\mathbf{\Gamma}^{(k)})\asymp n^{-1}$ for all $\ell\in[d]$, we have $\lambda_\ell((\mathbf{\Gamma}^{(k)})^{-1})\asymp n$ for all $\ell\in[d]$.
We thus have $\zeta_k\asymp n \|\hat{\mathbf{y}}_{k}\|^2_F$, i.e., $n \|\hat{\mathbf{y}}_{k}\|^2_F$ is bounded in probability.
Now recall the definition of $T_k$ in Theorem~\ref{thm:HT}.
We then by Lemma~\ref{lemma:sigma} have
\begin{equation}\label{eq:Tk-zeta_k}
	\begin{aligned}
	|T_k-\zeta_k|
	&\leq \|\mw
(\mathbf{\Gamma}^{(k)})^{-1}
\mw^\top-(\hat{\mathbf{\Gamma}}^{(k)})^{-1}\|
\cdot \|\hat{\mathbf{y}}_{k}\|^2_F\\
&\lesssim [(n\rho_n)^{-1/2}\log^{1/2}n]\cdot [n\|\hat{\mathbf{y}}_{k}\|^2_F]
\xrightarrow{p} 0
\end{aligned}
\end{equation}
provided that {\color{black} $n\rho_n=\omega(\log n)$}.
Therefore, by Slutsky's theorem we have $T_k\rightsquigarrow \chi_{d}^2$ under $\mathbb{H}_0$.

We now consider the case where $\mathbf{y}_k\neq\mathbf{0}$ satisfies a local alternative hypothesis where 
\begin{equation}\label{eq:eta}
	\mathbf{y}_k^\top (\mathbf{\Gamma}^{(k)})^{-1} \mathbf{y}_k\to\eta_k
\end{equation}
for some finite $\eta>0$.
Define $\xi_k=(\mathbf{\Gamma}^{(k)})^{-1/2} \mathbf{y}_k$, and then Eq.~\eqref{eq:eta} means $\|\xi_k\|^2\to\eta_k$.
Recall Theorem~\ref{thm:CLT}. In particular we have
$$
(\mathbf{\Gamma}^{(k)})^{-1/2}
\mw^\top \hat{\mathbf{y}}_k
-\xi_k \rightsquigarrow \mathcal{N}(\mathbf{0},\mi).
$$
We therefore have $\zeta_k\rightsquigarrow \chi_{d}^2 (\eta_k)$, where $\zeta_k$ is defined at the beginning of the current proof. As $\eta_k$ is finite, we conclude that $\zeta_k$ is bounded in probability. Finally, using the same argument as that for deriving Eq.~\eqref{eq:Tk-zeta_k} under $\mathbb{H}_0$, we also have $|T_k-\zeta_k|\xrightarrow{p} 0$ under the local alternative in Eq.~\eqref{eq:eta} and hence $T_k\rightsquigarrow \chi_{d}^2 (\eta_k)$ as desired.

\subsection{Proof of Theorem~\ref{thm:max}}

Consider any fixed $k$ and $\ell$ such that vertices $k$ and $\ell$ are unshifted.
Recall Lemma~\ref{lemma:Xhat-XW} and let $\tilde\xi^{(i)}_k=\mw^{(i)\top}\hat{\mathbf{x}}^{(i)}_{k}-{\mathbf{x}}_{k}$ for $i=1,2$ and $k\in[n]$. We then have
\begin{equation}\label{eq:clt1_2}
\begin{aligned}
    \hat\Delta_{k,\ell} &=
	\hat\mpp^{(1)}_{k,\ell}-\hat\mpp^{(2)}_{k,\ell}\\
	&=\hat{\mathbf{x}}^{(1)\top}_{k}\hat{\mathbf{x}}^{(1)}_{\ell}
	-\hat{\mathbf{x}}^{(2)\top}_{k}\hat{\mathbf{x}}^{(2)}_{\ell}\\
	&=(\hat{\mathbf{x}}^{(1)\top}_{k}\mw^{(1)})(\mw^{(1)\top}\hat{\mathbf{x}}^{(1)}_{\ell})
	-(\hat{\mathbf{x}}^{(2)\top}_{k}\mw^{(2)})(\mw^{(2)\top}\hat{\mathbf{x}}^{(2)\top}_{\ell})\\
	&=({\mathbf{x}}^{(1)\top}_{k}+\tilde\xi_k^{(1)\top})({\mathbf{x}}^{(1)}_{\ell}+\tilde\xi_\ell^{(1)})
	-({\mathbf{x}}^{(2)\top}_{k}+\tilde\xi_k^{(2)\top})({\mathbf{x}}^{(2)}_{\ell}+\tilde\xi_\ell^{(2)})\\
	&=({\mathbf{x}}^{(1)\top}_{k}{\mathbf{x}}^{(1)}_{\ell}-{\mathbf{x}}^{(2)\top}_{k}{\mathbf{x}}^{(2)}_{\ell})
	+\tilde\xi_k^{(1)\top}{\mathbf{x}}^{(1)}_{\ell}
	+{\mathbf{x}}^{(1)\top}_{k}\tilde\xi_\ell^{(1)}
	-\tilde\xi_k^{(2)\top}{\mathbf{x}}^{(2)}_{\ell}
	-{\mathbf{x}}^{(2)\top}_{k}\tilde\xi_\ell^{(2)}
	+\tilde\xi_k^{(1)\top}\tilde\xi_\ell^{(1)}-\tilde\xi_k^{(2)\top}\tilde\xi_\ell^{(2)}\\
	&=\Delta_{k,\ell}+\tilde\theta_{k,\ell}+\tilde r_{k,\ell},
\end{aligned}
\end{equation}
where we define
$$
\begin{aligned}
	\tilde\theta_{k,\ell}:&=\mathbf{e}^{(1)\top}_k\mx^{(1)}(\mx^{(1)\top}\mx^{(1)})^{-1}\mathbf{x}_\ell^{(1)}
	+\mathbf{e}^{(1)\top}_\ell\mx^{(1)}(\mx^{(1)\top}\mx^{(1)})^{-1}\mathbf{x}_k^{(1)}
    \\&-\mathbf{e}^{(2)\top}_k\mx^{(2)}(\mx^{(2)\top}\mx^{(2)})^{-1}\mathbf{x}_\ell^{(2)}
	-\mathbf{e}^{(2)\top}_\ell\mx^{(2)}(\mx^{(2)\top}\mx^{(2)})^{-1}\mathbf{x}_k^{(2)}\\
	&=\mathbf{e}^{(1)\top}_k\bm\Pi^{(1)}_\ell
	+\mathbf{e}^{(1)\top}_\ell\bm\Pi^{(1)}_k
    -\mathbf{e}^{(2)\top}_k\bm\Pi^{(2)}_\ell
	-\mathbf{e}^{(2)\top}_\ell\bm\Pi^{(2)}_k,\\
\tilde r_{k,\ell}
:&=\mathbf{r}^{(1)\top}_k\mathbf{x}_\ell^{(1)}
+\mathbf{x}_\ell^{(1)\top}\mathbf{r}^{(1)}_k
-\mathbf{r}^{(2)\top}_k\mathbf{x}_\ell^{(2)}
-\mathbf{x}_\ell^{(2)\top}\mathbf{r}^{(2)}_k
+\tilde\xi_k^{(1)\top}\tilde\xi_\ell^{(1)}-\tilde\xi_k^{(2)\top}\tilde\xi_\ell^{(2)}.
\end{aligned}
$$
Here, $\bm{\Pi}^{(i)}_k$, $\mathbf{e}^{(i)}_k$, and $\mathbf{r}^{(i)}_k$ denote the $k$th rows of $\bm{\Pi}^{(i)}$, $\me^{(i)}$, and $\mr^{(i)}$, respectively, and $\bm{\Pi}^{(i)}$ is defined in Eq.~\eqref{eq:Pi}.
For $\tilde r$, by Lemma~\ref{lemma:Xhat-XW} we have
\begin{equation}\label{eq:r_kl}
	\begin{aligned}
	|\tilde r_{k,\ell}|
	&\lesssim \|\mr^{(1)}\|_{2\to\infty}\cdot\|\mx^{(1)}\|_{2\to\infty}+\|\mx^{(1)}\|_{2\to\infty}\cdot\|\mr^{(1)\top}\|_{2\to\infty}
\\&+\|\mr^{(2)}\|_{2\to\infty}\cdot\|\mx^{(2)}\|_{2\to\infty}+\|\mx^{(2)}\|_{2\to\infty}\cdot\|\mr^{(2)}\|_{2\to\infty}
\\&+\|\xi_1\|_{2\to\infty}\cdot\|\xi_1\|_{2\to\infty}+\|\xi_2\|_{2\to\infty}\cdot\|\xi_2\|_{2\to\infty}\\
&\lesssim n^{-1/2}(n\rho_n)^{-1/2}\log n\cdot n^{-1/2}(n\rho_n)^{1/2}
+(n^{-1/2}\log^{1/2}n)^2
\lesssim n^{-1}\log n
\end{aligned}
\end{equation}
with high probability, where $\xi_i$ is defined in the proof of Theorem~\ref{thm:X1W-X2}.
Notice $\operatorname{Var}\big[\tilde \theta_{k,\ell}\big]= {\mathbf{\Upsilon}}_{k,\ell}$ where the $n\times n$ matrix $ {\mathbf{\Upsilon}}$ defined in Eq.~\eqref{eq;tilde Upsilon}, as $\operatorname{Var}[\me^{(i)}_{st}] = \mpp^{(i)}_{st}(1 - \mpp^{(i)}_{st})$, and $\me^{(1)},\me^{(2)}$ are independent symmetric matrices, where the upper triangular entries are independent. 

We now show that $\tilde{r}_{k,\ell}$ is asymptotically normal. We will analyze the case for $k \neq \ell$. And for the case where $k = \ell$, by the observation that
$
\tilde{\theta}_{k,k} := 2 \mathbf{e}^{(1)\top}_k \mx^{(1)} \big(\mx^{(1)\top} \mx^{(1)}\big)^{-1} \mathbf{x}_k^{(1)} 
- 2 \mathbf{e}^{(2)\top}_k \mx^{(2)} \big(\mx^{(2)\top} \mx^{(2)}\big)^{-1} \mathbf{x}_k^{(2)},
$
the analysis for this case is similar.
Let
$$
\begin{aligned}
	\my^{(k,\ell,1)}_{t}:=\me^{(1)}_{k,t}\bm\Pi^{(1)}_{\ell,t},
	\quad \my^{(k,\ell,2)}_{t}:=-\me^{(2)}_{k,t}\bm\Pi^{(2)}_{\ell,t}
	\text{ for any }t\in[n],
\end{aligned}
$$
and let
$$
\tilde\my^{(k,\ell,i)}_{t}:=( {\mathbf{\Upsilon}}_{k,\ell})^{-1/2}\my^{(k,\ell,i)}_{t}
\text{ for }i=1,2\text{ and any }t\in[n].
$$
We then have
$$
\begin{aligned}
( {\mathbf{\Upsilon}}_{k,\ell})^{-1/2}\tilde\theta_{k,\ell}
&=( {\mathbf{\Upsilon}}_{k,\ell})^{-1/2}\Big[\sum_{i\in\{1,2\}}\sum_{t\in[n]}\my^{(k,\ell,i)}_{t}+\my^{(\ell,k,i)}_{t}\Big]\\
&=\sum_{i\in\{1,2\}}\sum_{t\neq \ell}\tilde\my^{(k,\ell,i)}_{t}
+\sum_{i\in\{1,2\}}\sum_{t\neq k}\tilde\my^{(\ell,k,i)}_{t}
+\sum_{i\in\{1,2\}}(\tilde\my^{(k,\ell,i)}_{\ell}+\tilde\my^{(\ell,k,i)}_k)
=\sum_{\tilde\my\in\mathcal{M}}\tilde\my,
\end{aligned}
$$
where we define 
$${\mathcal{M}}:=\{\tilde\my^{(k,\ell,i)}_{t}\}_{i\in\{1,2\},t\neq \ell}\cup \{\tilde\my^{(\ell,k,i)}_{t}\}_{i\in\{1,2\},t\neq k}\cup \{(\tilde\my^{(k,\ell,i)}_{\ell}+\tilde\my^{(\ell,k,i)}_k)\}_{i\in\{1,2\}},$$
and notice that $\mathcal{M}$ contains mutually independent zero-mean random variables.
We can bound the spectral norm of $\tilde\my\in\mathcal{M}$. 
For any $t\neq \ell$ we have
\begin{equation}\label{eq:Yk1l_2}
	\begin{aligned}
		|\tilde\my^{(k,\ell,i)}_{t}|
		&=|( {\mathbf{\Upsilon}}_{k,\ell})^{-1/2}|
		\cdot |\me^{(i)}_{k,t}|
		\cdot\|(\mx^{(i)\top}\mx^{(i)})^{-1}\|
		\cdot\|\mx^{(i)}\|_{2\to\infty}^2
		\\& \lesssim n(n\rho_n)^{-1/2} \cdot 1\cdot (n\rho_n)^{-1}\cdot (n^{-1/2}(n\rho_n)^{1/2})^2
		\lesssim (n\rho_n)^{-1/2}
	\end{aligned}
\end{equation}
almost surely.
Then for any fixed $\epsilon>0$, Eq.~\eqref{eq:Yk1l_2} implies that, for sufficiently large $n$, we have $|\tilde\my^{(k,\ell,i)}_{t}|\leq \epsilon$ almost surely; by similar analysis we also have $|\tilde\my^{(\ell,k,i)}_{t}|\leq \epsilon$ for any $t\neq k$, and $|(\tilde\my^{(k,\ell,i)}_{\ell}+\tilde\my^{(\ell,k,i)}_k)|\leq \epsilon$ almost surely.
And thus
$$
\begin{aligned}
	\lim_{n \to \infty}
	&\sum_{\tilde\my\in\mathcal{M}}\mathbb{E}\big[|\tilde\my|^2\cdot\mathbb{I}\{|\tilde\my|>\epsilon\}\big]
=0.
\end{aligned}
$$
As $\epsilon>0$ is fixed but arbitrary, the collection $\mathcal{M}$ satisfies the condition of the Lindeberg-Feller central limit theorem (see e.g.,
Proposition~2.27 in \cite{van2000asymptotic}) and hence
\begin{equation}\label{eq:clt2_2}
	( {\mathbf{\Upsilon}}_{k,\ell})^{-1/2}\tilde\theta_{k,\ell}
		\rightsquigarrow \mathcal{N}(0, 1)
\end{equation}
as $n\to\infty$.
By Eq.~\eqref{eq:r_kl}, we also have
\begin{equation*}
\begin{aligned}
	|( {\mathbf{\Upsilon}}_{k,\ell})^{-1/2}\tilde r_{k,\ell}|
	\lesssim \|( {\mathbf{\Upsilon}}_{k,\ell})^{-1/2}\|\cdot |\tilde r_{k,\ell}|
	\lesssim n(n\rho_n)^{-1/2}\cdot n^{-1}\log n
   \lesssim (n\rho_n)^{-1/2}\log n
\end{aligned}
\end{equation*}
with high probability. Thus we have
\begin{equation}\label{eq:clt3_2}
	( {\mathbf{\Upsilon}}_{k,\ell})^{-1/2}\tilde r_{k,\ell}  \xrightarrow{p} 0
\end{equation}
as $n\to\infty$ provided that {\color{black}$n\rho_n=\omega(\log^2 n)$}.
Combining Eq.~\eqref{eq:clt1_2}, Eq.~\eqref{eq:clt2_2}, Eq.~\eqref{eq:clt3_2}, and applying Slutsky's theorem, then the desired result is obtained. 

\subsection{Proof of Theorem~\ref{thm:max_HT}}

We first bound $|(\hat{\bm\Upsilon}_{k,\ell})^{-1/2}-(\bm\Upsilon_{k,\ell})^{-1/2}|$.
By Lemma~\ref{lemma:lambda W} and
$$
\begin{aligned}
	\hat{\bm\Pi}^{(i)}-{\bm\Pi}^{(i)}
	&=\hat\muu^{(i)}\hat\muu^{(i)\top}- \muu^{(i)} \muu^{(i)\top} \\
	&=\hat\muu^{(i)}\mw^{(i)}\mw^{(i)\top}\hat\muu^{(i)\top}-\ \muu^{(i)} \muu^{(i)\top}\\
	&=(\hat\muu^{(i)}\mw^{(i)}-\muu^{(i)})(\hat\muu^{(i)}\mw^{(i)}-\muu^{(i)})^\top
	+(\hat\muu^{(i)}\mw^{(i)}-\muu^{(i)})\muu^{(i)\top}
	+\muu^{(i)}(\hat\muu^{(i)}\mw^{(i)}-\muu^{(i)})^\top,
\end{aligned}
$$
we have
\begin{equation}\label{eq:hatPi-Pi}
	\begin{aligned}
		\|\hat{\bm\Pi}^{(i)}-{\bm\Pi}^{(i)}\|_{\max}
		&\leq \|\hat\muu^{(i)}\mw^{(i)}-\muu^{(i)}\|_{2\to\infty}^2
		+2\|\muu^{(i)}\|_{2\to\infty}\cdot \|\hat\muu^{(i)}\mw^{(i)}-\muu^{(i)}\|_{2\to\infty}\\
		&\lesssim (n^{-1/2}(n\rho_n)^{-1/2} \log^{1/2} n)^2
		+n^{-1/2}\cdot n^{-1/2}(n\rho_n)^{-1/2} \log^{1/2} n\\
		&\lesssim n^{-1}(n\rho_n)^{-1/2} \log^{1/2} n
	\end{aligned}
\end{equation}
with high probability. By the similar analysis of $\|\mh_1\|$ in the proof of Lemma~\ref{lemma:sigma}, Eq.~\eqref{eq:H1_2}, and Eq.~\eqref{eq:H1_3}, we have
$$
\begin{aligned}
	&\|\hat{\bm\Pi}^{(i)}\hat{\mathbf{\Xi}}^{(k,i)}\hat{\bm\Pi}^{(i)}-\bm\Pi^{(i)}\mathbf{\Xi}^{(k,i)}\bm\Pi^{(i)}\|_{\max}\\
	&\leq \|\hat\muu^{(i)}\mw^{(i)}-\muu^{(i)}\|_{2\to\infty}\cdot\|{\mathbf{\Xi}}^{(k,i)}\|\cdot \|\hat\muu^{(i)}\|_{2\to\infty}
	+\|\muu^{(i)}\|_{2\to\infty}\cdot\|{\mathbf{\Xi}}^{(k,i)}\|\cdot \|\hat\muu^{(i)}\mw^{(i)}-\muu^{(i)}\|_{2\to\infty}\\
	&+\|\hat\muu^{(i)}\|_{2\to\infty}^2\cdot\|\hat{\mathbf{\Xi}}^{(k,i)}-{\mathbf{\Xi}}^{(k,i)}\|\\
	&\lesssim n^{-1/2}(n\rho_n)^{-1/2} \log^{1/2} n\cdot n^{-1}(n\rho_n)\cdot n^{-1/2}
	+(n^{-1/2})^2\cdot n^{-1}(n\rho_n)^{1/2}\log^{1/2} n\\
	&\lesssim n^{-2}(n\rho_n)^{1/2} \log^{1/2} n
\end{aligned}
$$
with high probability,
and thus
\begin{equation}\label{eq:hatPsi-Psi}
\|\hat{\mathbf{\Psi}}^{(k)}-\mathbf{\Psi}^{(k)}\|_{\max}
\leq \sum_{i\in\{1,2\}}\|\hat{\bm\Pi}^{(i)}\hat{\mathbf{\Xi}}^{(k,i)}\hat{\bm\Pi}^{(i)}-\bm\Pi^{(i)}\mathbf{\Xi}^{(k,i)}\bm\Pi^{(i)}\|_{\max}
\lesssim n^{-2}(n\rho_n)^{1/2} \log^{1/2} n
\end{equation}
with high probability.
By Eq.~\eqref{eq:hatPi-Pi}, Eq.~\eqref{eq:hatPsi-Psi}, Eq.~\eqref{eq:H1_2}, and Eq.~\eqref{eq:H1_3}, and $\|\hat{\bm\Pi}^{(i)}\|_{\max}\asymp \|{\bm\Pi}^{(i)}\|_{\max} \lesssim n^{-1}$ we have
\begin{equation}\label{eq:hat Upsilon- Upsilon}
    |\hat{\bm\Upsilon}_{k,\ell}-\bm\Upsilon_{k,\ell}|
    \lesssim n^{-2}(n\rho_n)^{1/2} \log^{1/2} n
\end{equation}
with high probability.
Recall the assumption that $\bm\Upsilon_{k,\ell}\gtrsim n^{-2}(n\rho_n)$.
Then for $n\rho_n\gg \log n$, by Eq.~\eqref{eq:hat Upsilon- Upsilon} we have
\begin{equation}\label{eq:Upsilon-1/2}
	(\bm\Upsilon_{k,\ell})^{-1/2}\lesssim n(n\rho_n)^{-1/2},
	\quad (\hat{\bm\Upsilon}_{k,\ell})^{-1/2}\lesssim n(n\rho_n)^{-1/2}
\end{equation}
with high probability.
Since $|a^{-1/2}-b^{-1/2}|=\frac{|a-b|}{a^{1/2}b^{1/2}(a^{1/2}+b^{1/2})}$ for any $a,b>0$.
We have by Eq.~\eqref{eq:hat Upsilon- Upsilon} and Eq.~\eqref{eq:Upsilon-1/2} that
\begin{equation}
	|(\hat{\bm\Upsilon}_{k,\ell})^{-1/2}-(\bm\Upsilon_{k,\ell})^{-1/2}|
	\lesssim n^{-2}(n\rho_n)^{1/2} \log^{1/2} n\cdot (n(n\rho_n)^{-1/2})^3
	\lesssim n (n\rho_n)^{-1}\log^{1/2}n
\end{equation}
with high probability.

We define
$$
\tilde\zeta_{k,\ell}:=
({\bm\Upsilon}_{k,\ell})^{-1/2}
\hat\Delta_{k,\ell}.
$$
Under $\mathbb{H}_0:(k,\ell)\in\mathcal{U}\times \mathcal{U}$ we have $\tilde\zeta_{k,\ell}\rightsquigarrow \mathcal{N}(0,1)$ by Theorem~\ref{thm:max}. 
Thus $\tilde\zeta_{k,\ell}$ is bounded in probability.
By the assumption $\bm\Upsilon_{k,\ell}\asymp n^{-2}(n\rho_n)$, we have $\tilde\zeta_{k,\ell}\asymp n(n\rho_n)^{-1/2}$.
We thus have $\zeta_k\asymp n(n\rho_n)^{-1/2} |\hat\Delta_{k,\ell}|$, i.e., $n(n\rho_n)^{-1/2} |\hat\Delta_{k,\ell}|$ is bounded in probability.
Now recall the definition of $\tilde T_{k,\ell}$ in Theorem~\ref{thm:max_HT}.
We then by Lemma~\ref{lemma:sigma} have
\begin{equation}\label{eq:Tk-zeta_k2}
	\begin{aligned}
	|\tilde T_{k,\ell}-\tilde\zeta_{k,\ell}|
	&\leq |(\hat{\bm\Upsilon}_{k,\ell})^{-1/2}-(\bm\Upsilon_{k,\ell})^{-1/2}|
\cdot |\hat\Delta_{k,\ell}|\\
&\lesssim [(n\rho_n)^{-1/2}\log^{1/2}n]\cdot [n(n\rho_n)^{-1/2}|\hat\Delta_{k,\ell}|]
\xrightarrow{p} 0
\end{aligned}
\end{equation}
provided that {\color{black} $n\rho_n=\omega(\log n)$}.
Therefore, by Slutsky's theorem we have $\tilde T_{k,\ell}\rightsquigarrow \mathcal{N}(0,1)$ under $\mathbb{H}_0$.

We now consider the case where $(k,\ell)\notin\mathcal{U}\times \mathcal{U}$ satisfies a local alternative hypothesis where 
\begin{equation}\label{eq:eta2}
	\tilde\xi_{k,\ell}:=(\bm\Upsilon_{k,\ell})^{-1/2}\Delta_{k,\ell}\to\tilde\eta_{k,\ell}
\end{equation}
for some finite $\eta>0$.
Recall Theorem~\ref{thm:max}. In particular we have
$$
({\bm\Upsilon}_{k,\ell})^{-1/2}
\hat\Delta_{k,\ell}
-\tilde\xi_{k,\ell} \rightsquigarrow \mathcal{N}(0,1).
$$
We therefore have $\tilde\zeta_{k,\ell}\rightsquigarrow \mathcal{N}(\tilde\eta_{k,\ell},1)$. As $\tilde\eta_{k,\ell}$ is finite, we conclude that $\tilde\zeta_{k,\ell}$ is bounded in probability. Finally, using the same argument as that for deriving Eq.~\eqref{eq:Tk-zeta_k2} under $\mathbb{H}_0$, we also have $|\tilde T_{k,\ell}-\tilde\zeta_{k,\ell}|\xrightarrow{p} 0$ under the local alternative in Eq.~\eqref{eq:eta2} and hence $\tilde T_{k,\ell}\rightsquigarrow \mathcal{N}(\tilde\eta_{k,\ell},1)$ as desired.

\section{Technical lemmas}
\label{sec:lemmas for thm:X1W-X2}

\begin{lemma}
  \label{lemma|E|_2|UEV|F}
  Consider the setting in Theorem~\ref{thm:X1W-X2}.
   Then 
for $i,j\in\{1,2\}$ we have
  $$
\begin{aligned}
&\|\mathbf{E}^{(i)}\|\lesssim (n \rho_{n})^{1 / 2},\quad
\|\muu^{(i)\top} \mathbf{E}^{(i)}\muu^{(i)}\| \lesssim ( \rho_{n} \log n)^{1/2},\\
&\|\me^{(i)}\muu^{(i)}  \|_{2\to\infty}\lesssim (\rho_n \log n)^{1/2}
\end{aligned}
$$
with high probability,
and 
$$
\|\muu^{(i)\top}_{\mathcal{S}} \mathbf{E}^{(j)}_{\mathcal{S}}\muu^{(j)}\| \lesssim ( \rho_{n} \log n)^{1/2}\cdot |\mathcal{S}|^{1/2}n^{-1/2}
$$
with high probability.
\end{lemma}

\ref{lemma|E|_2|UEV|F} is derived using the same arguments as Lemma~2 in \cite{zheng2022limit}. While Lemma~2 in \cite{zheng2022limit} is stated for asymmetric matrices, a little extra care is required in the proof of \ref{lemma|E|_2|UEV|F}, as the dependency among the entries of $\mathbf{E}^{(j)}$ leads to slightly more involved book-keeping.

\begin{lemma}
  \label{lemma:|lambda|}
  Consider the setting in Theorem~\ref{thm:X1W-X2}.
   Then 
for $i,j\in\{1,2\}$ we have
  $$
\begin{aligned}
&\lambda_k(\ma^{(i)})\asymp n\rho_n\text{ for }k=1,\dots,d,\\
&\lambda_k(\ma^{(i)})\lesssim (n\rho_n)^{1/2}\text{ for }k=d+1,\dots,n
\end{aligned}
$$
with high probability.
\end{lemma}

Using \ref{lemma|E|_2|UEV|F} and Weyl's inequality, we obtain the desired results for the eigenvalues of $\ma^{(i)}$.

\begin{lemma}
  \label{lemma:lambda W}
  Consider the setting in Theorem~\ref{thm:X1W-X2}.
   Then 
for $i,j\in\{1,2\}$ we have
  $$
\begin{aligned}
&\|\hat\muu^{(i)}\mw^{(i)}-\muu^{(i)}\|\lesssim (n\rho_n)^{-1/2},\\
&\|\hat\muu^{(i)}\mw^{(i)}-\muu^{(i)}\|_{2\to\infty}\lesssim n^{-1/2}(n\rho_n)^{-1/2} \log^{1/2} n,\\
&\|\mw^{(i)}(\mLambda^{(i)})^{-1/2}-(\hat\mLambda^{(i)})^{-1/2}\mw^{(i)}\|
    \lesssim n^{-1/2}(n\rho_n)^{-1}\log^{1/2}n
\end{aligned}
$$
with high probability.
\end{lemma}

Using \ref{lemma|E|_2|UEV|F} and Lemma~\ref{lemma:|lambda|}, and following the proof of Lemma~B.3 in \cite{xie2024entrywise}, we derive the results for Lemma~\ref{lemma:lambda W}.

\subsection{Proof of Lemma~\ref{lemma:WWW-W}}
\label{sec:proof of lemma:WWW-W}
Note that
$$
\begin{aligned}
	&\mw^{(1)\top}\hat\mw^{(1,2)}\mw^{(2)}
	= \argmin\limits_{\mo \in \mathcal{O}_d} \|\hat\mx^{(1)}_{\mathcal{S}}\mw^{(1)}\mo-\hat\mx^{(2)}_{\mathcal{S}}\mw^{(2)}\|_F
	,
	\\ 
	&\mw^{(1,2)}
	=\underset{\mo\in \mathcal{O}_d}{\arg\min}
	\|\mx_{\mathcal{S}}^{(1)}\mo-\mx_{\mathcal{S}}^{(2)}\|_F
	.
\end{aligned}
$$
Denote \[\mathbf{F}:=\mw^{(1)\top}\hat\mx^{(1)\top}_{\mathcal{S}}\hat\mx^{(2)}_{\mathcal{S}}\mw^{(2)}
		-\mx_{\mathcal{S}}^{(1)\top}\mx_{\mathcal{S}}^{(2)}.\]
We therefore have, by perturbation bounds for polar decompositions, that
\begin{equation}
 \label{eq:rencang} 
  \|\mw^{(1)\top}\hat\mw^{(1,2)}\mw^{(2)} -\mw^{(1,2)}\| \leq \frac{2\|\mathbf{F}\|}{\sigma_{\min}(\mx_{\mathcal{S}}^{(1)\top}\mx_{\mathcal{S}}^{(2)})}. 
\end{equation}
Indeed, 
from \ref{assum:main}, $\mx_{\mathcal{S}}^{(1)\top}\mx_{\mathcal{S}}^{(2)}$ is invertible.
Now suppose $\|\mathbf{F}\| <  \sigma_{\min}(\mx_{\mathcal{S}}^{(1)\top}\mx_{\mathcal{S}}^{(2)})$. Then 
$\hat\mx^{(1)\top}_{\mathcal{S}}\hat\mx^{(2)}_{\mathcal{S}}$ is also invertible and hence, by Theorem~1 in \cite{rencang} we have
$$
\begin{aligned}
	\|\mw^{(1)\top}\hat\mw^{(1,2)}\mw^{(2)} -\mw^{(1,2)}\| &\leq \frac{2\|\mathbf{F}\|}{\sigma_{\min}(\hat\mx^{(1)\top}_{\mathcal{S}}\hat\mx^{(2)}_{\mathcal{S}}) + \sigma_{\min}(\mx_{\mathcal{S}}^{(1)\top}\mx_{\mathcal{S}}^{(2)})} \leq \frac{2\|\mathbf{F}\|}{\sigma_{\min}(\mx_{\mathcal{S}}^{(1)\top}\mx_{\mathcal{S}}^{(2)})}.
\end{aligned}
 $$
Otherwise if $\|\mathbf{F}\| \geq  \sigma_{\min}(\mx_{\mathcal{S}}^{(1)\top}\mx_{\mathcal{S}}^{(2)})$
then, as $\|\mw^{(1)\top}\hat\mw^{(1,2)}\mw^{(2)} -\mw^{(1,2)}\| \leq 2$, Eq.~\eqref{eq:rencang} holds trivially. 

We now bound the spectral norm of $\mathbf{F}$.
First note that
$$
\begin{aligned}
	\mf
	&=(\hat\mx^{(1)}_{\mathcal{S}}\mw^{(1)}-\mx^{(1)}_{\mathcal{S}})^\top
	(\hat\mx^{(2)}_{\mathcal{S}}\mw^{(2)}-\mx^{(2)})
	+(\hat\mx^{(1)}_{\mathcal{S}}\mw^{(1)}-\mx^{(1)}_{\mathcal{S}})^\top\mx^{(2)}
	+\mx^{(1)\top}_{\mathcal{S}}(\hat\mx^{(2)}_{\mathcal{S}}\mw^{(2)}-\mx^{(2)}_{\mathcal{S}}).
\end{aligned}
$$
Next, by Eq.~\eqref{eq:hatXiWi-Xi} in Lemma~\ref{lemma:Xhat-XW}, we have
$$
\hat\mx^{(i)}_{\mathcal{S}}\mw^{(i)}-\mx^{(i)}_{\mathcal{S}}
=\me^{(i)}_{\mathcal{S}}\mx^{(i)}(\mx^{(i)\top}\mx^{(i)})^{-1}
    	+\mr^{(i)}_{\mathcal{S}}
=\me^{(i)}_{\mathcal{S}}\muu^{(i)}(\mLambda^{(i)})^{-1/2}
    	+\mr^{(i)}_{\mathcal{S}},
$$
where $\me^{(i)}_{\mathcal{S}}$ and $\mr^{(i)}_{\mathcal{S}}$ contain the rows in $\me^{(i)}$ and $\mr^{(i)}$ corresponding to $\mathcal{S}$, respectively; note that $\mx^{(i)}_{\mathcal{S}}=\muu^{(i)}_{\mathcal{S}}(\mLambda^{(i)})^{1/2}$.
We therefore have
$$
\begin{aligned}
	\mf
	&=
	(\hat\mx^{(1)}_{\mathcal{S}}\mw^{(1)}-\mx^{(1)}_{\mathcal{S}})^\top
	(\hat\mx^{(2)}_{\mathcal{S}}\mw^{(2)}-\mx^{(2)})
	\\
    	&
    	+
    	(\mLambda^{(1)})^{-1/2}
    	\muu^{(1)\top}
    	\me^{(1)\top}_{\mathcal{S}}
       	\muu^{(2)}_{\mathcal{S}}
       	(\mLambda^{(2)})^{1/2}
    	+
    	(\mLambda^{(1)})^{1/2}
    	\muu^{(1)\top}_{\mathcal{S}} 	
    	\me^{(2)}_{\mathcal{S}}
    	\muu^{(2)}
    	(\mLambda^{(2)})^{-1/2}
    	\\
    	&+
    	\mr^{(1)\top}_{\mathcal{S}}
    	\muu^{(2)}_{\mathcal{S}}
       	(\mLambda^{(2)})^{1/2}
    	+
    	(\mLambda^{(1)})^{1/2}
    	\muu^{(1)\top}_{\mathcal{S}}	
    	\mr^{(2)}_{\mathcal{S}}
    	.
\end{aligned}
$$
Note for any matrix $\mm\in\mathbb{R}^{d_1\times d_2}$, we have $\|\mm\|\leq\|\mm\|_F\leq d_1^{1/2}\|\mm\|_{2\to\infty}$. 
Then by Lemma~\ref{lemma:Xhat-XW} and \ref{lemma|E|_2|UEV|F} we have
$$
\begin{aligned}
	\|\mf\|
	&\leq |\mathcal{S}|^{1/2}
    \cdot \|\hat\mx^{(1)}_{\mathcal{S}}\mw^{(1)}-\mx^{(1)}_{\mathcal{S}}\|_{2\to\infty}
    \cdot |\mathcal{S}|^{1/2}
    \cdot \|\hat\mx^{(2)}_{\mathcal{S}}\mw^{(2)}-\mx^{(2)}_{\mathcal{S}}\|_{2\to\infty}
    \\&+\|(\mLambda^{(1)})^{-1/2}\|
    \cdot \|\muu^{(1)\top}
    	\me^{(1)\top}_{\mathcal{S}}
       	\muu^{(2)}_{\mathcal{S}}\|
    \cdot \|(\mLambda^{(2)})^{1/2}\|
    +\|(\mLambda^{(1)})^{1/2}\|
    \cdot \|\muu^{(1)\top}_{\mathcal{S}} 	
    	\me^{(2)}_{\mathcal{S}}
    	\muu^{(2)}\|
    \cdot \|(\mLambda^{(2)})^{-1/2}\|
    \\&+|\mathcal{S}|^{1/2}
    \|\mr^{(1)}_{\mathcal{S}}\|_{2\to\infty}
    \cdot|\mathcal{S}|^{1/2}
    \|\muu^{(2)}_{\mathcal{S}}\|_{2\to\infty}
    \cdot \|(\mLambda^{(2)})^{1/2}\|
    +\|(\mLambda^{(1)})^{1/2}\|
    \cdot|\mathcal{S}|^{1/2}
    \|\muu^{(1)}_{\mathcal{S}}\|_{2\to\infty}
    \cdot |\mathcal{S}|^{1/2}
    \|\mr^{(2)}_{\mathcal{S}}\|_{2\to\infty}
    \\
    &\lesssim |\mathcal{S}|^{1/2}
    \cdot n^{-1/2}\log^{1/2}n
    \cdot |\mathcal{S}|^{1/2}
    \cdot n^{-1/2}\log^{1/2}n
    +(n\rho_n)^{-1/2}
    \cdot (( \rho_{n} \log n)^{1/2}\cdot |\mathcal{S}|^{1/2}n^{-1/2})
    \cdot (n\rho_n)^{1/2}
    \\&
    +(|\mathcal{S}|^{1/2}
    \cdot n^{-1/2}(n\rho_n)^{-1/2}\log n)
    \cdot (|\mathcal{S}|^{1/2}
    \cdot n^{-1/2})
    \cdot (n\rho_n)^{1/2}
    \\&
    \lesssim |\mathcal{S}|n^{-1}\log n
    +|\mathcal{S}|^{1/2} n^{-1/2}( \rho_{n} \log n)^{1/2}
    +|\mathcal{S}| n^{-1}\log n
    \\&
    \lesssim  |\mathcal{S}|^{1/2}n^{-1} (n\rho_{n})^{1/2} \log^{1/2} n
    + |\mathcal{S}|n^{-1}\log n
\end{aligned}
$$
with high probability.
Note 
$
\sigma_{d}(\mx_{\mathcal{S}}^{(1)\top}\mx_{\mathcal{S}}^{(2)})
=\sigma_{d}(\mx_{\mathcal{S}}^{(1)\top}\mx_{\mathcal{S}}^{(1)})
=(\sigma_{d}(\mx_{\mathcal{S}}^{(1)}))^2
\gtrsim\frac{|\mathcal{S}|}{n}\lambda_d(\mpp^{(1)})
\asymp |\mathcal{S}|\rho_n
$.
Finally by Eq.~\eqref{eq:rencang} we obtain
$$
\begin{aligned}
	\|\mw^{(1)\top}\hat\mw^{(1,2)}\mw^{(2)} -\mw^{(1,2)}\|
	\lesssim \frac{\|\mf\|}{\sigma_{d}(\mx_{\mathcal{S}}^{(1)\top}\mx_{\mathcal{S}}^{(2)})}
	&\lesssim \frac{|\mathcal{S}|^{1/2}n^{-1} (n\rho_{n})^{1/2} \log^{1/2} n
    + |\mathcal{S}|n^{-1}\log n}
    {|\mathcal{S}|\rho_n}
	\\&\lesssim |\mathcal{S}|^{-1/2}(n\rho_n)^{-1/2}\log^{1/2} n
    + (n\rho_n)^{-1}\log n
\end{aligned}
$$
with high probability.

\subsection{Proof of Lemma~\ref{lemma:sigma}}

We first derive an upper bound of $\|\mw
\mathbf{\Gamma}^{(k)}
\mw^\top-\hat{\mathbf{\Gamma}}^{(k)}\|$.
Let 
$$
\begin{aligned}
	\mh_1
	:=&\mw^{(2)}(\mx^{(2)\top}\mx^{(2)})^{-1}\mx^{(2)\top}\mathbf{\Xi}^{(k,2)}\mx^{(2)}(\mx^{(2)\top}\mx^{(2)})^{-1}\mw^{(2)\top}
    -(\hat\mx^{(2)\top}\hat\mx^{(2)})^{-1}\hat\mx^{(2)\top}\hat{\mathbf{\Xi}}^{(k,2)}\hat\mx^{(2)}(\hat\mx^{(2)\top}\hat\mx^{(2)})^{-1},\\
    \mh_2
    :=&\mw^{(2)}\mw^{(1,2)\top}(\mx^{(1)\top}\mx^{(1)})^{-1}\mx^{(1)\top}\mathbf{\Xi}^{(k,1)}\mx^{(1)}(\mx^{(1)\top}\mx^{(1)})^{-1}\mw^{(1,2)}\mw^{(2)\top}
\\&-\hat\mw^{(1,2)\top}(\hat\mx^{(1)\top}\hat\mx^{(1)})^{-1}\hat\mx^{(1)\top}\hat{\mathbf{\Xi}}^{(k,1)}\hat\mx^{(1)}(\hat\mx^{(1)\top}\hat\mx^{(1)})^{-1}\hat\mw^{(1,2)}.
\end{aligned}
$$
Then $\mw
\mathbf{\Gamma}^{(k)}
\mw^\top
-\hat{\mathbf{\Gamma}}^{(k)}=\mh_1+\mh_2$. We now bound $\mh_1$ and $\mh_2$, respectively.
For $\mh_1$, we have
\begin{equation}\label{eq:H1}
	\begin{aligned}
	\|\mh_1\|
	&\leq \|\mw^{(2)}(\mx^{(2)\top}\mx^{(2)})^{-1}\mx^{(2)\top}-(\hat\mx^{(2)\top}\hat\mx^{(2)})^{-1}\hat\mx^{(2)\top}\|
	\cdot\|\mathbf{\Xi}^{(k,2)}\|
	\cdot\|\mx^{(2)}(\mx^{(2)\top}\mx^{(2)})^{-1}\|
	\\&+\|(\hat\mx^{(2)\top}\hat\mx^{(2)})^{-1}\hat\mx^{(2)\top}\|
	\cdot\|\mathbf{\Xi}^{(k,2)}\|
	\cdot\|\mx^{(2)}(\mx^{(2)\top}\mx^{(2)})^{-1}\mw^{(2)\top}-\hat\mx^{(2)}(\hat\mx^{(2)\top}\hat\mx^{(2)})^{-1}\|
	\\&+\|(\hat\mx^{(2)\top}\hat\mx^{(2)})^{-1}\hat\mx^{(2)\top}\|
	\cdot\|\mathbf{\Xi}^{(k,2)}-\hat{\mathbf{\Xi}}^{(k,2)}\|
	\cdot\|\hat\mx^{(2)}(\hat\mx^{(2)\top}\hat\mx^{(2)})^{-1}\|.
\end{aligned}
\end{equation}
We now bound the terms in Eq.~\eqref{eq:H1} respectively.
For $\mw^{(2)}(\mx^{(2)\top}\mx^{(2)})^{-1}\mx^{(2)\top}-(\hat\mx^{(2)\top}\hat\mx^{(2)})^{-1}\hat\mx^{(2)\top}$, we have
$$
\begin{aligned}
	&\mw^{(2)}(\mx^{(2)\top}\mx^{(2)})^{-1}\mx^{(2)\top}-(\hat\mx^{(2)\top}\hat\mx^{(2)})^{-1}\hat\mx^{(2)\top}\\
	&=\mw^{(2)}(\mLambda^{(2)})^{-1/2}\muu^{(2)\top}
	-(\hat\mLambda^{(2)})^{-1/2}\hat\muu^{(2)\top}\\
	&=[\mw^{(2)}(\mLambda^{(2)})^{-1/2}-(\hat\mLambda^{(2)})^{-1/2}\mw^{(2)}]\muu^{(2)\top}
	+(\hat\mLambda^{(2)})^{-1/2}\mw^{(2)}[\muu^{(2)}
	-\hat\muu^{(2)}\mw^{(2)}]^\top.
\end{aligned}
$$
Then by Lemma~\ref{lemma:|lambda|} and Lemma~\ref{lemma:lambda W} we have
\begin{equation}\label{eq:H1_1}
	\begin{aligned}
		&\|\mw^{(2)}(\mx^{(2)\top}\mx^{(2)})^{-1}\mx^{(2)\top}-(\hat\mx^{(2)\top}\hat\mx^{(2)})^{-1}\hat\mx^{(2)\top}\|\\
		&\leq \|\mw^{(2)}(\mLambda^{(2)})^{-1/2}-(\hat\mLambda^{(2)})^{-1/2}\mw^{(2)}\|
		+\|(\hat\mLambda^{(2)})^{-1/2}\|
		\cdot \|\hat\muu^{(2)}\mw^{(2)}-\muu^{(2)}\|\\
		&\lesssim n^{-1/2}(n\rho_n)^{-1}\log^{1/2}n
		+(n\rho_n)^{-1/2}\cdot (n\rho_n)^{-1/2}
		\\&\lesssim (n\rho_n)^{-1}
	\end{aligned}
\end{equation}
with high probability.
For $\mathbf{\Xi}^{(k,2)}-\hat{\mathbf{\Xi}}^{(k,2)}$, using the same arguments as that for Eq.~(C.21) in \cite{zheng2022limit} we have
\begin{equation}\label{eq:H1_2}
	\begin{aligned}
		\|\mathbf{\Xi}^{(k,i)}-\hat{\mathbf{\Xi}}^{(k,i)}\|
		\lesssim n^{-1}(n\rho_n)^{1/2}\log^{1/2}n
	\end{aligned}
\end{equation}
with high probability.
We also have
\begin{equation}\label{eq:H1_3}
	\begin{aligned}
		&\|\mathbf{\Xi}^{(k,i)}\|
		\leq \|\mpp^{(i)}\|_{\max}
		\leq \|\muu^{(i)}\|_{2\to\infty}^2
		\cdot\|\mLambda^{(i)}\|
		\lesssim n^{-1}(n\rho_n),\\
		& \|\mx^{(i)}(\mx^{(i)\top}\mx^{(i)})^{-1}\|
		\leq \|(\mLambda^{(i)})^{-1/2}\|
		\lesssim (n\rho_n)^{-1/2},\\
		& \|(\hat\mx^{(i)\top}\hat\mx^{(i)})^{-1}\hat\mx^{(i)\top}\|
		\leq \|(\hat\mLambda^{(i)})^{-1/2}\|
		\lesssim (n\rho_n)^{-1/2}
	\end{aligned}
\end{equation}
with high probability by Lemma~\ref{lemma:|lambda|}.
Combining Eq.~\eqref{eq:H1}, Eq.~\eqref{eq:H1_1}, Eq.~\eqref{eq:H1_2}, and Eq.~\eqref{eq:H1_3} we have
$$
\begin{aligned}
	\|\mh_1\|
	&\lesssim (n\rho_n)^{-1}
	\cdot n^{-1}(n\rho_n)
	\cdot (n\rho_n)^{-1/2}
	+[(n\rho_n)^{-1/2}]^{2}
	\cdot n^{-1}(n\rho_n)^{1/2}\log^{1/2}n\\
	&\lesssim n^{-1}(n\rho_n)^{-1/2}
	+n^{-1}(n\rho_n)^{-1/2}\log^{1/2}n
	\lesssim n^{-1}(n\rho_n)^{-1/2}\log^{1/2}n
\end{aligned}
$$
with high probability.

For $\mh_2$, we have
\begin{equation}\label{eq:H2}
	\begin{aligned}
		\mh_2&=\underbrace{\mw^{(2)}(\mw^{(1,2)\top}-\mw^{(2)\top}\hat\mw^{(1,2)\top}\mw^{(1)})(\mx^{(1)\top}\mx^{(1)})^{-1}\mx^{(1)\top}\mathbf{\Xi}^{(k,1)}\mx^{(1)}(\mx^{(1)\top}\mx^{(1)})^{-1}\mw^{(1,2)}\mw^{(2)\top}}_{\mh_{2,1}}
\\&
+\underbrace{\hat\mw^{(1,2)\top}\mw^{(1)}(\mx^{(1)\top}\mx^{(1)})^{-1}\mx^{(1)\top}\mathbf{\Xi}^{(k,1)}\mx^{(1)}(\mx^{(1)\top}\mx^{(1)})^{-1}(\mw^{(1,2)}-\mw^{(1)\top}\hat\mw^{(1,2)}\mw^{(2)})\mw^{(2)\top}}_{\mh_{2,2}}
\\&+\underbrace{\hat\mw^{(1,2)\top}\mw^{(1)}(\mx^{(1)\top}\mx^{(1)})^{-1}\mx^{(1)\top}\mathbf{\Xi}^{(k,1)}\mx^{(1)}(\mx^{(1)\top}\mx^{(1)})^{-1}\mw^{(1)\top}\hat\mw^{(1,2)}}_{\mh_{2,3}}
\\&+\underbrace{[-\hat\mw^{(1,2)\top}(\hat\mx^{(1)\top}\hat\mx^{(1)})^{-1}\hat\mx^{(1)\top}\hat{\mathbf{\Xi}}^{(k,1)}\hat\mx^{(1)}(\hat\mx^{(1)\top}\hat\mx^{(1)})^{-1}\hat\mw^{(1,2)}]}_{\mh_{2,4}}.
	\end{aligned}
\end{equation}
With the identical analysis of $\mh_1$ we have 
\begin{equation}\label{eq:H2_1}
	\begin{aligned}
		\|\mh_{2,3}+\mh_{2,4}\|\lesssim n^{-1}(n\rho_n)^{-1/2}\log^{1/2}n
	\end{aligned}
\end{equation}
with high probability.
For $\mh_{2,1}$ and $\mh_{2,2}$, by Lemma~\ref{lemma:WWW-W} and Eq.~\eqref{eq:H1_3} for $i=1$ we have
\begin{equation}\label{eq:H2_2}
	\begin{aligned}
		\|\mh_{2,1}\|
		&\leq \|\mw^{(1)\top}\hat\mw^{(1,2)}\mw^{(2)}-\mw^{(1,2)}\|
		\cdot \|\mx^{(1)}(\mx^{(1)\top}\mx^{(1)})^{-1}\|^2
		\cdot \|\mathbf{\Xi}^{(k,1)}\|\\
		&\lesssim [|\mathcal{S}|^{-1/2}(n\rho_n)^{-1/2}\log^{1/2}n+(n\rho_n)^{-1}\log n]
		\cdot (n\rho_n)^{-1}
		\cdot n^{-1}(n\rho_n)\\
		&\lesssim |\mathcal{S}|^{-1/2}n^{-1}(n\rho_n)^{-1/2}\log^{1/2}n+n^{-1}(n\rho_n)^{-1}\log n,\\
		\|\mh_{2,2}\|
		&\lesssim |\mathcal{S}|^{-1/2}n^{-1}(n\rho_n)^{-1/2}\log^{1/2}n+n^{-1}(n\rho_n)^{-1}\log n
	\end{aligned}
\end{equation}
with high probability.
Combining Eq.~\eqref{eq:H2}, Eq.~\eqref{eq:H2_1}, and Eq.~\eqref{eq:H2_2} we have
$$
\begin{aligned}
	\|\mh_2\|
	&\lesssim |\mathcal{S}|^{-1/2}n^{-1}(n\rho_n)^{-1/2}\log^{1/2}n+n^{-1}(n\rho_n)^{-1}\log n
	+n^{-1}(n\rho_n)^{-1/2}\log^{1/2}n\\
	&\lesssim n^{-1}(n\rho_n)^{-1/2}\log^{1/2}n
\end{aligned}
$$
with high probability.
By the above bounds for $\mh_1$ and $\mh_2$ we thus have
\begin{equation}\label{eq:WGammaW-hatGamma}
	\begin{aligned}
	\|\mw
\mathbf{\Gamma}^{(k)}
\mw^\top
-\hat{\mathbf{\Gamma}}^{(k)}\|
\lesssim n^{-1}(n\rho_n)^{-1/2}\log^{1/2}n
\end{aligned}
\end{equation}
with high probability.
Recall the assumption that $\lambda_{\ell}(\mathbf{\Gamma}^{(k)})\asymp n^{-1}$ for all $\ell\in[d]$ in Theorem~\ref{thm:CLT}. Then by Weyl's theorem and Eq.~\eqref{eq:WGammaW-hatGamma} we also have $\lambda_{\ell}(\hat{\mathbf{\Gamma}}^{(k)})\asymp n^{-1}$ for all $\ell\in[d]$ whenever {\color{black}$n\rho_n\gg \log n$}.
Therefore we have
\begin{equation}\label{eq:A-1,B-1}
	\begin{aligned}
		\|\mw
(\mathbf{\Gamma}^{(k)})^{-1}
\mw^\top\|\lesssim n,\quad\|(\hat{\mathbf{\Gamma}}^{(k)})^{-1}\|
\lesssim n
	\end{aligned}
\end{equation}
with high probability.
Since $\|\ma^{-1}-\mb^{-1}\|\leq \|\ma^{-1}\|\cdot\|\ma-\mb\|\cdot\|\mb^{-1}\|$ for any invertible matrices $\ma$ and $\mb$, we have by Eq.~\eqref{eq:WGammaW-hatGamma} and Eq.~\eqref{eq:A-1,B-1} that
$$\|\mw
(\mathbf{\Gamma}^{(k)})^{-1}
\mw^\top-(\hat{\mathbf{\Gamma}}^{(k)})^{-1}\|\lesssim n (n\rho_n)^{-1/2}\log^{1/2} n
$$
with high probability.

\subsection{Results with weaker assumptions}

\label{sec:weaker}

In \ref{assum:main}, we assume a lower bound on $\lambda_{d}(\mpp^{(1)}_{\mathcal{S},\mathcal{S}})$ for the seed set to ensure that the estimated orthogonal transformation $\hat{\mw}^{(1,2)}$, based on the latent positions of the seed set, is sufficiently accurate for the true orthogonal transformation between the two networks $\mw^{(1,2)}$; see the proof of Lemma~\ref{lemma:WWW-W} for details. 
Specifically, we assume the lower bound as $\lambda_{d}(\mpp^{(1)}_{\mathcal{S},\mathcal{S}}) \gtrsim \frac{|\mathcal{S}|}{n} \lambda_{d}(\mpp^{(1)})\gtrsim \frac{|\mathcal{S}|}{n}n\rho_n\gtrsim |\mathcal{S}|\rho_n$, as it holds with high probability when $\mathcal{S}$ is drawn uniformly, which is also consistent with the seed set generation method in Algorithm~\ref{alg:without seeds}. This gives us an intuition for how large the seed size $|\mathcal{S}|$ and the average degree of the network $n \rho_n$ need to be in a general scenario.

We now present results without this assumption by directly addressing $\lambda_{d}(\mpp^{(1)}_{\mathcal{S},\mathcal{S}})$. Using the same arguments in the proof of Lemma~\ref{lemma:WWW-W}, we have
\[
\begin{aligned}
	\|\mw^{(1)\top}\hat\mw^{(1,2)}\mw^{(2)} - \mw^{(1,2)}\|
	\lesssim \frac{|\mathcal{S}|^{1/2} n^{-1} (n \rho_{n})^{1/2} \log^{1/2} n
    + |\mathcal{S}| n^{-1} \log n}
    {\lambda_{d}(\mpp^{(1)}_{\mathcal{S},\mathcal{S}})}
\end{aligned}
\]
with high probability. It follows that, in Theorem~\ref{thm:X1W-X2}, the bound for the remainder becomes
\[
\begin{aligned}
	\|\mr\|_{2\to\infty}
	&\lesssim n^{-1/2}(n \rho_n)^{-1/2} \log n
	+ \lambda_{d}^{-1}(\mpp^{(1)}_{\mathcal{S},\mathcal{S}})
	|\mathcal{S}|^{1/2} n^{-3/2} (n \rho_{n}) \log^{1/2} n \\
	& + \lambda_{d}^{-1}(\mpp^{(1)}_{\mathcal{S},\mathcal{S}})
    |\mathcal{S}| n^{-3/2} (n \rho_n)^{1/2} \log n
\end{aligned}
\]
with high probability. 
Then the error bound $\|\hat{\my} \mw - \my\|_{2\to\infty} \lesssim n^{-1/2} \log^{1/2} n$ is still valid when
\[
\lambda_{d}(\mpp^{(1)}_{\mathcal{S},\mathcal{S}}) = \Omega\big(\max\{|\mathcal{S}|^{1/2} \rho_n, |\mathcal{S}| n^{-1} (n \rho_n)^{1/2} \log^{1/2} n\}\big),
\]
and the condition for the normal approximations in Theorem~\ref{thm:CLT} and the test in Theorem~\ref{thm:HT} becomes $n\rho_n = \omega(\log^2 n)$ and
\[
\lambda_{d}(\mpp^{(1)}_{\mathcal{S},\mathcal{S}})
= \omega\big(\max\{|\mathcal{S}|^{1/2} n^{-1} (n \rho_n) \log^{1/2} n,
|\mathcal{S}| n^{-1} (n \rho_n)^{1/2} \log n\}\big).
\]

\end{sloppypar}

\end{document}